\documentclass[aps, onecolumn, notitlepage, floatfix, nofootinbib, amsmath, letterpaper,amssymb, showpacs,superscriptaddress, preprintnumbers]{revtex4-1} 
\pdfoutput=1
\usepackage{graphicx}
\usepackage{amssymb}
\usepackage{epstopdf}
\usepackage{slashed}
\usepackage{amsmath}
\usepackage{enumerate}

\usepackage{url}
\usepackage{color}
\usepackage{listings}

\newcommand{\ignore}[1]{}
\newcommand{\be}{\begin{equation}} \newcommand{\ee}{\end{equation}}
\newcommand{\ba}{\begin{eqnarray}} \newcommand{\ea}{\end{eqnarray}}
\newcommand{\nn}{\nonumber} \renewcommand{\bf}{\textbf}

\newcommand{\MET}{\sl{E_T}\,\,\,}
\newcommand{\ASV}{A_{t\bar t}^{\rm SV} }

\newcommand{\ADTT}{ATLAS-$d_{12}$ tagger}
\newcommand{\cOO}{\mathcal{O}}
\newcommand{\GeV}{{\rm\ GeV}}

\newcommand{\TeV}{{\rm\ TeV}}

\definecolor{red1}{cmyk}{0,1,1,0.1}

\def\slashb#1{\setbox0=\hbox{$#1$}#1\hskip-\wd0\dimen0=5pt\advance
        \dimen0 by-\ht0\advance\dimen0 by\dp0\lower0.5\dimen0\hbox
          to\wd0{\hss\sl/\/\hss}}

\def\sl#1{#1 \!\!\! \!\!   \!\! \slash}
\input{epsf}

\begin{document}

\preprint{\scriptsize CERN-PH-TH/2013-260\vspace*{.1cm}}
\preprint{\scriptsize SISSA 52/2013/FISI}

\title{Measuring Boosted Tops in Semi-leptonic $t\bar{t}$  Events \\ for the Standard Model and Beyond}

\author{Mihailo Backovi\'{c}} \email{mihailo.backovic@weizmann.ac.il} \affiliation{Department of Particle Physics and Astrophysics, \\ Weizmann Institute of Science, Rehovot 76100, Israel} 
\author{Ofir Gabizon} \email{ofir.gabizon@weizmann.ac.il} \affiliation{Department of Particle Physics and Astrophysics, \\ Weizmann Institute of Science, Rehovot 76100, Israel} 
\author{Jos\'e Juknevich} \email{jose.juknevich@sissa.it} \affiliation{Department of Particle Physics and Astrophysics, \\ Weizmann Institute of Science, Rehovot 76100, Israel}  \affiliation{SISSA/ISAS,  I-34136 Trieste, Italy}
\author{Gilad Perez}\email{gilad.perez@weizmann.ac.il} \affiliation{Department of Particle Physics and Astrophysics, \\ Weizmann Institute of Science, Rehovot 76100, Israel} \affiliation{CERN, Theory Division, CH1211 Geneva 23, Switzerland}
\author{Yotam Soreq} \email{yotam.soreq@weizmann.ac.il} \affiliation{Department of Particle Physics and Astrophysics, \\ Weizmann Institute of Science, Rehovot 76100, Israel} 

\begin{abstract}

We present a procedure for tagging  boosted semi-leptonic $t\bar{t}$ events based on the Template Overlap Method. We introduce a new formulation of the template overlap function for leptonically decaying boosted tops and show that it can be used to compensate for the loss of background rejection due to reduction of $b$-tagging efficiency at high $p_T$. A study of asymmetric top pair production due to higher order effects shows that our approach improves the resolution of  the truth level kinematic distributions. We show that the hadronic top overlap is weakly susceptible to pileup up to 50 interactions per bunch crossing, while leptonic overlap remains impervious to pileup to at least 70 interactions. A case study of Randall-Sundrum Kaluza-Klein gluon production suggests that the new formulation of semi-leptonic template overlap can extend the projected exclusion of the LHC  $\sqrt{s}= 8 \TeV$ run to Kaluza-Klein gluon masses of $2.7 \TeV$, using the leading order signal cross section.

\end{abstract}

\maketitle

\section{Introduction}

Boosted massive jets are becoming increasingly important at the LHC, both in searches for new physics (NP) and in the Standard Model (SM) 
related measurements. With the LHC pushing the energy frontier forward, and 
no physics beyond the SM showing up at $\cOO (1 \TeV),$ the experimental searches are entering 
a kinematic regime in which a significant fraction of heavy SM particles are produced 
at ultra high $p_T$. Of particular interest are boosted top quarks, as many models of 
new physics that address the hierarchy problem predict  resonances or top partners with mass of $\cOO(1 \TeV)$ and a 
large decay rate to top quark pairs~(see for example Refs.~\cite{Harris:1999ya,Hill:1993hs,Agashe:2006hk,Lillie:2007ve,Lillie:2007yh,Contino:2008hi,Contino:2006qr}).
Boosted top quarks are also significant for the measurements of 
the SM differential cross sections at high transverse momentum or at high invariant $t\bar t$ mass, as well as precision measurements 
of the total top production cross section.

Traditional jet reconstruction techniques are inadequate to fully describe the decays 
of heavy boosted objects at high transverse momentum. Small angular scales in the lab frame, 
which characterize the decays of massive boosted particles, make it difficult to distinguish them from the background of light parton QCD jets 
or electro-weak events using only jet mass and $p_T$. Additional information about energy distribution within the jet,  
commonly referred to as jet substructure, allows for more efficient identification of heavy boosted objects. 
The leading order (LO) three prong decay structure of a boosted top and the correlations therein can be employed to distinguish 
top quark jets from, say,  light parton QCD jets, which typically have a two prong topology. 

A myriad of techniques to reconstruct and identify boosted massive jets have been developed over the 
past decade (see {\it e.g.} Refs.~\cite{Ellis:2007ib,Abdesselam:2010pt,Salam:2009jx,Nath:2010zj,Almeida:2011ud,Plehn:2011tg,Altheimer:2012mn,Soper:2011cr, Soper:2012pb} for reviews),
many of which can be grouped into two broad categories. The first category employs jet shape observables to probe the energy flow inside jets. 
 These include the angular correlation functions studied in Ref.~\cite{Jankowiak:2011qa}, as well as the sphericity tensors or planar flow of Refs.~\cite{Bjorken:1969wi, Thaler:2008ju,Almeida:2008yp}.

The second category makes use of the fact that the signal events are coming from decay of genuine massive particles and are thus characterized by spikes of energy which to leading order would correspond to massive particles daughter products. 
This category can be broken down to methods that incorporate
Filtering~\cite{Butterworth:2008iy} (see also Refs.~\cite{Krohn:2009th,Ellis:2009me}) and the Template Overlap Method~\cite{Almeida:2010pa}. 
Filtering algorithms act on the list of jet constituents by removing the soft components based on some measure which defines the ``hard'' part of the jet. The remaining constituents are then reclustered into the ``filtered'' jet. 
 The Template Overlap Method, to be further exploited in this paper, does not manipulate the jet constituent list, nor does it require a special clustering algorithm for substructure analysis. Instead, the method compares the jet to a set of parton level states built according to
a fixed-order distribution of signal jets called templates.  The comparison makes use of an ``overlap
function'' which evaluates the level of agreement between each measured jet and a set of templates.

Methods that employ elements from both categories or use other ingredients also exist. Jet dipolarity~\cite{Hook:2011cq} and N-subjettiness~\cite{Thaler:2010tr,Thaler:2011gf} are examples of hybrid jet shapes which study jet energy flow with
respect to directions of candidate subjets identified by the above mentioned techniques. More recently, shower deconstruction method~\cite{Soper:2011cr,Soper:2012pb} appeared as a variant of the matrix element method~\cite{Abazov:2004cs,Aaltonen:2010pe,Artoisenet:2010cn} to classify jets with the help of approximations to hard matrix elements and the parton shower. 

 Here we discuss the performance of the Template Overlap Method (TOM) as a tagger of semi-leptonic $t\bar{t}$ events. 
 TOM is a jet substructure tool which aims to match the energy distribution of a jet to a parton level structure of heavy particle decay, with all the relevant kinematic constraints. Reference~\cite{Almeida:2010pa} showed that TOM is a powerful hadronic top tagger, with rejection power over QCD background of $\cOO(100)$ being possible when the method is used in conjunction with other jet substructure correlations.  A consequent ATLAS study of Ref.~\cite{Aad:2012qa} validated TOM in experimental conditions at $\sqrt{s} = 7 \TeV$ in the fully hadronic channel. The results were used to set a useful bound on a Randall-Sundrum (RS)~\cite{Randall:1999ee} Kaluza-Klein (KK) gluon mass.
 
References~\cite{Almeida:2011aa, Backovic:2012jj} also studied TOM in the context of boosted Higgs decays to $b\bar{b}$. The results showed that a combination of leading order (two-body) and beyond leading order (three-body) template analysis could significantly improve the signal to background ratio, at a cost to signal efficiency. Reference~\cite{Backovic:2012jj} also demonstrated that TOM is robust against pileup contamination. The study with twenty interactions per bunch crossing showed that pileup has little effect on TOM, with impact on rejection power being a $10 \%$ effect, thus reducing the need for pileup correction or subtraction.

In our current work, we consider the task of boosted top tagging in two ways. On one hand we are interested in quantifying the capacity of TOM to both tag and measure boosted tops. The ability to efficiently tag boosted tops and improve the top sample purity is important both in measurements of the $t\bar{t}$ differential distributions as well as in discriminating the non-top backgrounds in BSM physics searches. 
On the other hand, we study the ability of TOM to determine whether a $t\bar{t}$ event as a whole originated from an interesting BSM signal (\textit{e.g}. an $s$-channel resonance that decays to $t\bar{t}$), or whether it came from SM. Tagging the boosted $t\bar{t}$ event as a whole is of great importance in searches for BSM physics, as it provides a discriminant of SM $t\bar{t}$ events which would otherwise be considered an irreducible background. 
To that end,
we introduce a formulation of TOM for top decays with large missing energy, whereby we derive the ``leptonic top overlap,'' $Ov_3^{lep}$, from the standard definition of the peak template overlap. The leptonic overlap function differs from the hadronic top overlap is two ways. First, $Ov_3^{lep}$ requires to keep track of the identities of template partons, while the identical template sets can be used both for hadronic and leptonic top analysis.  Second, since only the transverse component of missing energy is available, we define the neutrino overlap function only in the transverse plane. 

In addition to extending the TOM algorithm, we address several challenges relevant to jet substructure studies both with the recent $\sqrt{s} = 8 \TeV$  LHC data, as well as the future runs. First, we are interested in the ability of TOM to accurately resolve the kinematic parameters of boosted tops.  At high energies, the top pairs are often not produced back to back. 
Higher order effects become prominent at high $p_T$, with gluon splitting to $t\bar{t}$ and hard gluon emission becoming non-negligible contributions to the total $t\bar{t}$ cross section. We show that TOM is able to distinguish back to back $t\bar{t}$ events from configurations in which the hadronic top does not recoil against the leptonically decaying top. As an illustration, we show that the resulting $p_T$ resolution of the top jet improves compared to the \ADTT~\cite{Aad:2013ttr}. 
The ability to reject the ``asymmetric''  $t\bar{t}$ events comes with the additional benefit of improving the signal to background ratio in heavy resonance searches, where we expect the fraction of asymmetric top events coming from new physics to be significantly lower than for SM tops. 

Next, we study the capability of TOM to tag SM semi-leptonic $t\bar{t}$ events and reject the relevant backgrounds over a wide range of fat jet $p_T$.  For the purpose of measuring the $t \bar{t}$ system within the SM, our main background channel consists of $W+\rm jets$ events, while multijet QCD background does not contribute significantly after requiring that a ``mini-isolated'' lepton exists~\cite{Rehermann:2010vq}. In the later sections which deal with BSM searches, we also consider the SM $t \bar{t}$ channel as one of the dominant backgrounds. We consider data from both MadGraph/MadEvent~\cite{Maltoni:2002qb} showered with Pythia~\cite{Sjostrand:2006za}, and Sherpa~\cite{Gleisberg:2008ta} to illustrate the effects of different showering algorithms and matching procedures on the analysis. For the signal we also provide comparison with the next to leading order (NLO) results from  POWHEG~\cite{Frixione:2007nw}.
Our analysis shows that hadronic template overlap, $Ov_3^{had}$, properly tags about 10 signal jets  for every 1 fake event at $60 \%$ top tagging efficiency and $p_T   \sim 500 \GeV$, with no additional cuts on the jet mass or $b$-tagging. The ability of $Ov_3^{had}$ to reject background events slowly decreases with $p_T$, due to the higher order effects becoming more prominent. Adding $Ov_3^{lep}$ to our analysis proves to be rewarding, as leptonic overlap's potential to reject background events can compensate for the reduction of $b$-tagging efficiency at high $p_T$ (assuming a tentative $b$-tagging efficiency of $50 \%$).

Pileup and underlying event provide much nuisance for jet substructure observables, and here we extend the study of the TOM's susceptibility to pileup contamination. In order to reduce the difficulties of estimating the fat jet $p_T$ in a pileup environment, we introduce a method of selecting template $p_T$ bins based on the scalar sum of the leptonic top decay products and the kinematics of top pair events. 
In accord with the study in Ref.~\cite{Backovic:2012jj}, we show that TOM is only mildly sensitive to pileup up to $\sim 50$ interactions per bunch crossing. At higher levels of pileup (i.e. $ \sim 70$ interactions per bunch crossing), the signal tagging remains unaffected, whereas the increase in amount of fake events becomes important.

A case study on the discovery potential of an RS KK gluon serves to illustrate the performance of semi-leptonic TOM in new physics searches. We analyze the common benchmark KK gluon model, which features a large coupling to $t\bar{t}$, in order to typify a resonance search, while an effective theory serves to illustrate the performance of TOM in searches where the signal $m_{t\bar{t}}$ distribution is characterized by depletion of the $t\bar t$ spectrum from its SM expectation. We show that a combination of $Ov_3^{had}$ and $Ov_3^{lep}$ can improve the analysis of the 20 ${\rm fb}^{-1}$ of data collected during the $\sqrt{s} = 8 \TeV$ run, extending the projected limits to KK gluon masses of $\approx 2.7 \TeV$.

Finally, we discuss technical aspects of TOM relevant for the experimental implementation of the method both for SM and NP measurements. 
Subjets of highly boosted tops ($p_T \sim 1 \TeV$) are characterized by sizable differences between the $p_T$ of the hardest and softest partons  (typically greater than a hundred GeV). In order to adequately capture the radiation pattern of all three leading subjets over a wide range of fat jet $p_T$, while at the same time not affecting the shape of the peak overlap distribution, we vary the template sub-cones radii according to their $p_T$. The variation is inspired by the jet-shape data~\cite{Aad:2011kq}, and we follow the scaling rule for the sub-cone radii from the boosted Higgs study in Ref.~\cite{Backovic:2012jj}. This results in a stable signal efficiency for a fixed cut on template overlap over a wide range of top jet $p_T$. A comparison with several values of fixed template sub-cone radii reveals a non-trivial fact that no single fixed radius provides stable signal efficiency for a fixed overlap cut.  

We further show that missing energy resolution has little effect on the results of the overlap analysis, as well as demonstrate that TOM is insensitive to the angular resolution of the template momenta, with 50 steps in $\eta, \phi$ and beyond providing adequate template phase space coverage.

We organized the paper in seven sections addressing the above-mentioned novelties and issues. In Section~\ref{Sec:Ov} we define the hadronic top template overlap following the longitudinally boost invariant notation of Ref.~\cite{Backovic:2012jj}, as well as introduce the leptonic top template overlap. Section~\ref{sec:preselection} addresses our data generation and describes the pre-selection cuts we use to define our data sets. In Section~\ref{sec:NLO} we address the issues of higher order effects and the ability of TOM to reject asymmetric $t \bar{t}$ events. In Section~\ref{sec:RP} we present our results on the rejection power of TOM for SM $t\bar{t}$ events over a wide range of fat jet $p_T$ values for both hadronic and leptonic overlap. Section~\ref{sec:pileup} is dedicated to pileup studies, ranging from $0-70$ interactions per bunch crossing. Finally, Section~\ref{sec:BSM} shows an example study of a search for new physics in a $t\bar{t}$ channel and illustrates the improvements TOM can provide for the analysis. The technical details of Template Overlap, such as the adequate number of templates, effects of missing energy resolution and template sub-cone scaling can be found in the Appendix.

\section{template overlap Method} \label{Sec:Ov}

\subsection{Hadronic Top Template Overlap}

Following the notation of Ref.~\cite{Backovic:2012jj}, here we consider the definition of \textit{hadronic peak template overlap} in terms of longitudinally boost invariant quantities: 
\begin{equation}
 Ov_3^{had} = \max_{\{f\}}\,\left\{ \exp\left[ -\sum_{a=1}^N \frac{1}{\sigma_a^2}\left( \epsilon \,  p_{T,a} -\sum_{i\in j} p_{T,i} \,F(\hat n_i,\hat n_a) \right)^2 \right] \right\} \, , \label{eq:ovHad}
\end{equation}
where $p_{T,a}$ is the transverse momentum of the $a^{th}$ template parton and $p_{T, i}$ is the transverse momentum of the $i^{th}$ jet constituent. The functional is maximized over ${f}$, a set of kinematically allowed decay configurations of the boosted top (templates). The weight $\sigma_a$ defines the energy resolution of the peak template overlap which we set to $1/3 p_{T,a}$, while the coefficient $\epsilon = 0.8$ serves to compensate for the radiation which falls outside the template sub-cones.  We define the kernel $F(\hat n_i,\hat n_a)$ as a step function
\begin{equation}
 F(\hat n_i,\hat n_a) = \left\{ \begin{array}{rl}
 1 &\mbox{ if $ \Delta R < r_a$} \\
  0 &\mbox{otherwise}
       \end{array} \right. \, , \label{eq:kernel}
\end{equation}
where $\hat{n}_{i,a}$ is the position vector of a jet constituent ($i$) or template parton ($a$) in the $\eta$, $\phi$ space,  and $\Delta R$ is the plain distance in $\eta, \phi$ between the $i^{th}$ jet constituent and the $a^{th}$ template parton.  We determine the size of the template sub-cone, $r_a$, according to a polynomial fit to the scaling rule of Ref.~\cite{Backovic:2012jj} (see Appendix~\ref{sec:temp_scaling} for details), in addition to requiring that the template partons be isolated such that 
\be
	\Delta R_{ab} > r_a + r_b\,, \label{eq:Rab}
\ee
for any two template partons $a$ and $b$.

\subsection{Leptonic Top Template Overlap}

So far, TOM has only been discussed in the context of fully hadronic decays of massive objects. It is also possible to define template overlap on heavy particle decays with missing energy such as the leptonically decaying boosted top~\footnote{Leptonic Overlap can be used both on muons and electrons with no loss of generality.}. The missing information about the longitudinal component of the missing energy makes the ``canonical'' overlap function definition of Eq.~\eqref{eq:ovHad} inappropriate to describe a leptonic top decay. We begin instead by defining the leptonic three body overlap function, $Ov_3^{lep}$, as a product of the overlap functions for the $b$ jet, the lepton and the neutrino: 
\begin{equation}
 Ov_3^{lep} =\max_{\{f\}}\, \left\{\exp\left[ \underbrace{  -\frac{1}{\sigma_b^2}\left( \epsilon \,  k_{T,b} -\sum_{i\in j} p_{T,i} \,F(\hat n_i,\hat n_a) \right)^2  }_{\text{b quark}} 
 		\,\, \underbrace{ -\frac{1}{\sigma_l^2}\left( \epsilon_l \,  k_{T,\ell} -p_{T,\ell}  \right)^2  }_{\text{lepton}}
		\,\, \underbrace{ -\frac{1}{\sigma_\nu^2}\left( \epsilon_\nu \,  k_{T,\nu} - \sl{E_T} \,\,\,F'(\phi_\nu, \phi_{\sl{E_T}}\,\,\,) \right)^2} _{\text{neutrino}}   \right]\right\} \, . \label{eq:lep_overlap}
\end{equation}
The first exponential in Eq.~\eqref{eq:lep_overlap} is the familiar overlap function of Eq.~\eqref{eq:ovHad} for a single template parton, the second exponential refers to the lepton, while the third exponential is associated with missing energy. We introduce coefficients $\epsilon_i$ to include effects of energy reconstruction efficiency of the top decay product as in the case of $Ov_3^{had}$. Other than $\epsilon_b = 0.8$, here we use $\epsilon_\ell = \epsilon_\nu = 1$. 
We also find that $\sigma_{b, \ell, \nu} = 1 / 3 k_T^{b, \ell, \nu}$ provides sufficient background rejection, 
while keeping the signal efficiency comparable to the fully hadronic case.\footnote{Notice that as the detector-level corrections to the lepton energy scale much smaller than the above width chosen for the template, one can in principle improve upon the above definition by reducing $\sigma_\ell$. However, as qualitatively we expect that the template resolution would be controlled by the missing energy resolution which is much worse than the leptonic one, this potential modifications of the leptonic template and its implications is left for future work.}
The optimization of overlap parameters is relatively straightforward, however it 
requires experimental input which is beyond the scope of our current work.

The maximization in Eq.~\eqref{eq:lep_overlap} is performed over a full set of templates, in the same fashion as $Ov_3^{had}$, and with the same sets of templates.

We keep the kernel function $F$ for the $b$ template the same as in Eq.~\eqref{eq:ovHad}, while we define the neutrino kernel as
\begin{equation}
F'(\phi_\nu, \phi_{\sl{E_T}}\,\,\,) = \left\{ \begin{array}{rl}
 1 &\mbox{ if $ \Delta \phi_{\nu, \sl{E_T}} < r_\nu$} \\
  0 &\mbox{otherwise}
       \end{array} \right. \, ,
\end{equation}
 where $\Delta \phi$ is the azimuthal distance between the template parton and the total $\sl{E_T}\,\,\,$, and
 $r_\nu =0.2 $ is the neutrino azimuthal bin size.

The main difference between $Ov_3^{lep}$ and $Ov_3^{had}$ is that leptonic overlap takes into account only the azimuthal component of missing energy. Since our overlap algorithm requires us to rotate the templates into the fat jet frame on an event by event basis, the absence of the longitudinal component of missing energy does not allow for a good enough reconstruction of the top axis. We choose instead to rotate the templates so that the second template parton is always aligned with the lepton, the first template is always the neutrino and the third template is the $b$-quark. Anchoring template states to the lepton also eliminates the need for a lepton kernel function. In addition, the fact that leptonic overlap deals with three different species of particles forces us to keep track of the identities of template partons on a template by template basis, a requirement which is absent in the case of the fully hadronic overlap. Since the identities of reconstructed objects are matched to the identities of template partons, we also do not impose the non-overlapping template subcone criteria of Eq.~\eqref{eq:Rab} for $Ov_3^{lep}$.

\section{Event Generation and Pre-Selection} \label{sec:preselection}

Before we begin to discuss the performance of TOM in a semi-realistic experimental setting, 
we take a moment to define our samples and the kinematic constraints we use in the event pre-selection. 
Our current analysis focuses on tagging the semi-leptonic $t\bar{t}$ events and rejecting the $W+\rm jets$ background at $\sqrt{s} = 8 \TeV$. 
Samples from MadGraph~/MadEvent~\cite{Maltoni:2002qb}~v.1.5.3 + Pythia~\cite{Sjostrand:2006za}~v.~6.426 (with MLM matching~\cite{Mangano:2006rw} to one extra jet), 
 Sherpa~\cite{Gleisberg:2008ta}~v.1.4.3 (with CKKW matching~\cite{Hoeche:2009rj} to one extra jet) and POWHEG~\cite{Alioli:2010xd,Frixione:2007nw} for the signal events, serve to illustrate the effects of different showering algorithms and matching procedures on the analysis. If not stated otherwise, the generated samples do not include pileup. In Section~\ref{sec:pileup} we simulate  pileup by overlaying a Poisson distributed random number of minimum-bias events from Pythia on top of our signal and background events.
We use the default tunes of Sherpa and Pythia for the hadronization and underlying event model parameters, 
while for the matching scale we use $Q_{\rm cut} = 30 \GeV$ in both cases.\footnote{For the parton separation parameter of the MLM matching procedure in MadGraph, 
we use $ \texttt{xqcut}= 20 \GeV$.}

The MadGraph generated samples serve as a benchmark dataset in all sections, with the exception of Section~\ref{sec:NLO}
where, as mentioned, we use samples generated with  the BOX~\cite{Alioli:2010xd} version of the POWHEG-hvq~\cite{Frixione:2007nw}
in order to capture the NLO effects in top quark pair production more accurately. 
Our MadGraph and Sherpa samples assume the {\sc CTEQ6L1}~\cite{Nadolsky:2008zw} parton distribution function sets while for  POWHEG we use {\sc CTEQ6M}.  We perform jet clustering using the {\sc Fastjet}~\cite{Cacciari:2011ma}
implementation of the anti-$k_T$ algorithm~\cite{Cacciari:2008gp}.

Our event selection begins with a requirement of exactly one lepton with 
\ba
{\rm mini-ISO} \equiv \frac{p_T^\ell}{p_T^{\rm cone}} > 0.95\, ,  \label{eq:lepSel} 
\ea 
where mini-ISO is the lepton isolation observable of Ref.~\cite{Kaplan:2008ie} and $p_T^{\rm cone} $ is scalar sum of all the charged tracks with $p_T>1\,$GeV, including the hard lepton, inside a cone of radius 
\be
r^{\rm mini-ISO} = \frac{10{\rm \, GeV}}{p_T^{\ell}}\,, 
\ee
where we used the scaling convention of Ref.~\cite{Aad:2013nca}. We label this lepton as coming from the leptonically decaying top. 
Next, we define the hardest anti-$k_T$ $r = 0.4$ jet within a distance $\Delta R_{j\ell} < 1.5$ from the lepton 
as the $b$-jet of the leptonically decaying top. For the purpose of this analysis we define a transverse missing energy vector 
to be the vector sum of all the neutrino transverse momenta in the event, while we postpone a detailed study of the effects of  $\sl{E_T} \,\,\,$ resolution 
until Appendix~\ref{app:MET}.

We identify the hardest anti-$k_T$ ``fat'' jet using three different large effective cone sizes $R$, defined on an event-by-event basis as
\begin{equation}
	R = \left\{ \begin{array}{ll} 1.0, &\,\,\,\,\, 500 \GeV <h_T \le 700 \GeV\\ 
						0.8, & \,\,\,\,\, 700 \GeV < h_T  \le 900 \GeV \\
						0.6, & \,\,\,\,\, 900 \GeV< h_T
						 \end{array} \right. \, ,  \label{eq:Rscale}
\end{equation}
where $h_T$ is the scalar $p_T$ sum of the leptonic top decay products,
\be
	h_T = \sum_{i = \ell, b, \nu} p^i_T\, , \label{eq:ht0}
\ee
and it serves as an estimator of the top fat jet $p_T$ with a weak susceptibility to pileup contamination. We find that for fat jet $p_T > 500 \GeV$, the $p_T$ of the jet is well correlated with $h_T$ of the leptonic top. For more details on the criteria for correlating the fat jet parameters with the leptonically decaying top see Appendix~\ref{app:HTCorr}, while we present a detailed discussion of the NLO effects on the correlation in Section~\ref{sec:NLO}.

Continuing, including the previous requirements, namely the mini-isolated lepton and the $\Delta R_{j\ell}<1.5$, all events are subject to following \textit{Basic Cuts} (BC):
\ba
	p_T^{j\, R} > 500 \GeV   \,\,\,\,   &     \sl{E_T}\,\,\, > 40 \GeV \nn \\
	N_\ell^{\rm out} (p_T^\ell > 25 \GeV) =1             \,\,\,\,\,\,\,\,& N_j^{\rm out} (p_T^j > 25 \GeV) \ge 1 \nn \\
	\Delta \phi_{j\ell} > 2.3  \,\,\,\,\,\,\, & |\eta_{j, \,\ell}| < 2.5\,,
	\label{eq:bc} 
\ea
where $p_T^{j\, R}$ is the transverse momentum of the fat jet with radius $R$ and $N_j^{\rm out}$ is the number of $r=0.4$ jets with $\Delta R_{j\ell} < 1.5\,$. $N_\ell^{\rm out}$ refers to the lepton with selection criteria of Eq.~\eqref{eq:lepSel} in addition to $p_T > 25 \GeV$, $\Delta \phi_{j\ell}$ is the azimuthal distance between the mini-isolated lepton and the fat jet, and $\eta_{j,\,\ell} $ is the rapidity of the fat jet/ mini-isolated lepton.

Here we only consider $W$+jets as the dominant background to the semileptonic $t\bar{t}$ events at high transverse momentum. 
The multijet QCD contribution becomes negligible upon the mini-isolation requirement on the lepton (see Ref.~\cite{Kaplan:2008ie} for instance), while the single top cross section is already highly sub-leading compared to $t\bar{t}$ at pre-selection level~\cite{TheATLAScollaboration:2013kha}.

For the overlap analysis in the following sections of this paper (both in the context of hadronically and leptonically decaying tops) we use the TOM implementation of the TemplateTagger code \cite{Backovic:2012jk}.

\section{Asymmetric $t \bar{t}$ Production from Higher Order Corrections} \label{sec:NLO}

Standard Model top pair production involves non-trivial kinematic configurations of the final states which go beyond the simple back-to-back top quark pairs topologies.  At LO in perturbation theory, the top and the anti-top are produced back to back with equal $p_T$. Yet, high energy events often involve extra hard radiation as well as a non-negligible gluon splitting function to heavy flavors (i.e. NLO effects), all of which can result in an imbalance between the transverse momenta of the $t$ and  $\bar{t}$. 

 \begin{figure}[tb]
\begin{center}
\begin{tabular}{ccc}
\includegraphics[width=2in]{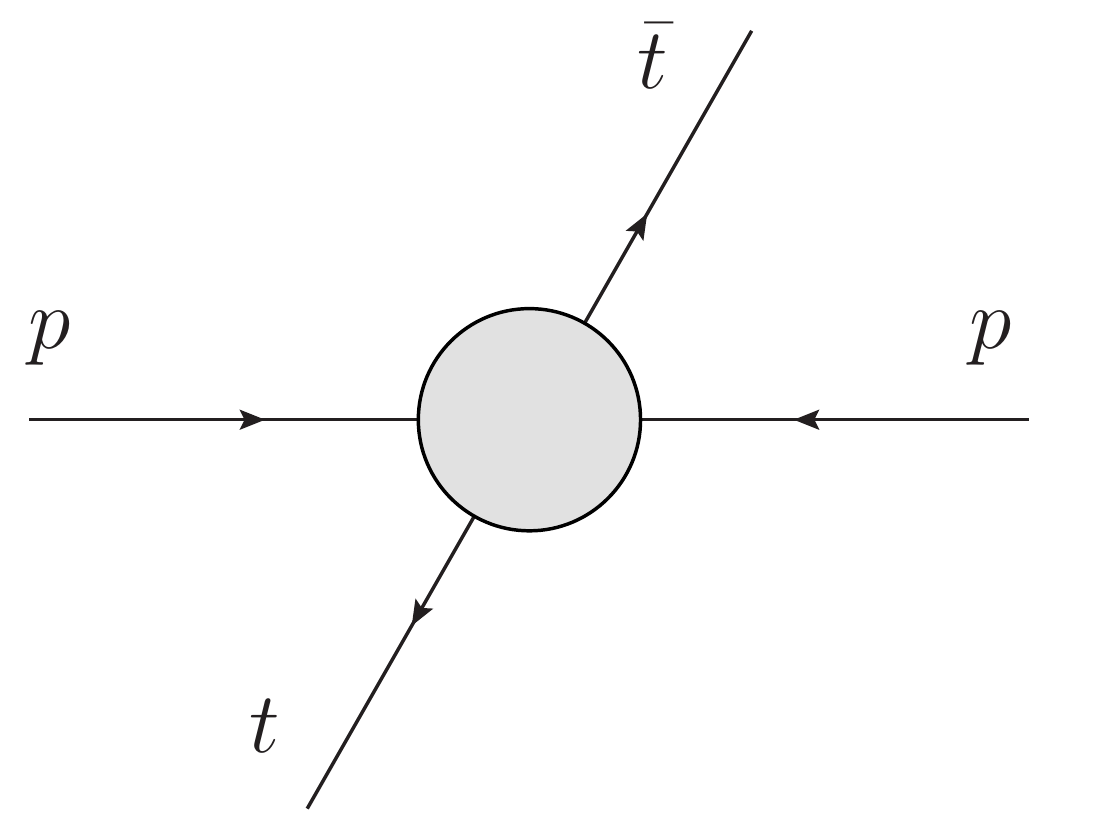} \,\,\,\,\,\,\,\,\,\,\,\,\,\,\,\,&
\includegraphics[width=2in]{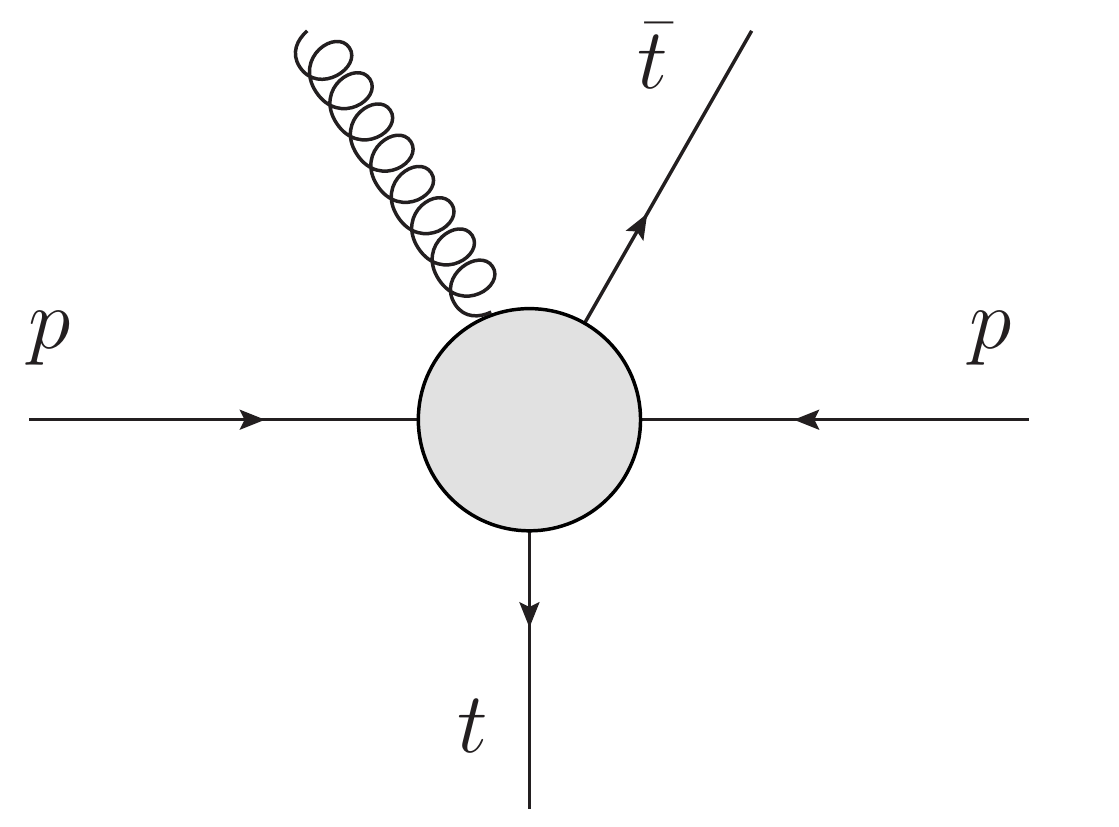} \,\,\,\,\,\,\,\,\,\,\,\,\,\,\,\,\,&
\includegraphics[width=2in]{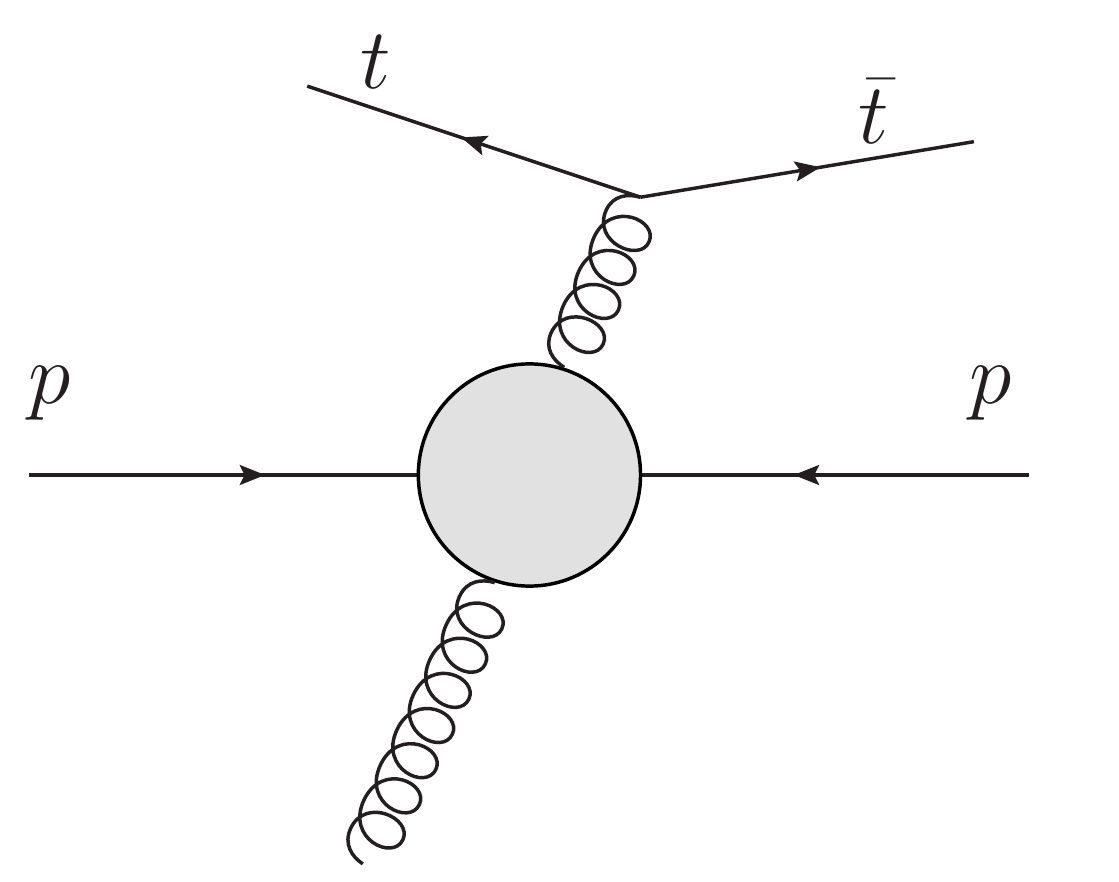} \nonumber \\
(i) \,\,\,\,\,\,\,\,& (ii) \,\,\,\,\,\,\,\,& (iii)
\end{tabular}
\caption{Three categories of $t\bar{t}$ events.}
\label{fig:feyn_diags}
\end{center}
\end{figure}

The fraction of events in which the top-antitop system
is not back to back is not only significant but increases with the $H_T$ of the event (here we define $H_T = \sum_j p_T^j,$ 
where $j$ runs over all final state particles in the event). The effect leads to a challenge for new physics searches at high $p_T$ due to difficulties in estimation of various $t\bar t$ differential distributions. 
The pre-selection of the ``top candidate'' 
as the hardest fat-jet in the event, combined with the selection criteria for the ``leptonic top'' object can result in mis-identifying a hard 
light-quark QCD jet for a top. 
Moreover, in the context of TOM,  
the imbalance in the transverse momenta of $t$ and $\bar{t}$ could lead to an inaccurate estimate of the top jet $p_T$ (based on the $h_T$ of the leptonically decaying top), and thus result in the use of a template $p_T$ bin which does not match the transverse momentum of the hadronically decaying top. 

In order to systematically study the NLO effects on performance of TOM, we first classify the SM top/antitop events into three different categories~\cite{Salam:2013talk}, depicted in Fig.~\ref{fig:feyn_diags}:
  \begin{enumerate}[(i)]
  \item Symmetric events, where the top and the anti-top are nearly back to back.
  \item Events with one central top and a forward one. 
  \item Events where the top and the anti-top come from a gluon splitting and recoil against a hard gluon or a light quark jet.  
  \end{enumerate}

\begin{figure}[htb]
\begin{center}
\begin{tabular}{cc}
\includegraphics[width=3.3in]{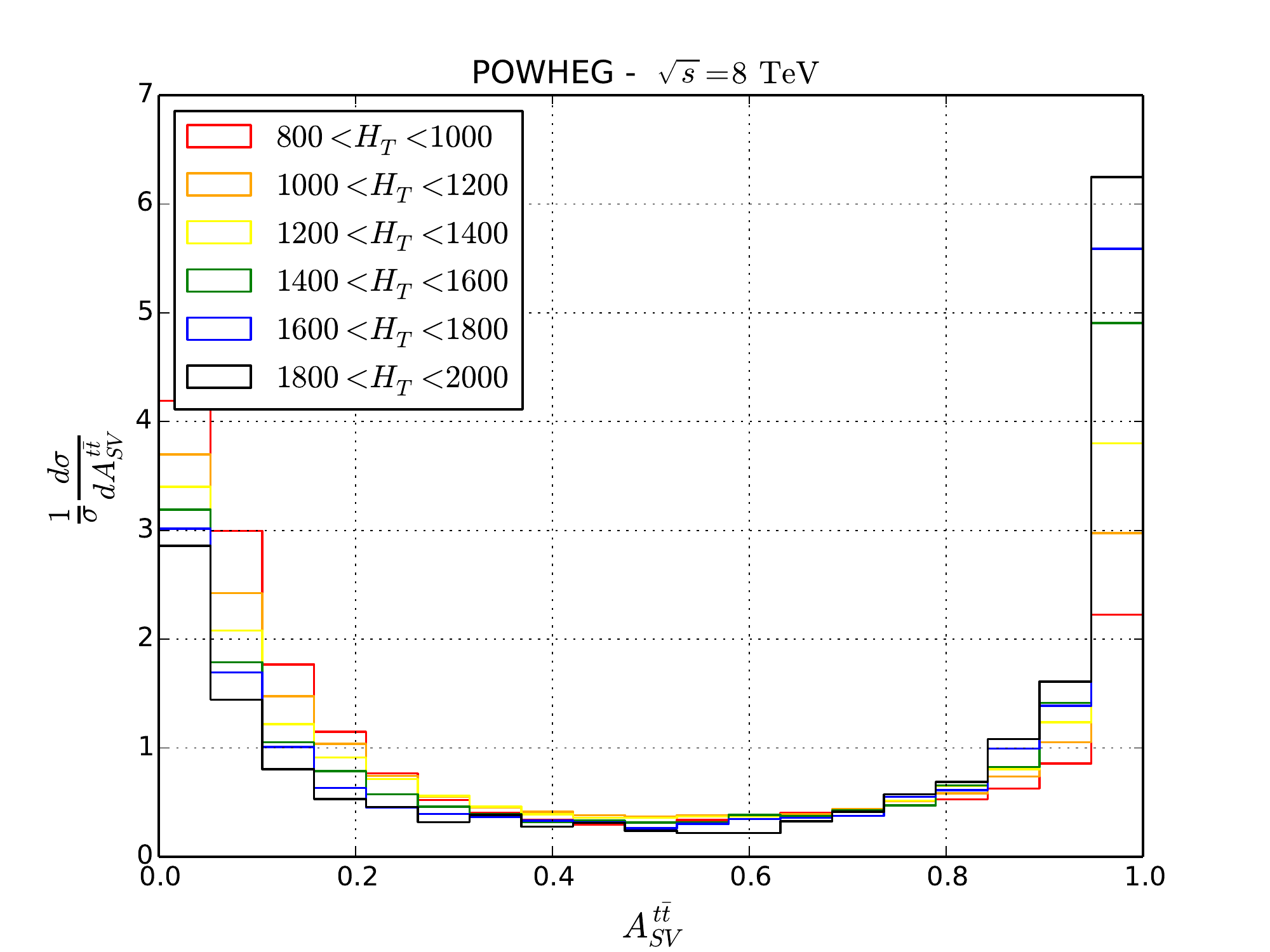} &
\includegraphics[width=3.3in]{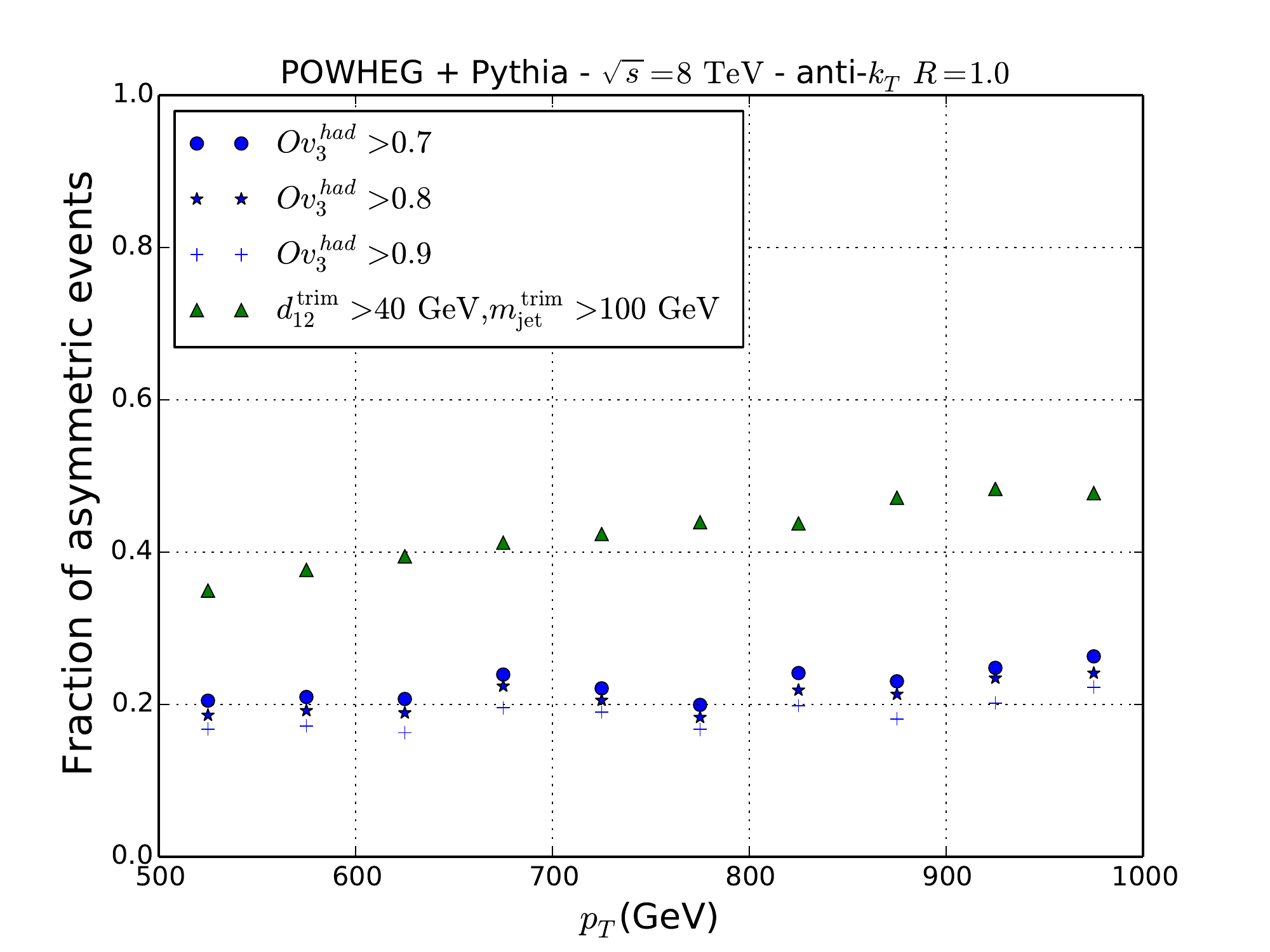}
\end{tabular}
\caption{Left: The Monte Carlo truth $t\bar t$ scalar-vector asymmetry, $A_{t\bar t}^{SV}$, for different $H_T$ bins. Right: the fraction of asymmetric events, $A_{t\bar t}^{SV} > 0.2$, which remain  after applying Top Template Tagger and \ADTT \  on the top reconstructed jets. }
\label{fig:asymm_ht}
\end{center}
\end{figure}

We quantify the top/anti-top $p_T$ imbalance by the following asymmetry between the vector sum and the scalar sum of the top transverse momenta:
\begin{equation}
\ASV = \frac{|\vec{ p_T}^t + \vec{p_T}^{\bar t}|}{|p_T^t |+ |p_T^{\bar t}|} \, ,
\end{equation}
where $\vec{p_T}^{t,\bar{t}}$ are the transverse momentum vectors of the top and the anti-top respectively, and we choose to study $\ASV$ on the truth level. The asymmetry vanishes for kinematic configurations in which the di-top system is back to back (i.e. large $m_{t\bar{t}}$), whereas the maximum occurs when the tops are parallel (i.e. $m_{t \bar{t}} \rightarrow 2\,m_t$). Hence, the events belonging to category~(i) are characterized by small asymmetry, roughly $\ASV\lesssim0.2$, while the two other categories have $\ASV\gtrsim0.2$ (with  class~(iii) events tending to $\ASV\to1$).

It is important to note that the events belonging to class~(ii) and~(iii) in the SM $t\bar{t}$ production come both as a blessing and a curse. 
For instance, if one is interested in measuring the SM top differential $p_T$ distribution,
the rejection of asymmetric events due to top-tagging will come at a cost of excluding a portion of relevant events.
Yet, including the asymmetric events into the event sample might lead to mis-identification of the hadronic top, 
and an inaccurate reconstruction of the event. Furthermore, top quark pairs produced in heavy resonance decays are typically symmetric. Hence,  rejecting asymmetric events implies that the SM $t\bar{t}$ is not an irreducible background anymore and a further improvement in signal to background can be achieved. 

A SM $t\bar{t}$ sample generated at NLO with POWHEG and showered with Pythia serves as a benchmark for studies of $\ASV$ in the context of TOM. We apply the same  pre-selection cuts as in  Section \ref{sec:preselection}, but with the requirement on $p_T$ of the fat jet lowered to $p_T > 300 \GeV$. We use template $p_T$ bins of $50 \GeV$
  for the overlap analysis, a preference which has little effect on the ability of TOM to tag jets, but it improves the $p_T$ resolution of the fat jet.
 
Figure~\ref{fig:asymm_ht} (left panel) shows the truth level $\ASV$ for a series of $H_T$ bins with two main features of the SM $t\bar{t}$ sample evident. First, the peak at $\ASV\to0$ is mainly due to the LO contribution, while the peak at $\ASV\to1$ corresponds in most part to events from class~(iii). The main contribution to the region of $\ASV$ in-between the two peaks, which spreads over large range of angles between the top and the anti-top, comes from category~(ii) and it is not seen as a peak. Second, it is evident that the fraction of asymmetric events increases with $H_T$ of the event sample, as both the phase space for hard gluon emission and the gluon splitting  to a $t\bar{t}$ pair increase with energy.

\begin{figure}[tp]
\begin{center}
\begin{tabular}{cc}
\includegraphics[width=3.3in]{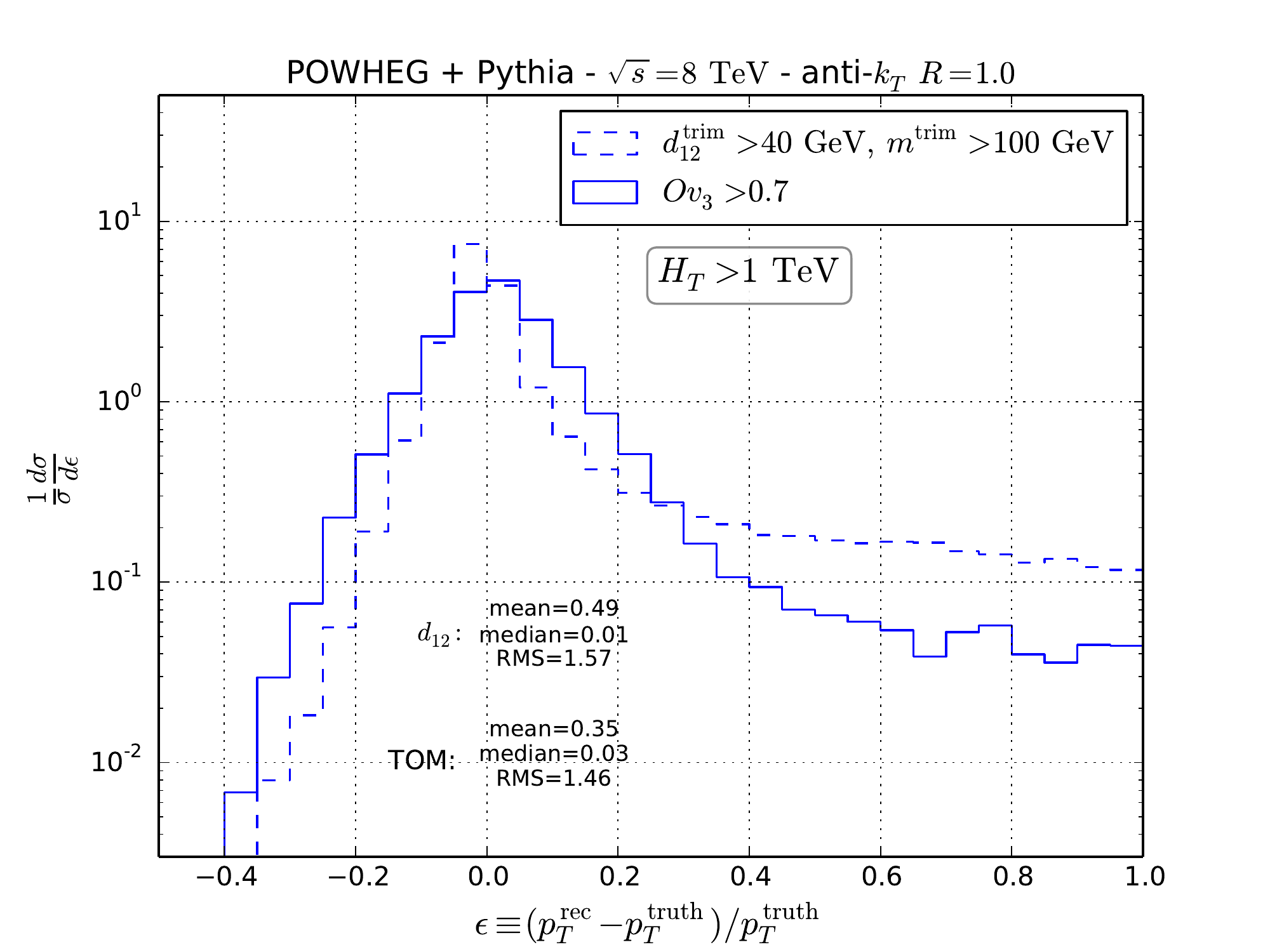} & \includegraphics[width=3.3in]{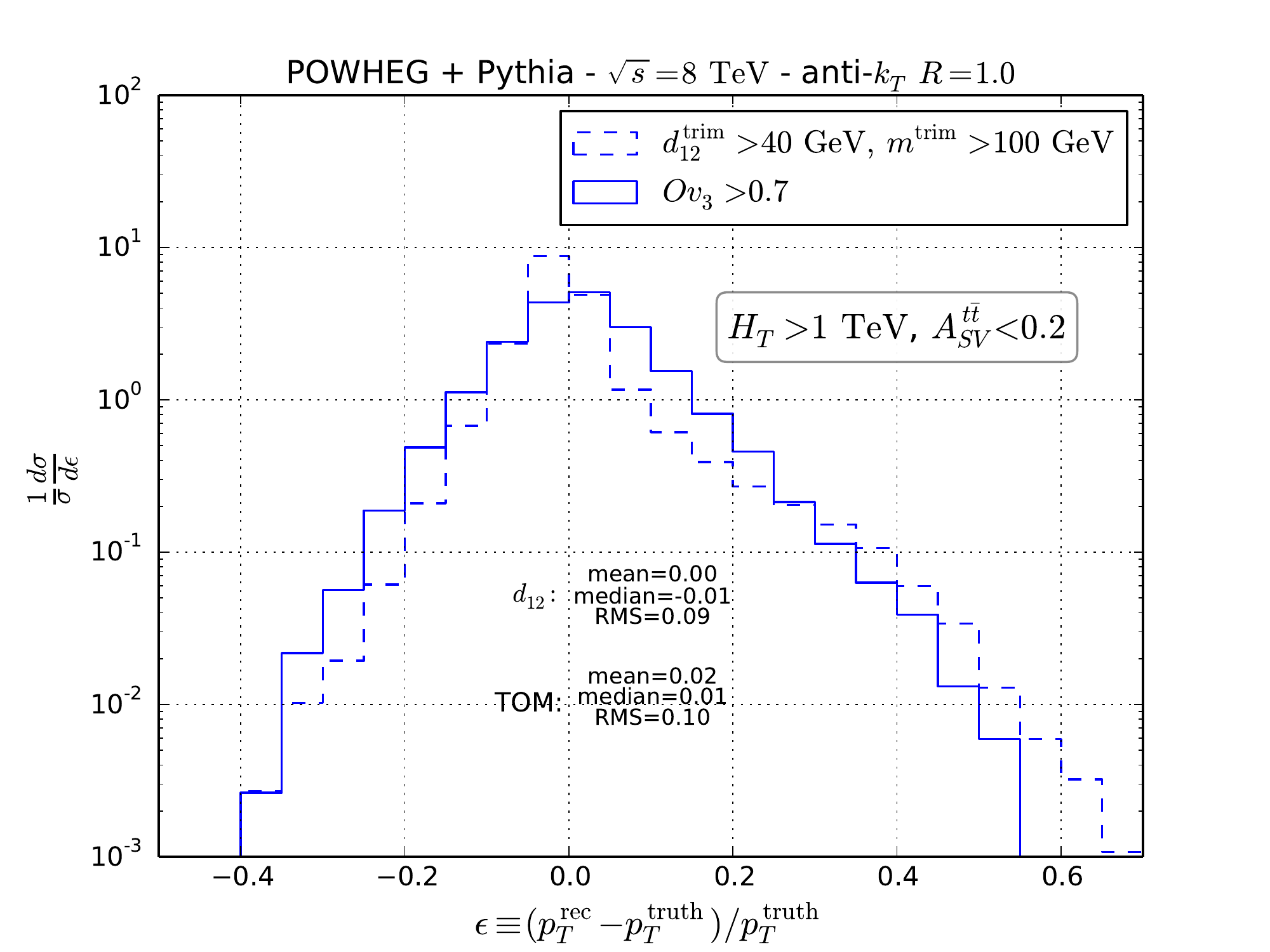} \\
\includegraphics[width=3.3in]{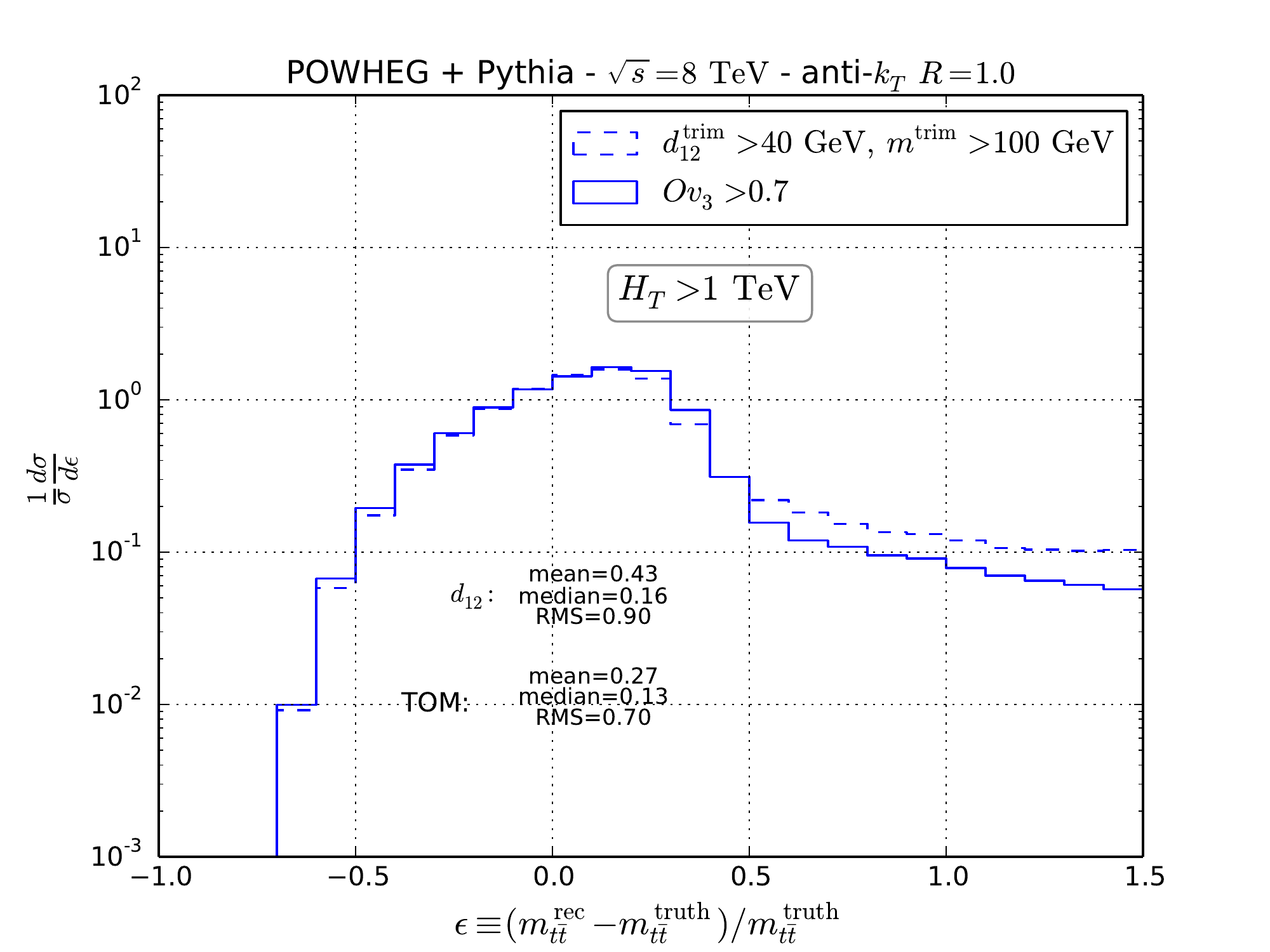} & \includegraphics[width=3.3in]{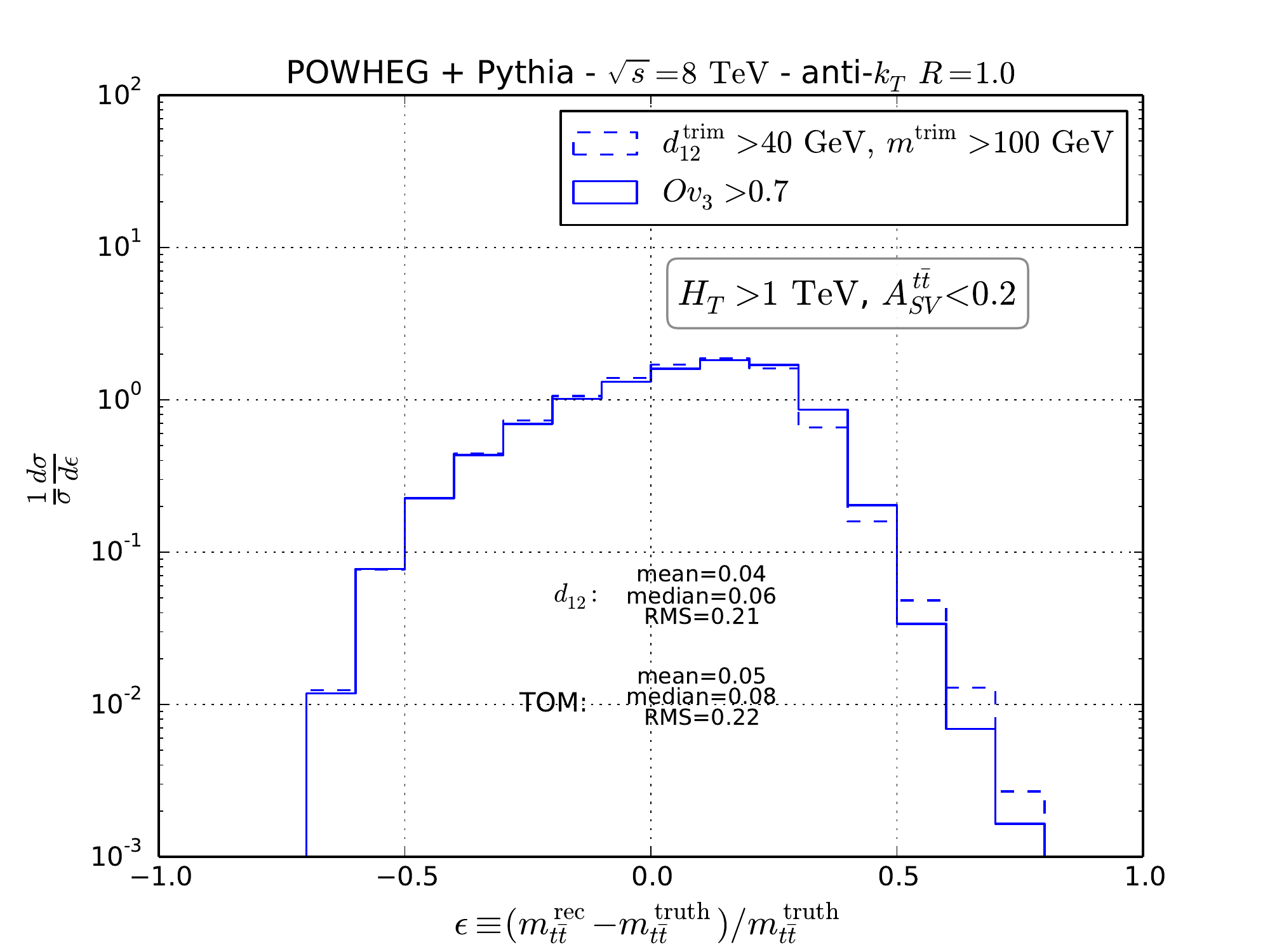}
\end{tabular}
\caption{Transverse momentum resolution (top) and di-top invariant mass resolution (bottom) of TOM compared to the \ADTT for events in the range $500 \GeV < p_T < 1000 \GeV$. $p_T^{\rm truth} $ and $m^{\rm truth}_{t\bar t}$ are always the truth top $p_T$ and  $m_{t\bar{t}}$, whereas $p_T^{\rm rec}$ and $m^{\rm rec}_{t\bar t}$ are the peak template transverse momentum and $m_{t\bar{t}}$, or the trimmed jet $p_T$ and  $m_{t\bar t}$.  The solid line shows the ability of TOM to resolve the $p_T$ and  $m_{t\bar t}$ of the parton level top with the cut of $Ov_3^{had} > 0.7$. The dashed line shows the corresponding $p_T$ and $m_{t\bar t}$ resolution using the \ADTT,  where the label ``trim'' refers to the trimmed jets with the trimming parameter $f = 0.05$ and $d_{12}$ is the $k_T$ splitting scale at the last step of fat-jet clustering. The left panel is for events with $H_T>1\TeV$ while in the right panel only symmetric events are considered, i.e. $\ASV<0.2$. All events assume the Basic Cuts of Eq.~\eqref{eq:bc} in addition to the cuts specified.  }
\label{fig:Resolution}
\end{center}
\end{figure}

How well can TOM distinguish the back to back $t\bar{t}$ events from the events with large $\ASV$? 
The ability of TOM to reject asymmetric events is highly correlated with the ability to reject light parton QCD jets, which we discuss in detail in Section~\ref{sec:RP}.  For the purpose of comparison, here we also include results using the \ADTT~\cite{Aad:2013ttr}, which consists of the following cuts:
\be
	\sqrt{d_{12}} > 40.0 \GeV , \,\,\,\,\, m_j^{\rm trim} > 100 \GeV \label{eq:d12}\,.
\ee
The $m_j^{\rm trim}$ in Eq.~\eqref{eq:d12} is the trimmed fat jet mass  with the trimming parameters $R_{\rm trim} = 0.3$ and $f = 0.05$ 
(see Ref.~\cite{Krohn:2009th} for more details), and $d_{12}$ is the $k_T$ measure at the last step of large-$R$ jet clustering with a  $k_T$ algorithm:
\be
	\sqrt{d_{12}}  = {\rm min}(p_{T,1}, p_{T,2}) \times \Delta R_{12}\,.
\ee
The values $p_{T,i}$ appearing in the last equation are the transverse momenta of the two subjets at the last step of fat jet 
clustering and $\Delta R_{12}$ is the plain distance between them. Boosted top quark decays are characterized
by symmetric splittings $d_{12} \approx m_{t}/2$, whereas background QCD jets tend to have much smaller $d_{12}$. 

The right panel of Fig.~\ref{fig:asymm_ht} shows the comparison of TOM and $d_{12}$ in their ability to reject asymmetric events. The blue points represent  the fraction of asymmetric events, $A_{t\bar t}^{SV} > 0.2$, which remain after applying various cuts on $Ov_3^{had}$ as a function of the peak template $p_T$. The green triangles show the analogous fraction of asymmetric events after the \ADTT, as a function of the trimmed fat-jet $p_T$. Our analysis shows that TOM is able to reject the asymmetric events over a wide range of $p_T$, by a factor of 2 better than the default \ADTT. 

Higher order effects can have a significant impact on the ability to experimentally resolve the underlying parton level distributions of the top kinematic observables. The issue of resolution is inseparable from the problem of signal purity, as misidentifying a light parton QCD jet for a top will lead to an incorrect estimate of the kinematic properties of the truth level objects. Figure~\ref{fig:Resolution} shows an example of the top transverse momentum and $t\bar t$ invariant mass resolution for TOM and the \ADTT. The left panels show the resolution of transverse momentum (top panel) and $m_{t\bar{t}}$ (lower panel) for the whole data set, with the \ADTT \ results shown in the dashed lines and TOM in solid lines. The selection of $Ov_3^{had} > 0.7$ and the \ADTT \ lead to a $p_T$ resolution distributions with the following mean ($\mu$), median ($\bar{\mu}$) and standard deviation (SD): 
\ba
	\mu^{p_T}_{\rm TOM} = 0.35 \,,  \quad &  \bar{\mu}^{p_T}_{\rm TOM} = 0.03 \,,  \quad &  {\rm SD}^{p_T}_{\rm TOM} = 1.46 \, , \nonumber \\  
	\mu^{p_T}_{ d_{12}} = 0.49 \, , \quad & \bar{\mu}^{p_T}_{d_{12}} = 0.01 \, , \quad & {\rm SD}^{p_T}_{d_{12}} = 1.57\, .
\ea
 Similarly, the distribution of the $m_{t\bar{t}}$ resolution parameter shows 
\ba
	\mu^{m_{t\bar{t}}}_{\rm TOM} = 0.27 \, ,  \quad& \bar{\mu}^{m_{t\bar{t}}}_{\rm TOM} = 0.13 \, , \quad & {\rm SD}^{m_{t\bar{t}}}_{\rm TOM} = 0.70 \, ,  \nonumber \\
	\mu^{m_{t\bar{t}}}_{ d_{12}} = 0.43 \, , \quad& \bar{\mu}^{m_{t\bar{t}}}_{d_{12}} = 0.16 \, , \quad& {\rm SD}^{m_{t\bar{t}}}_{d_{12}} = 0.90 \, .
\ea
Hence, we find that TOM is able to resolve the $p_T$ and $m_{t\bar{t}}$ of the truth level tops for events which pass the overlap selection criteria better than the \ADTT.  This finding is in accord with the right panel of Fig.~\ref{fig:asymm_ht}, as TOM is more efficient at  rejecting events in which a light jet is pre-selected as the top candidate.  

For completeness Fig.~\ref{fig:Resolution} also shows the $p_T$ and $m_{t\bar{t}}$ resolution for symmetric events only (right panels). In both cases, we find that the resolution obtained from TOM is comparable to the \ADTT, with $d_{12}$ slightly overestimating the $p_T$ and the $m_{t\bar{t}}$ compared to TOM. For the $p_T$ distribution we find that
\ba
	\mu^{p_T,\ASV<0.2}_{\rm TOM} = 0.02 \, ,  \quad& \bar{\mu}^{p_T,\ASV<0.2}_{\rm TOM} =  0.01 \, , \quad& {\rm SD}^{p_T,\ASV<0.2}_{\rm TOM} = 0.10 \, ,  \nonumber \\
         \mu^{p_T,\ASV<0.2}_{ d_{12}} = 0.0 \, , \quad& \bar{\mu}^{p_T,\ASV<0.2}_{d_{12}} = 0.01 \, ,  \quad & {\rm SD}^{p_T,\ASV<0.2}_{d_{12}} = 0.09 \, ,
\ea
while for the $m_{t\bar{t}}$ we obtain 
\ba
	\mu^{m_{t\bar{t}},\ASV<0.2}_{\rm TOM} = 0.05 \, ,   \quad & \bar{\mu}^{m_{t\bar{t}},\ASV<0.2}_{\rm TOM} = 0.08 \, , \quad& {\rm SD}^{m_{t\bar{t}},\ASV<0.2}_{\rm TOM} = 0.22 \, ,  \nonumber \\ 
	\mu^{m_{t\bar{t}},\ASV<0.2}_{ d_{12}} = 0.04 \, , \quad& \bar{\mu}^{m_{t\bar{t}},\ASV<0.2}_{d_{12}} = 0.06 \, , \quad& {\rm SD}^{m_{t\bar{t}},\ASV<0.2}_{d_{12}} = 0.21 \, .
\ea

A comparison of left and right panels of Fig.~\ref{fig:Resolution} reveals that most of the asymmetric $t\bar{t}$ events ($A_{t\bar{t}}^{SV} > 0.2 $) are characterized by the resolution parameters $ (m_{t\bar{t}}^{\rm rec} - m_{t\bar{t}}^{\rm truth} ) / m_{t\bar{t}}^{\rm truth} > 0.5$ and $ (p_{T}^{\rm rec} - p_{T}^{\rm truth} ) / p_{T}^{\rm truth} > 0.3$, implying that events with large $A_{t\bar{t}}^{SV} $ tend to over-estimate both $m_{t\bar{t}}$ and the $p_T$ of the fat jets. It is then reasonable that TOM results in distributions which resolve the truth level kinematic parameters to an improved degree, as TOM is more efficient at rejecting the events with large $A_{t\bar{t}}^{SV}.$

\section{Background Rejection Power} \label{sec:RP}

 \label{sec:RP0}

Previous work of Ref.~\cite{Almeida:2010pa} showed that TOM is able to efficiently reject the QCD background in cases where both the top and the anti-top are decaying hadronically at $p_T \sim 1 \TeV$. Tagging boosted tops in events with a hard lepton and missing $\MET$ constitutes a separate problem from the fully hadronic decays of $t\bar{t}$, due to differences in the background composition.
Namely, the dominant background to semi-leptonic decays of $t\bar{t}$ comes from $W$+jets, while the multijet contribution is already sub-leading after the lepton mini-isolation.

 In this section we focus on the performance of $Ov_3^{had}$ and $Ov_3^{lep}$ in rejecting $W$+jets with no contamination from soft radiation of minimum bias events, and postpone the discussion of effects of pileup and underlying event until Section~\ref{sec:pileup}. 

To quantify the ability of TOM to tag boosted tops against the $W$+jets background we study two observables
\be
		\epsilon_{\rm sig} =\frac{ \sigma(t\bar{t})^{\rm cuts} }{ \sigma(t\bar{t})^{\rm BC}}\, , \,\,\,\,\,\, \epsilon_{\rm bgd} = \frac{\sigma(Wjj)^{\rm cuts} }{ \sigma(Wjj)^{\rm BC}}\,,
\ee
where $cuts$ denotes all selection cuts including overlap, and BC denotes the Basic Cuts of Eq.~\eqref{eq:bc}. We then define the background rejection power (RP) relative to the Basic Cuts as
\be
		{\rm RP}  = \frac{\epsilon_{\rm sig} }{\epsilon_{\rm bgd}}\,.
\ee

We do not include an explicit $b$-tag in our analysis of RP, due to the experimental challenges of  $b$-tagging at high $p_T$ and high luminosity. Instead, we study $Ov_3^{lep}$ as an alternative and compare the rejection power obtained from a tentative $b$-tagging benchmark point to our results using leptonic top overlap.

\subsection{Rejection Power for Hadronically Decaying Tops at $\sqrt{s} = 8 \TeV$} \label{sec:rejpowOv3had}

\begin{figure}[htb]
\begin{center}
\begin{tabular}{ccc}
\includegraphics[width=2.2in]{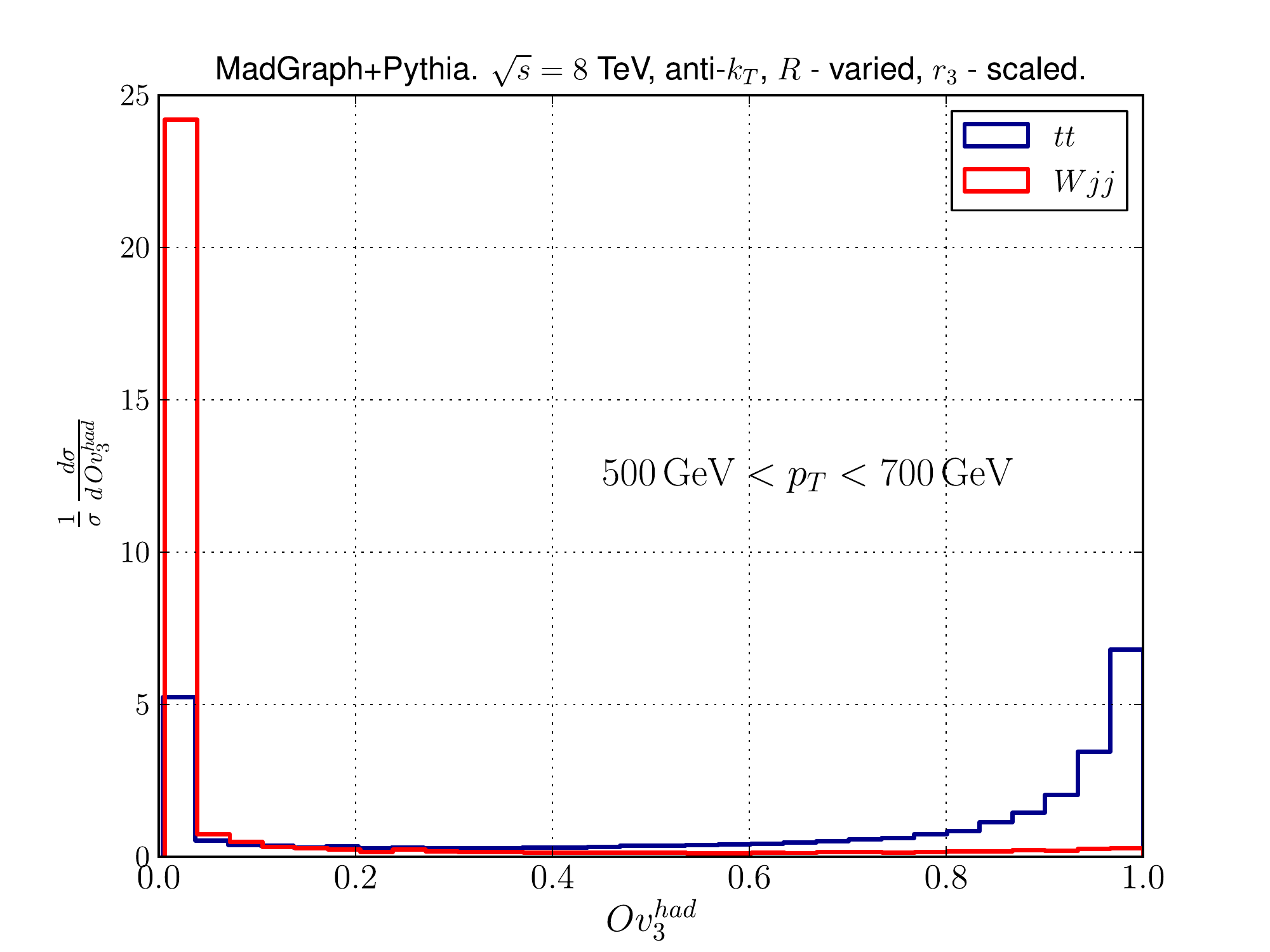} & \includegraphics[width=2.2in]{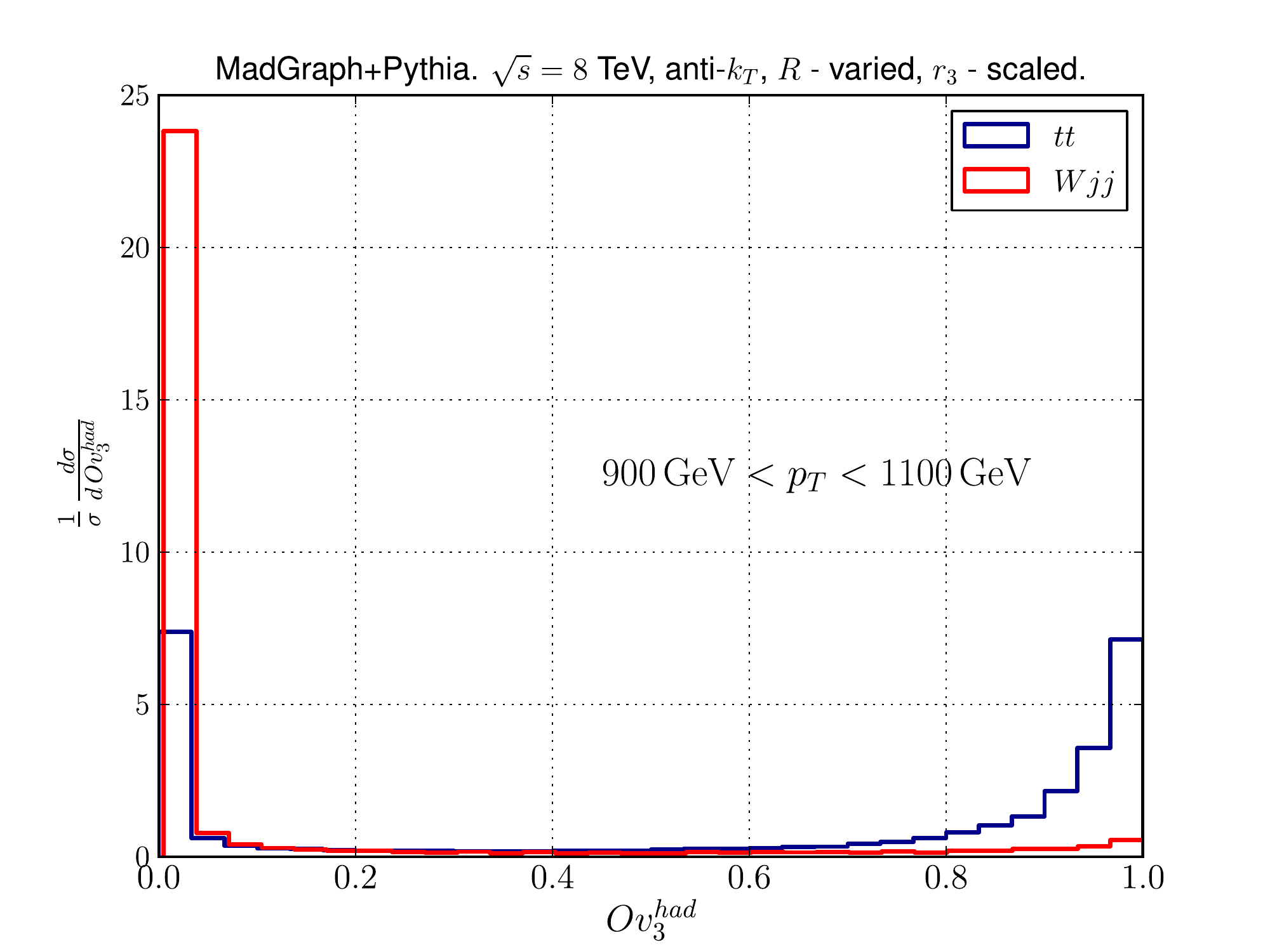} & \includegraphics[width=2.2in]{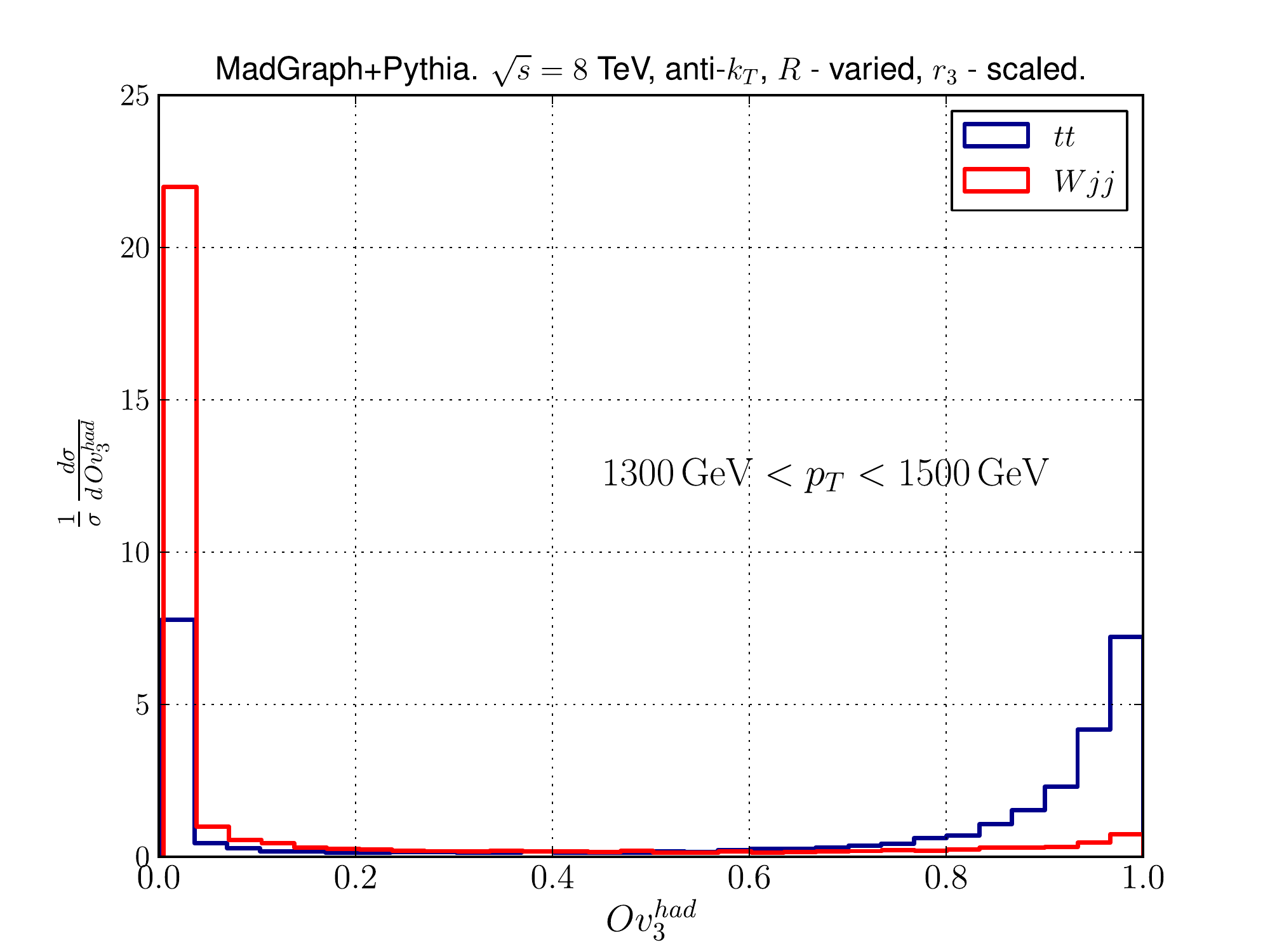}
\end{tabular}
\caption{Hadronic peak overlap distribution distributions for three different $p_T$ bins. The blue curves show the signal $t\bar{t}$ distributions whereas the red curves represent the $W$+jets background. All analyzed events assume the Basic Cuts of Eq.~\eqref{eq:bc} with no additional mass cut or $b$-tagging. The fat jet cone is varied according to the rule of Eq.~\eqref{eq:Rscale}, whereas the template sub cone radii are determined according to Eq.~\eqref{eq:temprscale}.} 
\label{fig:Ov3had}
\end{center}
\end{figure}

We perform the template overlap analysis on hadronically decaying tops according to the prescription of Eq.~\eqref{eq:ovHad}. 
Figure~\ref{fig:Ov3had} shows example distributions of $Ov_3^{had}$ for three different bins of fat jet transverse momenta. All plots assume the Basic Cuts of Eq.~\eqref{eq:bc}, with no additional mass cut. In all cases the distributions show clear separation of signal and background. Distribution of $W$+jets events sharply peaks at $Ov_3^{had} \approx 0$, while $t\bar{t}$ events occupying mostly the $Ov_3^{had} \rightarrow 1$ region, with a portion of characterized by low $Ov_3^{had}$ .  

The peak at $Ov_3^{had} \approx 0 $ in the signal distribution deserves some attention. The event pre-selection allows for many events in which one of the decay products of the top was not captured by the fat jet cone as well as asymmetric events discussed in Sec.~\ref{sec:NLO} to pass the cut. These events will likely have a low overlap score, due to having the wrong jet mass and/or substructure,  resulting in the peak at $Ov_3^{had} \approx 0$ in the signal distribution. A cut on hadronic overlap will efficiently remove such events in a systematic manner, without the need for additional customized cuts. 

\begin{figure}[htb]
\begin{center}
\begin{tabular}{cc}
\includegraphics[width=3.3in]{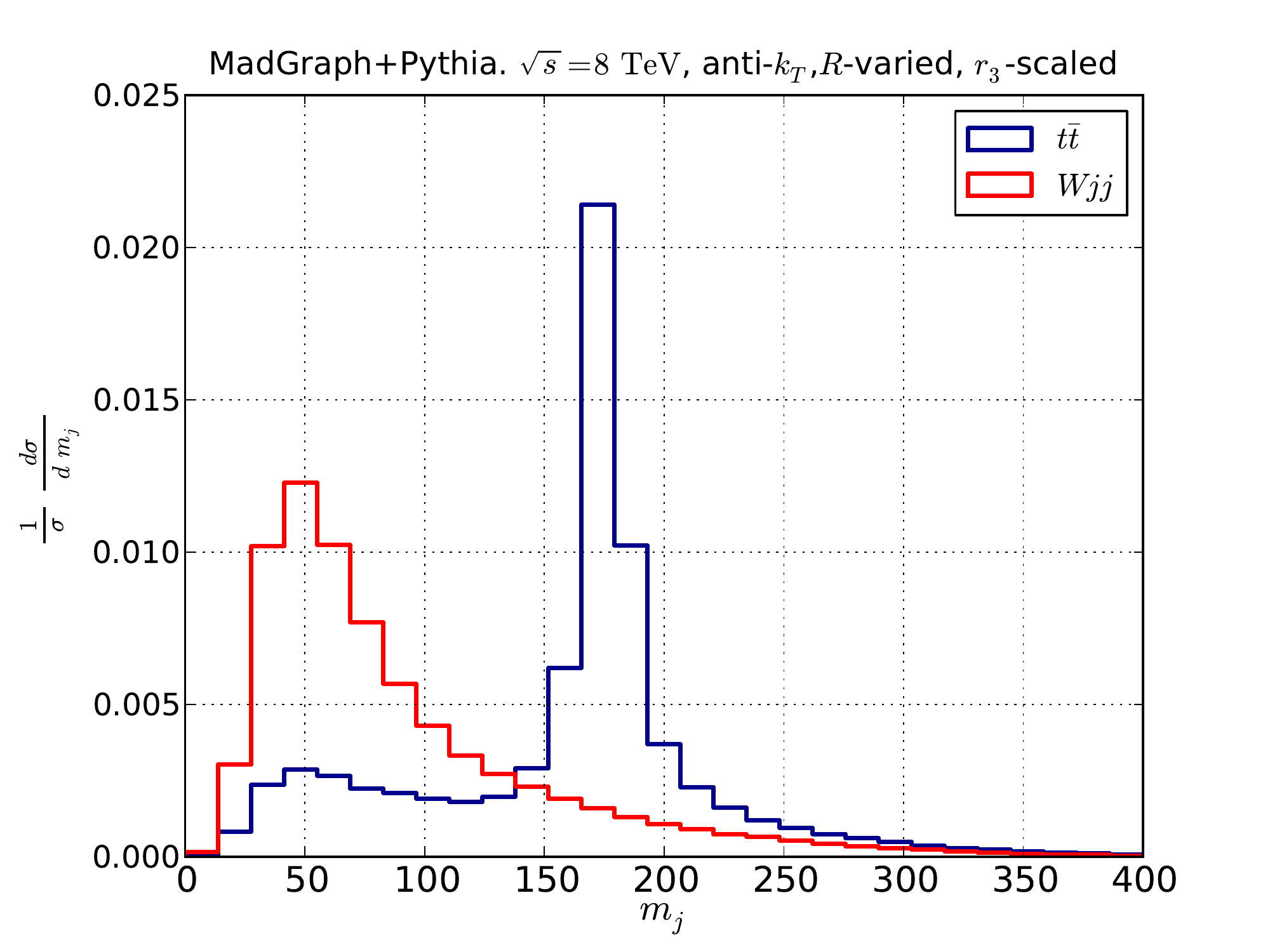} & \includegraphics[width=3.3in]{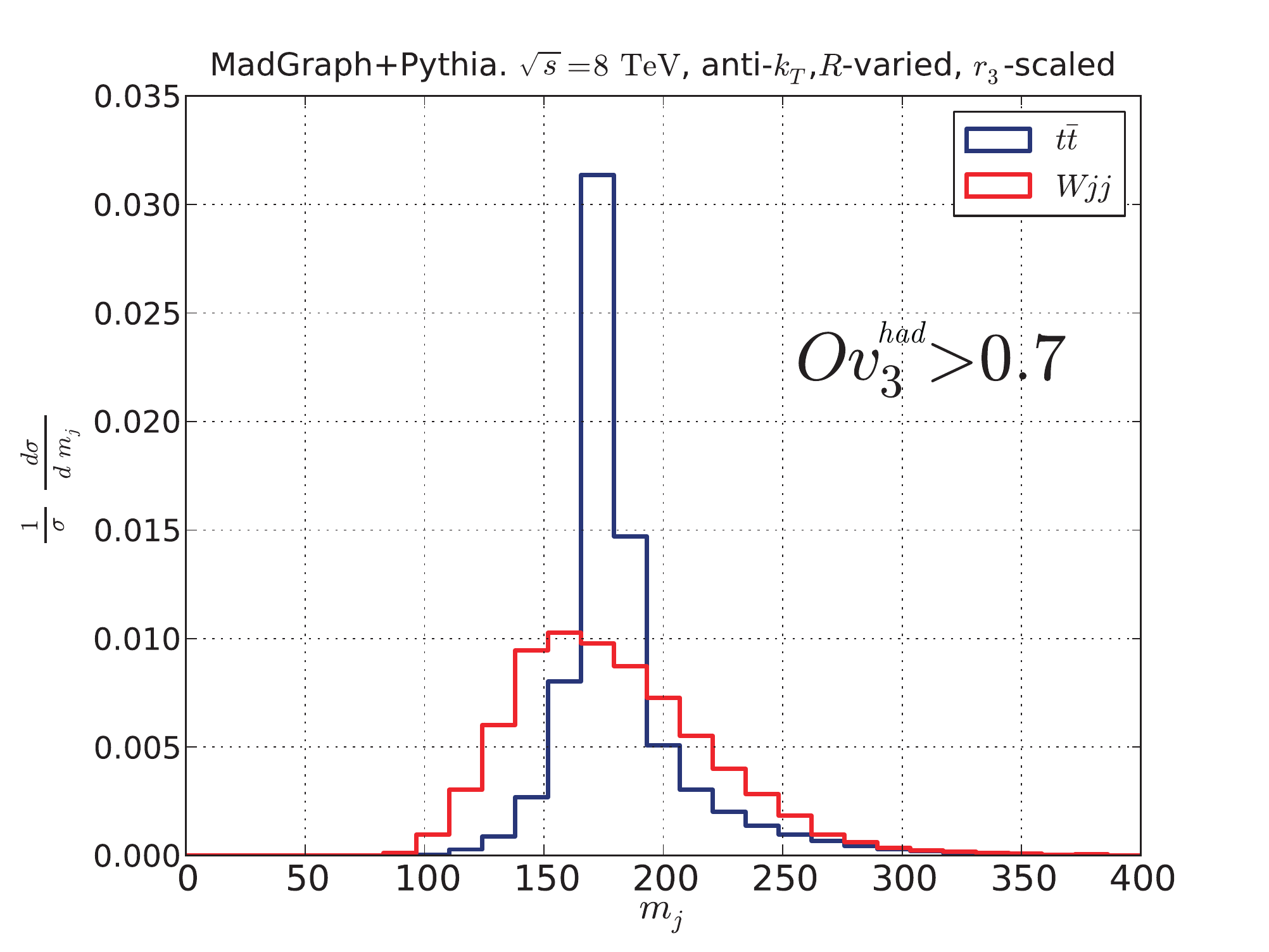} 
\end{tabular}
\caption{TOM as a mass filter. The panels show the fat jet mass distributions without (left panel) and with (right panel) a cut on $Ov_3^{had}$. All analyzed events assume the Basic Cuts of Eq.~\eqref{eq:bc} with no additional mass cut or $b$-tagging. The fat jet cone is varied according to the rule of Eq.~\eqref{eq:Rscale}, whereas the template sub cone radii are determined according to Eq.~\eqref{eq:temprscale}.}
\label{fig:Ov3_mass}
\end{center}
\end{figure}

As an example, consider the intrinsic feature of TOM mass filtering. A cut on hadronic peak overlap efficiently removes the low mass regions both in the signal and the background distributions as evident in Fig.~\ref{fig:Ov3_mass}. Implementing a mass cut via a cut on $Ov_3^{had}$ has a further advantage in a high pileup environment as TOM is much less susceptible to pileup contamination than the jet mass (see Section~\ref{sec:pileup} for more details).

We proceed to discuss the rejection power achievable with TOM at $\sqrt{s} = 8 \TeV$ over a wide range of fat jet $p_T$. Note that in the following, we assume signal events to be the SM $t\bar t$ events, including the events characterized by a large~$\ASV$.

\begin{figure}[!]
\begin{center}
\includegraphics[width=3.5in]{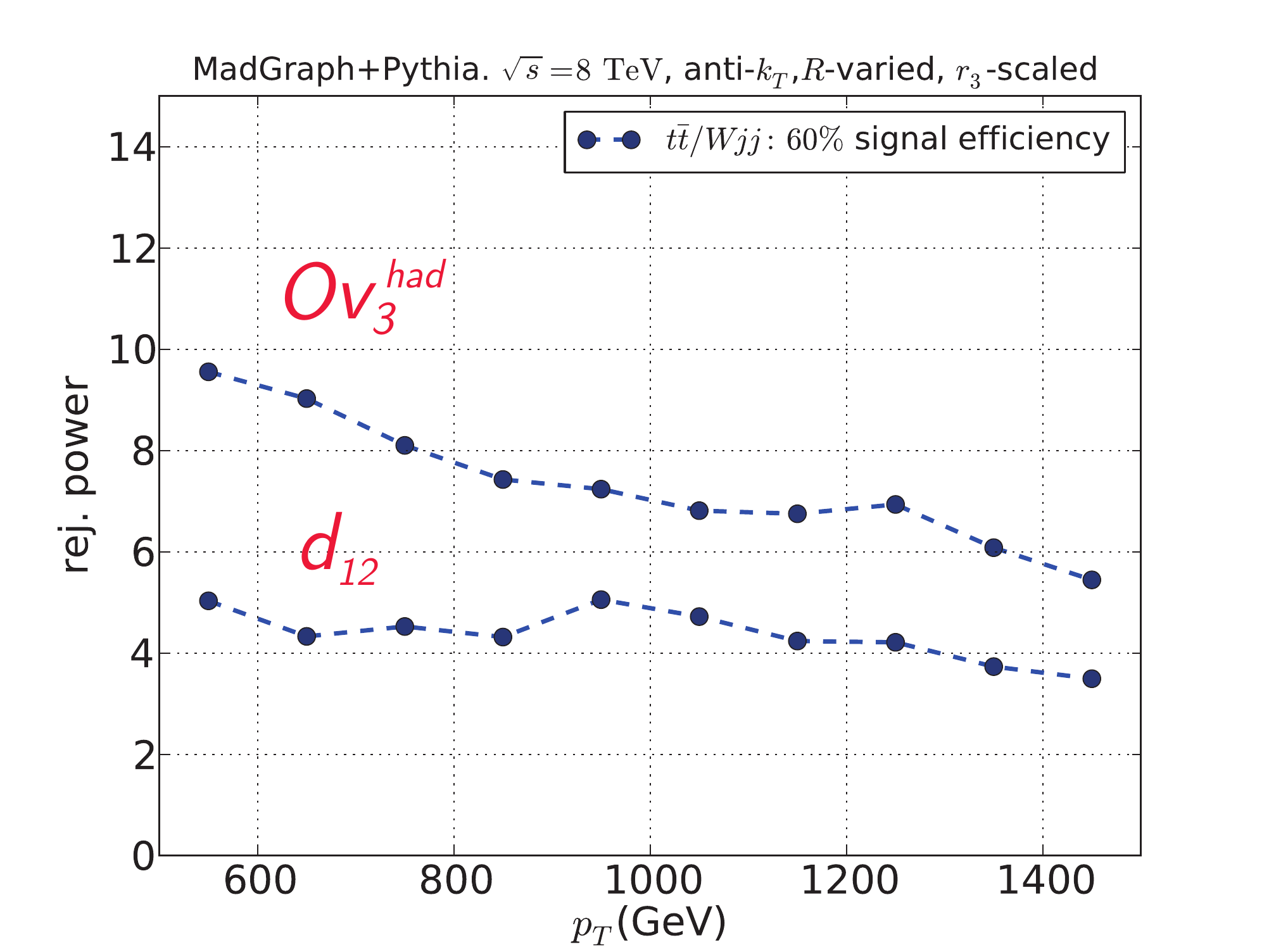}\includegraphics[width=3.5in]{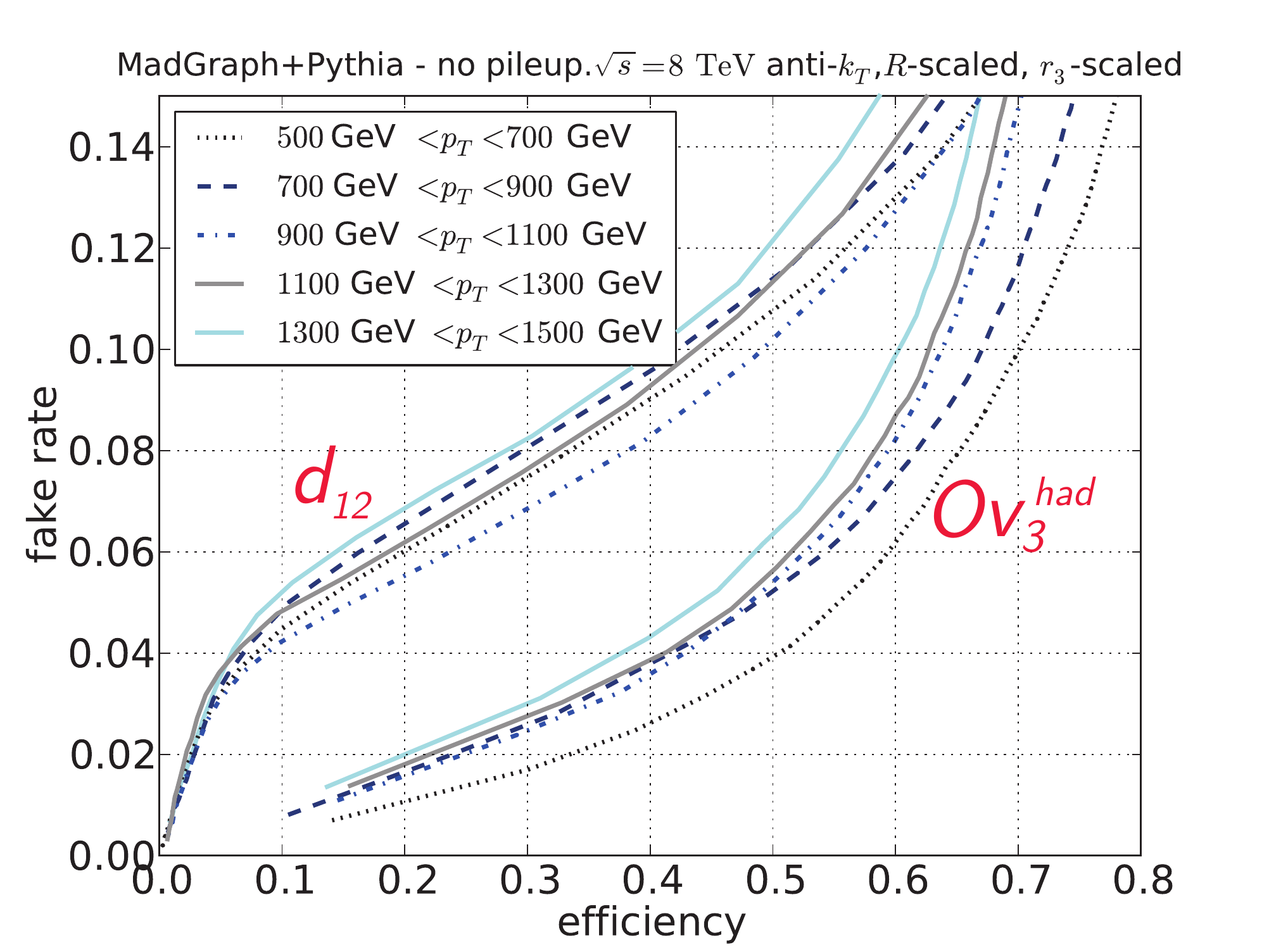}
\caption{Rejection power of TOM. The left panel shows dependence of the $W$+jets RP on the fat jet $p_T$ using TOM and $d_{12}$ as a background sicriminant. The points show RP at fixed signal efficiency of $60\%$ calculated relative to the Basic Cuts of Eq.~\eqref{eq:bc}.  The right panel shows the signal efficiency ($\epsilon_{\rm sig}$) and background fake rate ($\epsilon_{\rm bgd}$) as a function of the cut on $Ov_3^{had}$ and $d_{12}$ for various jet $p_T$ bins. The cut on $Ov_3^{had}$ and $d_{12}$ runs along the lines. All efficiencies are calculated relative to the Basic Cuts of Eq.~\eqref{eq:bc}. Both panels assume no a-priori cut on the mass of the fat jet.}
\label{fig:ov3had_rej}
\end{center}
\end{figure}

\begin{table}[!]
\begin{center}
\begin{tabular}{cc}
MG + Pythia \,\,&
\begin{tabular}{|c|c|c|c|c|c|c|}
\hline
 $Ov_3^{min}$ & $\epsilon_{\rm sig}(0.5 - 0.7 \TeV)$& ${\rm RP} $&$\epsilon_{\rm sig}(0.9 - 1.1 \TeV)$& ${\rm RP}$&$\epsilon_{\rm sig}(1.3 - 1.5 \TeV)$& ${\rm RP}$ \\
 \hline
 0.9 & 0.40& 16.7 &0.43 &11.3 &0.45 & 9.0 \\
 0.75 & 0.56 & 11.1 &0.56 & 8.4&  0.59 & 6.3  \\
 0.6 & 0.63&  8.8 &  0.62 & 6.9 & 0.64 & 5.7\\ 
 0.45 &  0.68 & 7.6  & 0.65  & 5.9 & 0.66  & 4.8  \\ \hline
\end{tabular}
\\
Sherpa \,\,&
\begin{tabular}{|c|c|c|c|c|c|c|}
\hline
 $Ov_3^{min}$ & $\epsilon_{\rm sig}(0.5 - 0.7 \TeV)$& ${\rm RP}$&$\epsilon_{\rm sig}(0.9 - 1.1 \TeV)$& ${\rm RP}$&$\epsilon_{\rm sig}(1.3 - 1.5 \TeV)$& ${\rm RP}$ \\
 \hline
 0.9 & 0.31 &\,\,6.1\, & 0.31 &\,4.6\,\,& 0.36 & 5.0 \\
 0.75 & 0.41 & 4.7 & 0.40 & 3.7 &  0.45 & 3.7 \\
 0.6 & 0.47 & 3.9 & 0.45 & 3.3 & 0.49 & 2.7 \\
0.45 & 0.51&  3.6 &0.49 & 2.9  & 0.52 & 2.5 \\
 \hline
 \end{tabular}
 \\
 POWHEG+Pythia\,\,&
\begin{tabular}{|c|c|c|c|c|c|c|}
\hline
 $Ov_3^{min}$ & $ \epsilon_{\rm sig}(0.5 - 0.7 \TeV)$& ${\rm RP}$&$\epsilon_{\rm sig}(0.9 - 1.1 \TeV)$&${\rm RP} $&$\epsilon_{\rm sig}(1.3 - 1.5 \TeV)$& ${\rm RP}$\\
 \hline
 
  0.9 & 0.34 &\,\,\,\,-\,\,\,& 0.38 &\,\,\,-\,\,\,\,& - &\,\,\,\,-\,\,\,\,  \\
 0.75 & 0.44 & - & 0.45 & - & -  & -  \\
 0.6 &  0.50 &  - & 0.49 & - & - & - \\
  0.45 & 0.55 & - & 0.52 &-& - & - \\
 
 \hline

\end{tabular}

\end{tabular}
\end{center}
\caption{Rejection power of $Ov_3^{had}$ for several benchmark cuts on $Ov_3^{had}$. All signal efficiencies and rejection powers are calculated relative to Basic Cuts of Eq. \eqref{eq:bc}, with no cut on the fat jet mass or $b$-tagging. Each RP corresponds to the signal efficiency and $p_T$ in the column before it. Also shown for comparison are the signal efficiencies for the POWHEG sample.}
\label{tab:Ov3hadRej}
\end{table}
 
Figure~\ref{fig:ov3had_rej} shows the rejection power of $Ov_3^{had}$ compared to the \ADTT. The left panel illustrates the dependence of rejection power on fat jet $p_T$ at a fixed signal efficiency of $60 \%$. We find that a rejection power of $\sim 10$ is possible at $p_T \approx 500 \GeV,$ while the ability to reject $W$+jets events reduces at higher $p_T$. We have checked that the decrease in rejection power with the increase in jet $p_T$ (dashed blue curve) is almost entirely due to the asymmetric events discussed in Section~\ref{sec:NLO}, since the proportion of $t\bar t$ events with large $\ASV$ increases with the $H_T$ of the event. 

The right panel of Fig.~\ref{fig:ov3had_rej} shows more complete
information on the ability of TOM and $d_{12}$ to reject background
events.
The curves represent the $W$+jets fake rate as a function of signal
efficiency, while the overlap cut runs along the curves. Each curve is
limited to a range of fat jet $p_T$ values. Notice that TOM clearly
outperforms $d_{12}$ for most efficiencies and the entire considered
$p_T$ range by roughly a factor of two.
Table \ref{tab:Ov3hadRej} summarizes results from different event
generators. Our results show that a significant gain in rejection
power can be obtained by tightening the $Ov_3^{had}$ cut.
Increase in the lower cut on the jet mass to $150 \GeV$ in the \ADTT \ only moderately improves the RP of $d_{12}$ with a
factor of $\approx 6$ achievable.
Lowering the signal efficiency to $40\%$ results roughly in a factor
of 2 improvement in rejection power over the previously discussed $60
\%$ benchmark efficiency point, as shown in Table \ref{tab:Ov3hadRej}.
 In order to better quantify uncertainties from different Monte~Carlo tools, the
efficiency of these cuts is also evaluated on the basis of full NLO
$t\bar t$ events from POWHEG for the $500-700 \GeV$ and
$900-1100 \GeV$ bins.  The level of agreement of Sherpa with POWHEG
corresponds to $5-10\%$ for most of the cuts, while MadGraph
shows differences of up to $20\%$.   The differences are not
surprising since both Sherpa and MadGraph work at leading
order in perturbation theory and use different matching procedures.

\subsection{ Rejection Power for Leptonically Decaying Tops $\sqrt{s} = 8 \TeV$}

In the previous section we showed that the rejection power of $\approx 10$ is possible at $60 \%$ signal efficiency relative to the Basic Cuts, considering only the hadronically decaying top quark. Leptonically decaying top contains additional information which can be used to discriminate against the backgrounds. Here we present results of the leptonic top overlap analysis, using the $Ov_3^{lep}$ implementation of Eq.~\eqref{eq:lep_overlap}. 

\begin{figure}[htb]
\begin{center}
\includegraphics[width=3.5in]{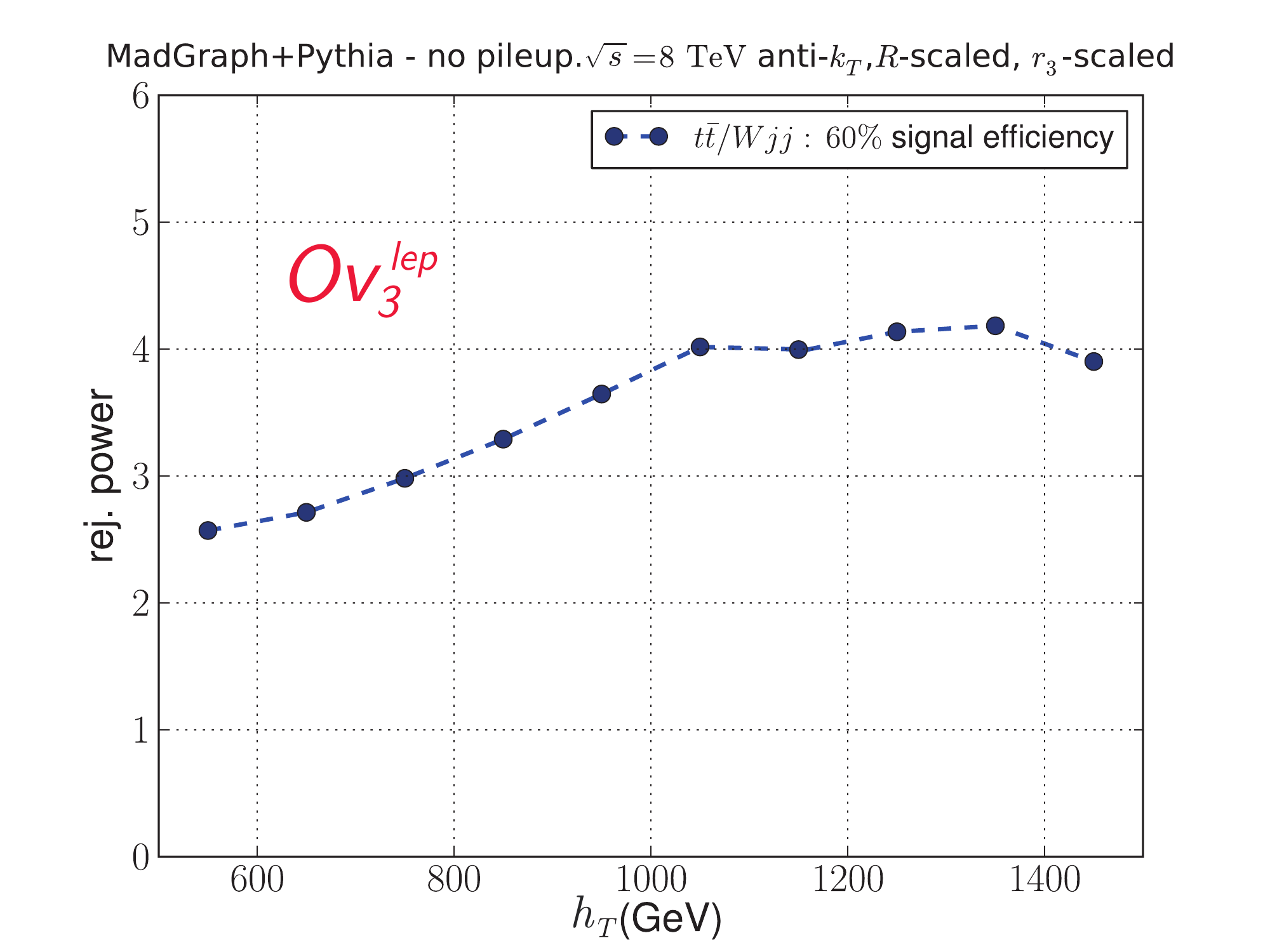}\includegraphics[width=3.5in]{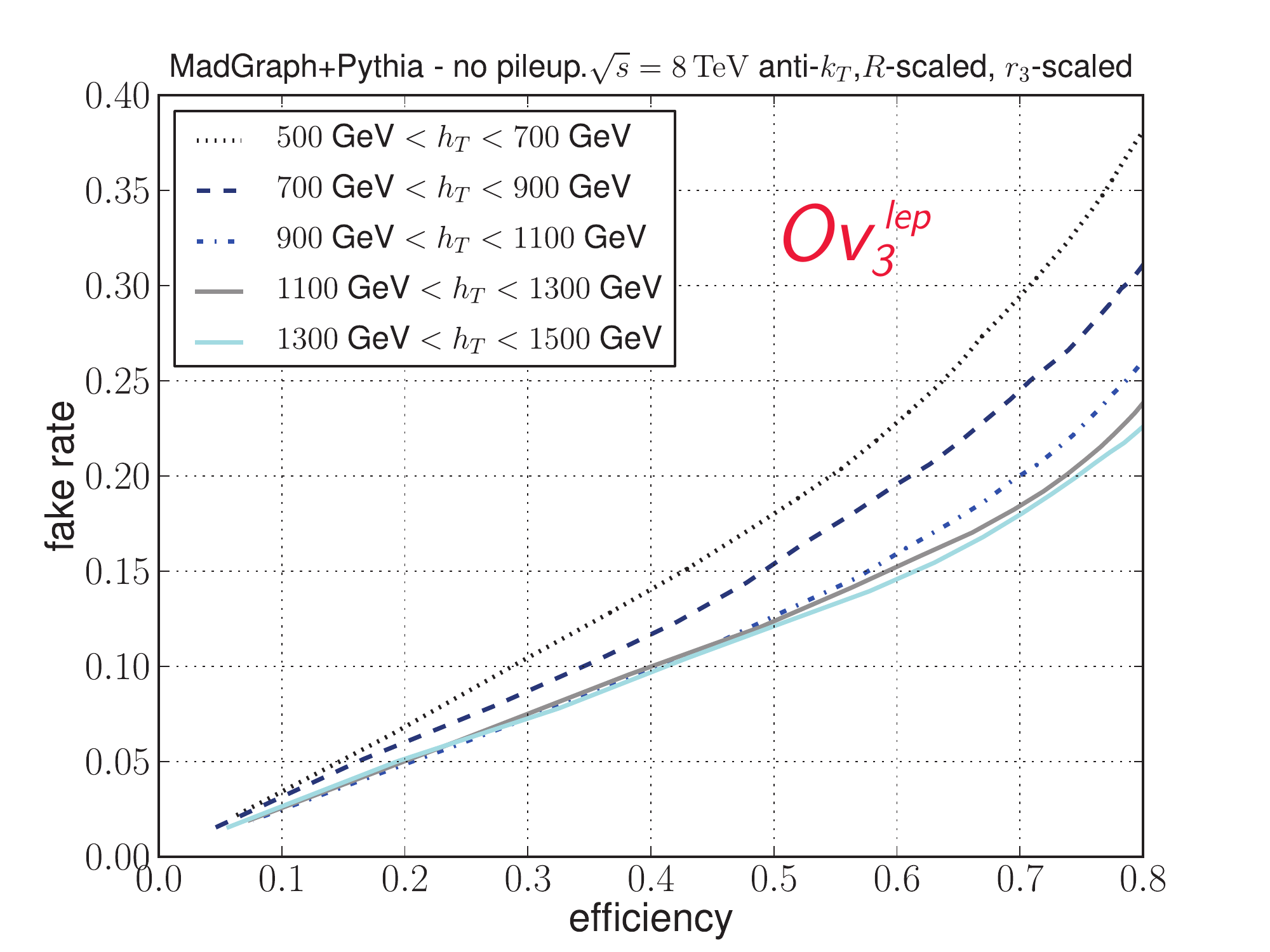}
\caption{Rejection power of $Ov_3^{lep}$ for $Wjj$. The left panel shows dependence of the $Wjj$ RP on the leptonic top $h_T$. The points show RP at fixed signal efficiency $(\epsilon_{\rm sig})$ of $60\%$.  The right panel shows the signal efficiency  $(\epsilon_{\rm sig})$ and background fake rate $(\epsilon_{\rm bgd})$ as a function of the cut on $Ov_3^{lep}$ for various jet $h_T$ bins. The cut on $Ov_3^{lep}$ runs along the line. All efficiencies are calculated relative to the Basic Cuts of Eq.~\eqref{eq:bc}. Both panels assume no a-priori cut on the mass of the fat jet or $b$-tagging.}
\label{fig:ov3_rej_lep}
\end{center}
\end{figure}

Figure~\ref{fig:ov3_rej_lep} shows our main results of this section, while Table~\ref{tab:Ov3l} summarizes the results from different event generators. The left panel shows the $h_T$ dependence of RP at fixed signal efficiency of $60 \%$ relative to the Basic Cuts. The RP achieved with $Ov_3^{lep}$ is lower than RP of $Ov_3^{had}$ at the same efficiency and $h_T$.  The reason for a lower RP compared to $Ov_3^{had}$ is that the kinematics of the object we construct from a jet, a lepton and missing energy in the $W$+jets events is by construction more similar to a boosted top decay at the pre-selection level. The object $Ov_3^{lep}$ is trying to distinguish from the leptonically decaying quark is typically of higher mass than a light jet in addition to the missing energy and the lepton already coming from a $W$ decay. The templates, which are designed to tag a $W$ and reconstruct the correct mass of the top quark (among other things) thus have a higher probability of mis-tagging such an object as a top. However, overlap analysis is extremely efficient in removing the pure QCD background which is of much higher rate, hence the analysis will result in a better sensitivity and reach.

Leptonic top implementation of TOM is able to reject $W$+jets events with RP $\approx 2.5$ for $h_T = 500 \GeV$, with the increase in rejection power to $\approx 4$ at higher values of $h_T$, as the pre-selection cuts are sufficient to relieve $Ov_3^{lep}$ from the higher order effects which plague the fat jet analysis. For completeness, we also summarize the rejection power analysis in Table~\ref{tab:Ov3l} for several cuts on $Ov_3^{lep}$. 

\begin{table}
\begin{center}
\begin{tabular}{cc}
MG+ Pythia &
\begin{tabular}{|c|c|c|c|c|c|c|}
\hline
 $Ov_3^{min}$ & $\epsilon_{\rm sig}(0.5 - 0.7 \TeV)$& ${\rm RP}$&$\epsilon_{\rm sig}(0.9 - 1.1 \TeV)$& ${\rm RP}$&$\epsilon_{\rm sig}(1.3 - 1.5 \TeV)$& ${\rm RP}$ \\
 \hline
 0.9 & 0.45 &  2.8 & 0.54  &  3.9 & 0.54 & 4.1 \\
 0.75 &0.67  & 2.5 & 0.74 & 3.4 & 0.77 & 3.6 \\
 0.6 & 0.76 &2.3 &0.81 &3.0   & 0.83 & 3.4  \\  
 0.45 & 0.82 & 2.0& 0.84 &2.8  &0.86 & 3.1 \\ \hline

\end{tabular} \\

Sherpa &
\begin{tabular}{|c|c|c|c|c|c|c|}
\hline
 $Ov_3^{min}$ & $\epsilon_{\rm sig}(0.5 - 0.7 \TeV)$& ${\rm RP}$&$\epsilon_{\rm sig}(0.9 - 1.1 \TeV)$& ${\rm RP}$&$\epsilon_{\rm sig}(1.3 - 1.5 \TeV)$& ${\rm RP}$ \\
 \hline
 0.9 & 0.56 & 2.2  & 0.59  &  2.8 & 0.58 & 3.1 \\
 0.75 & 0.72 &2.0  &0.72 & 2.5 & 0.73 &2.7  \\
 0.6 &0.79  &1.8 & 0.75& 2.3& 0.77 &2.6  \\  
 0.45 &  0.82 & 1.7&  0.78 & 2.2 &0.80 &2.5  \\ \hline

\end{tabular}

\end{tabular}

\caption{Rejection power of $Ov_3^{lep}$ in various $h_T$ bins. All signal efficiencies and rejection powers are calculated relative to Basic Cuts of Eq.~\eqref{eq:bc}, with no cut on the fat jet mass or $b$-tagging. Each RP corresponds to the signal efficiency and $h_T$ in the column before it. }

\label{tab:Ov3l}
\end{center}
\end{table}

\subsection{Leptonic Top Overlap as a $b$-tagging Alternative}

Tagging of $b$-quarks at high $p_T$ (\textit{i.e.} $ > 300 \GeV$) is an experimentally challenging task. Any alternative method which could at least compensate for the background rejection power provided by the $b$-tagging procedure could be a valuable asset in boosted top analyses. In the previous section we already discussed the rejection power which can be achieved by $Ov_3^{lep}.$ Here, we ask whether the achievable rejection power is sufficient to compensate for the reduction in the $b$-tagging efficiency.

The details of $b$-tagging involve an elaborate analysis of the detector level data (including both the tracking and calorimeter information), which is beyond the scope of this analysis. Here, we use a  semi-realistic $b$-tagging procedure, whereby the parton level information from the Monte Carlo hard process provides a ``tag'' for the showered jets. If an $r=0.4$ anti-$k_T$ jet is within $\Delta R = 0.4$ from a hard-process $b$ or $c$ quark, we assign a $b$-tag to the jet. Otherwise, the jet is tagged as a light jet. We then weigh the number of $b$, $c$ and light jets by the efficiencies for identifying each category as an actual $b$-jet. For the purpose of this analysis, we use the benchmark point of 
\be
	\epsilon_b = 0.5\, , \,\,\,\,\,\, \epsilon_c = 0.3\, , \,\,\,\,\,\,\, \epsilon_l = 0.1\, , \label{eq:btag}
\ee
where $\epsilon_{b, c, l}$ are efficiencies that a jet is identified as a $b$-jet for $b$, $c$ and light flavors respectively. 
Properly tagging the $b$-quark at high $p_T$ hence results in  the rejection power of roughly $5$  for light jets and $1.7$ for charm.

Table \ref{tab:btag} shows a comparison for a set of leptonic top $h_T$ values. The leptonic overlap performs slightly worse than $b$-tagging at high $p_T$ with the rejection power of $\approx 4$ achievable from $Ov_3^{lep}$. It is important to note that the results in the left column of  Table~\ref{tab:btag} reflect the optimistic values for $b$-tagging efficiencies of Eq.~\eqref{eq:btag}. In reality, the ability to properly tag the $b$ quarks deteriorates with the increase in energy, while the leptonic overlap rejection power increases. Hence we find that $Ov_3^{lep}$ could provide a useful substitute for the rejection power lost due to the reduction of $b$-tagging efficiency in an analysis. In addition, the information contained in $Ov_3^{lep}$ is complementary to $b$-tagging, and the combination of the two can be used to further increase the RP.

\begin{table}[htb]
\begin{tabular}{|c|c|c|c|}
\hline
$h_T$ &$ \epsilon_{\rm sig} $ & $b$-tag rejection & $Ov_3^{lep}$ RP \\
\hline
700 - 900 \GeV & 0.5 & 4.5 & 3.2 \\
900 - 1100 \GeV & 0.5 & 4.5 &3.9 \\
1100 - 1300 \GeV & 0.5 & 4.5 & 4.0 \\
1300 - 1500 \GeV & 0.5 & 4.5 & 4.2 \\
\hline

\end{tabular}

\caption{Comparison of rejection power obtained from $b$-tagging alone and $Ov_3^{lep}$ at various leptonic top $h_T$ and signal efficiency of $50 \%$. The table assumes the benchmark $b$-tagging efficiency of Eq.~\eqref{eq:btag} for all $h_T$ ranges with the light and charm flavors combined. All rejection powers are calculated relative to the Basic Cuts of Eq.~\eqref{eq:bc}.  }
\label{tab:btag}
\end{table}

\section{Effects of Pileup Contamination on TOM} \label{sec:pileup}

\begin{figure}[htb]
\begin{center}
\includegraphics[width=3.0in]{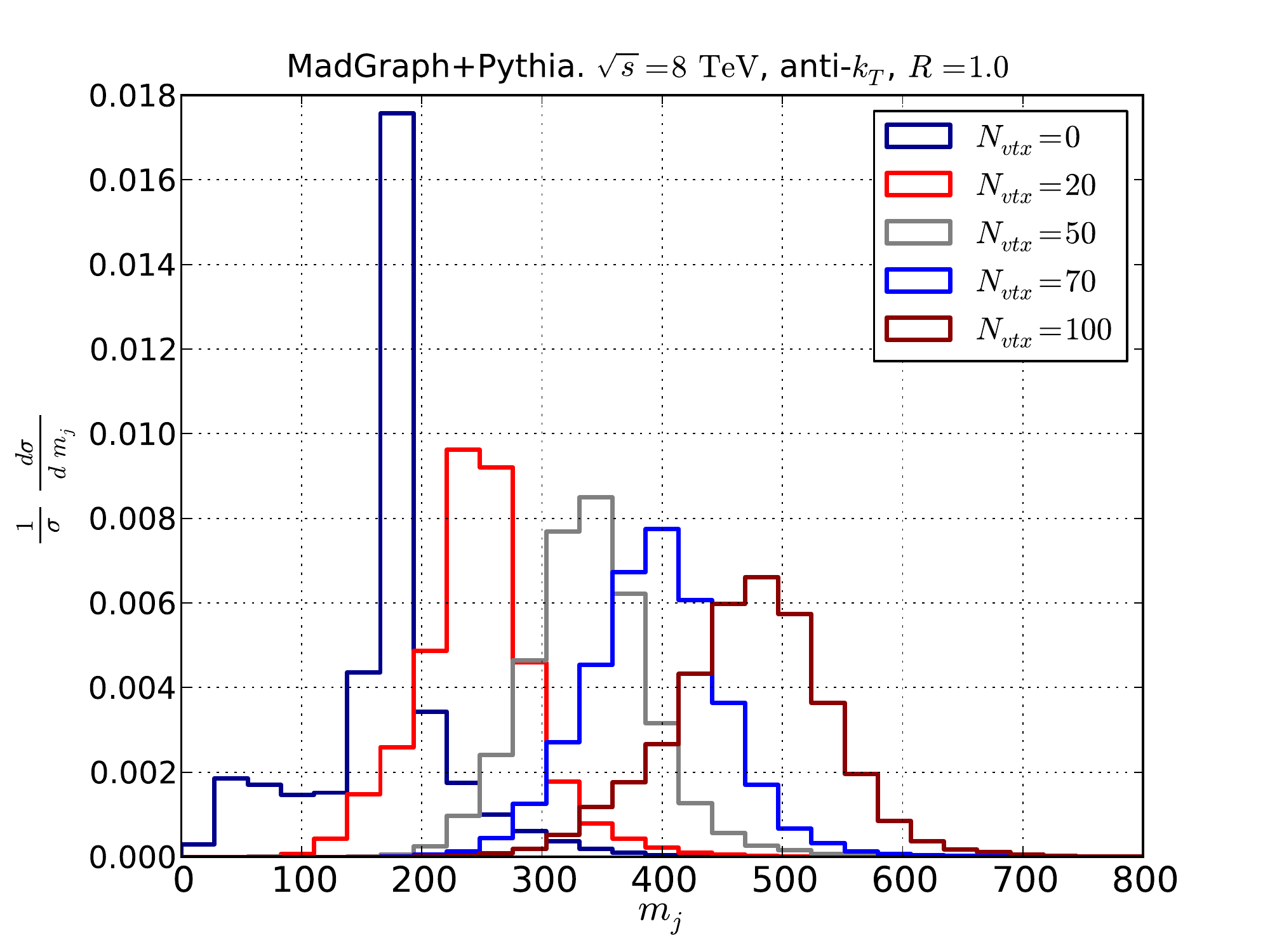} \includegraphics[width=3.0in]{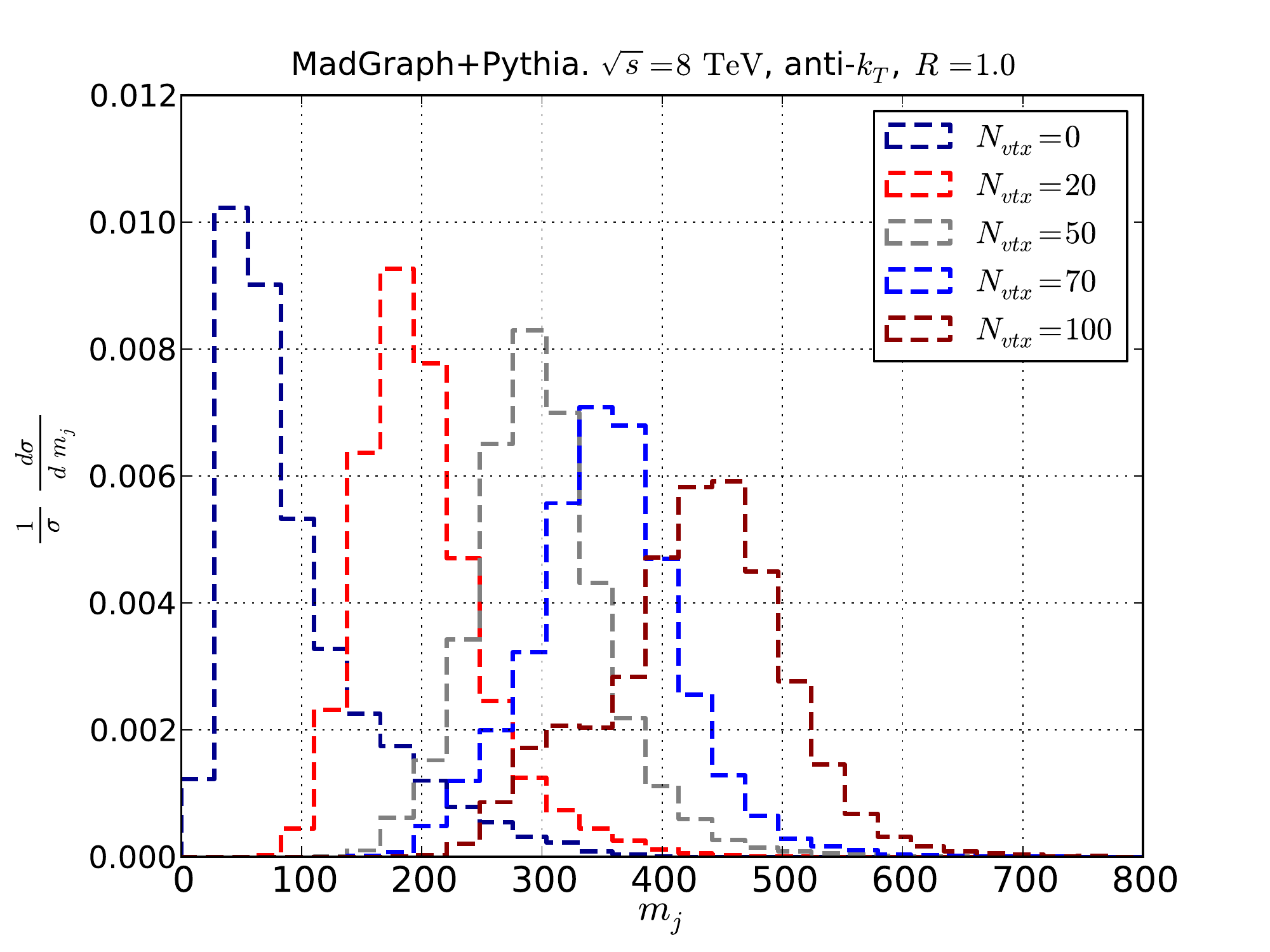} 
\caption{Mass distributions of hadronic top fat jet (left panel) and $W$+jets fat jet (right panel) at various levels of pileup contamination. Here we show only events with a leading jet of $500 \GeV < p_T < 600 \GeV$ and the Basic Cuts of Eq.~\eqref{eq:bc}.} 
\label{fig:mass_pileup}
\end{center}
\end{figure}

The high instantaneous luminosity characteristic of the LHC poses a serious problem for jet substructure physics. The current LHC run at $\sqrt{s} = 8 \TeV$ recorded an average $\langle N_{vtx} \rangle \approx 20$ interactions per bunch crossing,  with the projections that the future runs may result in as much as $\langle N_{vtx} \rangle \sim 100$~\cite{ATL-PHYS-PUB-2013-014}. Contamination due to diffuse radiation from pileup can significantly shift and broaden the jet kinematic distributions, sparking a need for methods to either subtract, or correct for large pileup effects. Figure~\ref{fig:mass_pileup} shows an example of effects of pileup on the boosted top and light quark QCD jet mass distribution. Pileup not only shifts the mass peak to the right, but significantly broadens the distributions as well. Imposing a fixed mass window on the fat jet distribution would thus result in decreased efficiency with the increase in pileup. The statement is true even after estimating the relative shift of the mass peak due to pileup, as the widening of the mass distribution is difficult to correct for. 

Algorithms such as Jet Trimming~\cite{Krohn:2009th} and Jet Pruning~\cite{Ellis:2009me} aim to remove the contamination of soft radiation from underlying event or pileup, which is important to improve the mass resolution for large jets. A data driven method of Ref.~\cite{Alon:2011xb} 
 focused on pileup correction for jet-shape variable at the differential level, say as a function of the jet mass, angularity and planar flow. Subsequent studies by CDF~\cite{CDFconfnote} and ATLAS~\cite{Aad:2012meb} collaborations provided qualitative validation of the method. In addition, the CMS collaboration employs track information to subtract pileup contamination coming from secondary vertices \cite{2011JInst...611002C}.  Reference~\cite{Cacciari:2007fd}  uses a jet area based method for pileup correction, whereby the effects of pileup are subtracted from jet observables such as $p_T$ and mass based on the data driven estimates of the pileup contamination per unit area. 
More recently, the authors of Ref.~\cite{Soyez:2012hv} proposed a method of subtracting the effects of pileup from jet shape variables using jet areas. The results were numerically shown to hold up to $\langle N_{vtx}  \rangle = 60$. 

Reference~\cite{Backovic:2012jj} showed that TOM is weakly affected by pileup, with boosted Higgs distributions of template overlap (and other template based observables) remaining mostly impervious to pileup at $\langle N_{vtx} \rangle = 20$ interactions per bunch crossing. The relative insensitivity of TOM to pileup comes from the fact that template sub-cones radii are typically of $\cOO(10^{-1})$ of the fat jet cone, yielding that the relative pileup contamination is only a few percent of the effect on jet observables such as fat jet mass or transverse momentum.  

In this section, we study the effects of pileup on top template overlap, at various top energies and levels of pileup contamination. For the purpose of our study, we choose to omit as many pileup sensitive observables as possible (such as the fat jet $p_T$ and mass). Instead, we focus on the intrinsic, pileup insensitive mass filtering property of TOM, as well as present the results in terms of $h_T$ instead of fat jet transverse momentum where appropriate.

To simulate the effects of pileup we add minimum bias events to each event we we wish to analyze, whereby the number of pileup events added is determined on an event-by-event basis, by drawing a random number from a Poisson distribution with the mean $\langle N_{vtx} \rangle$.

\subsection{Pileup Effects on Hadronic Peak Overlap ($Ov_3^{had}$)}

\begin{figure}[htb]
\begin{center}
\includegraphics[width=3.2in]{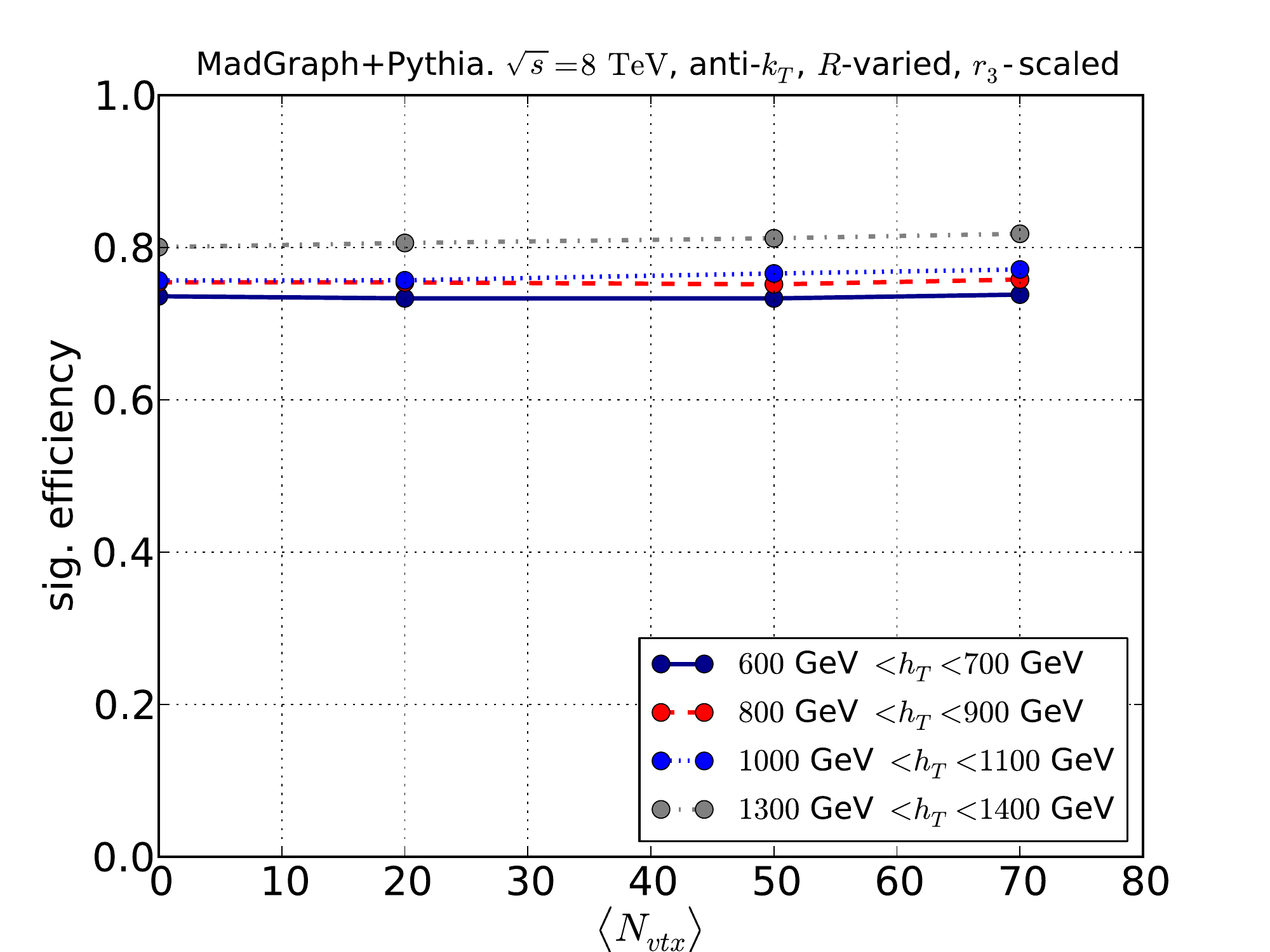}\includegraphics[width=3.2in]{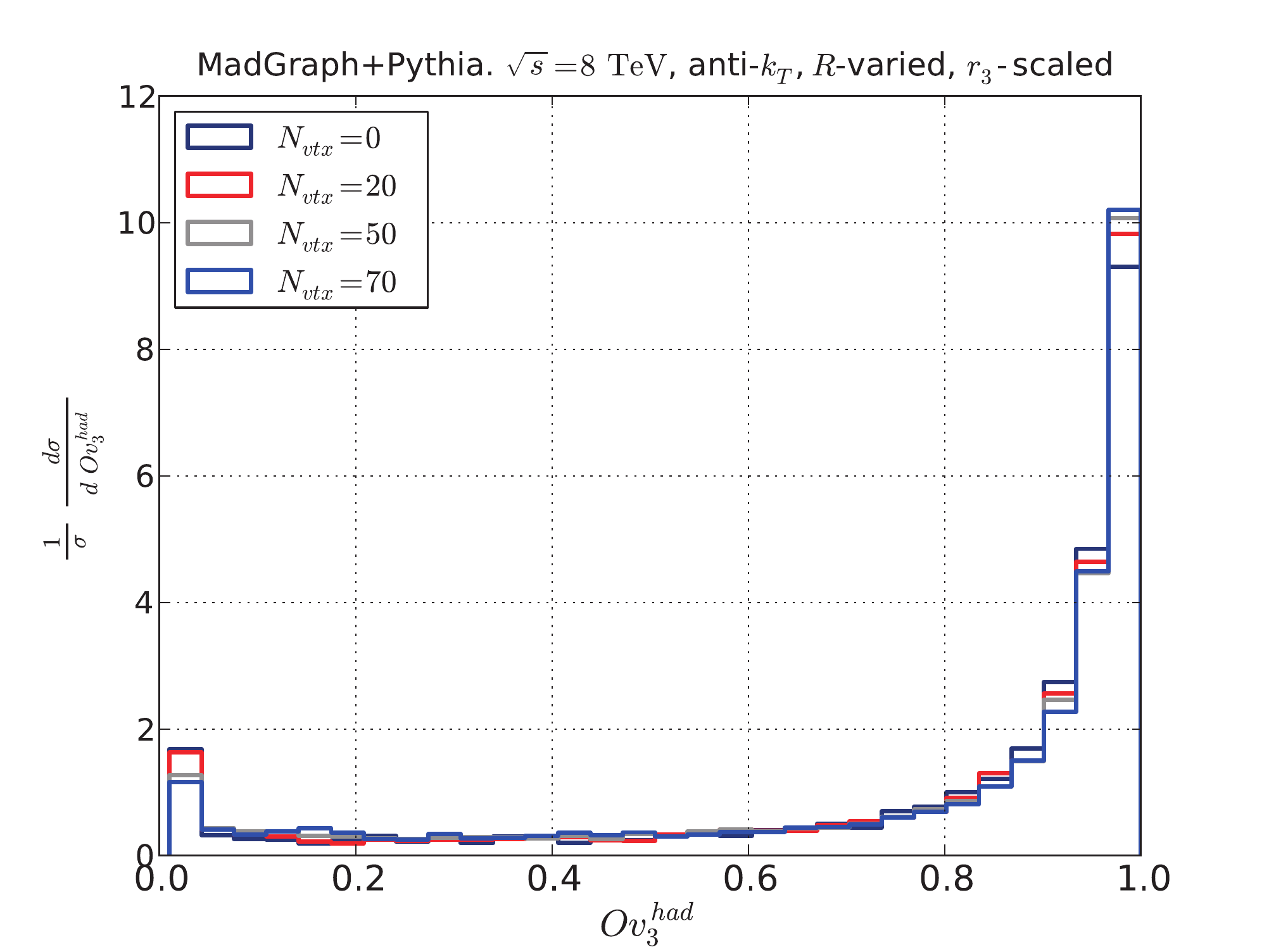}
\includegraphics[width=3.2in]{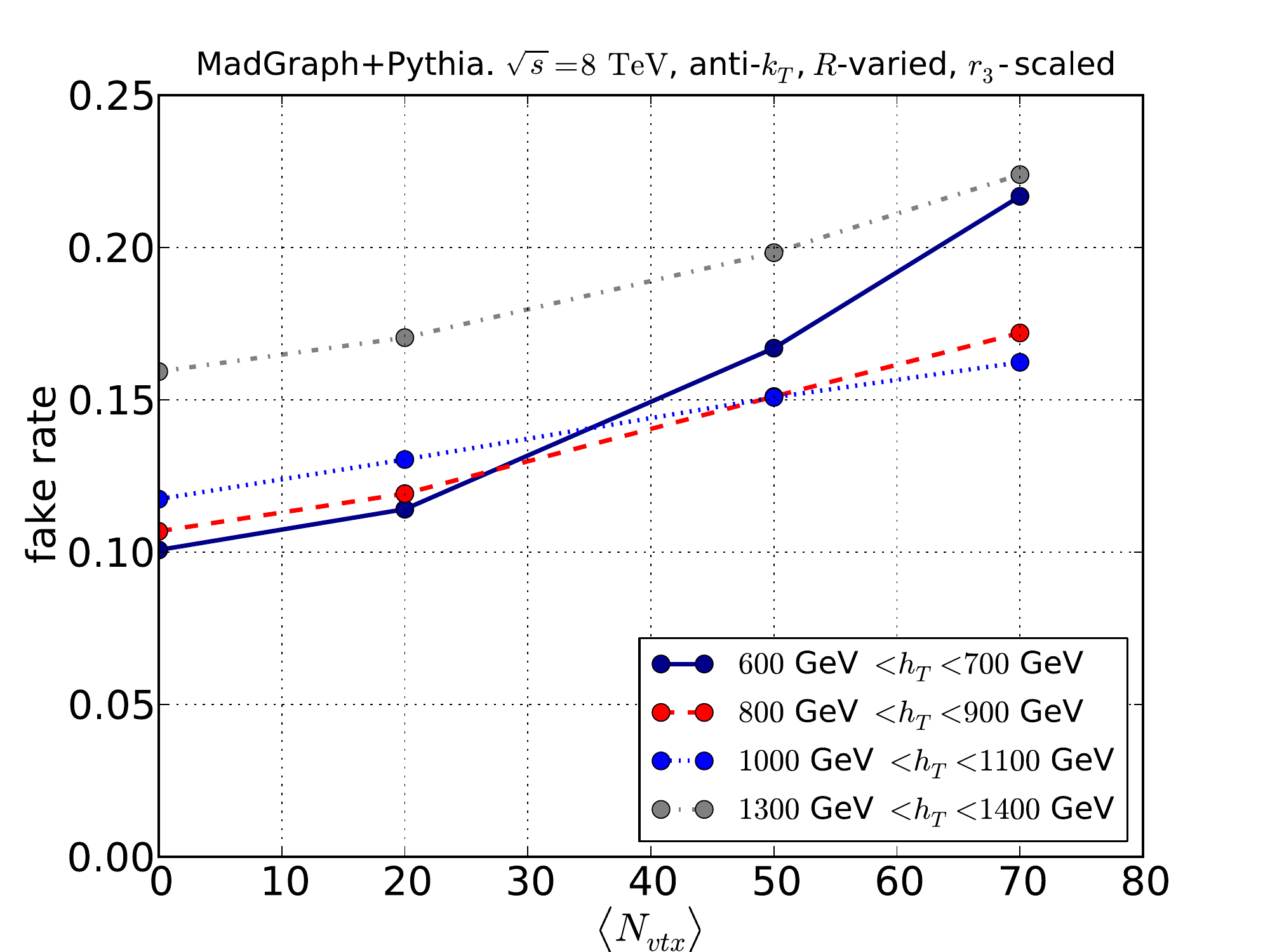}\includegraphics[width=3.2in]{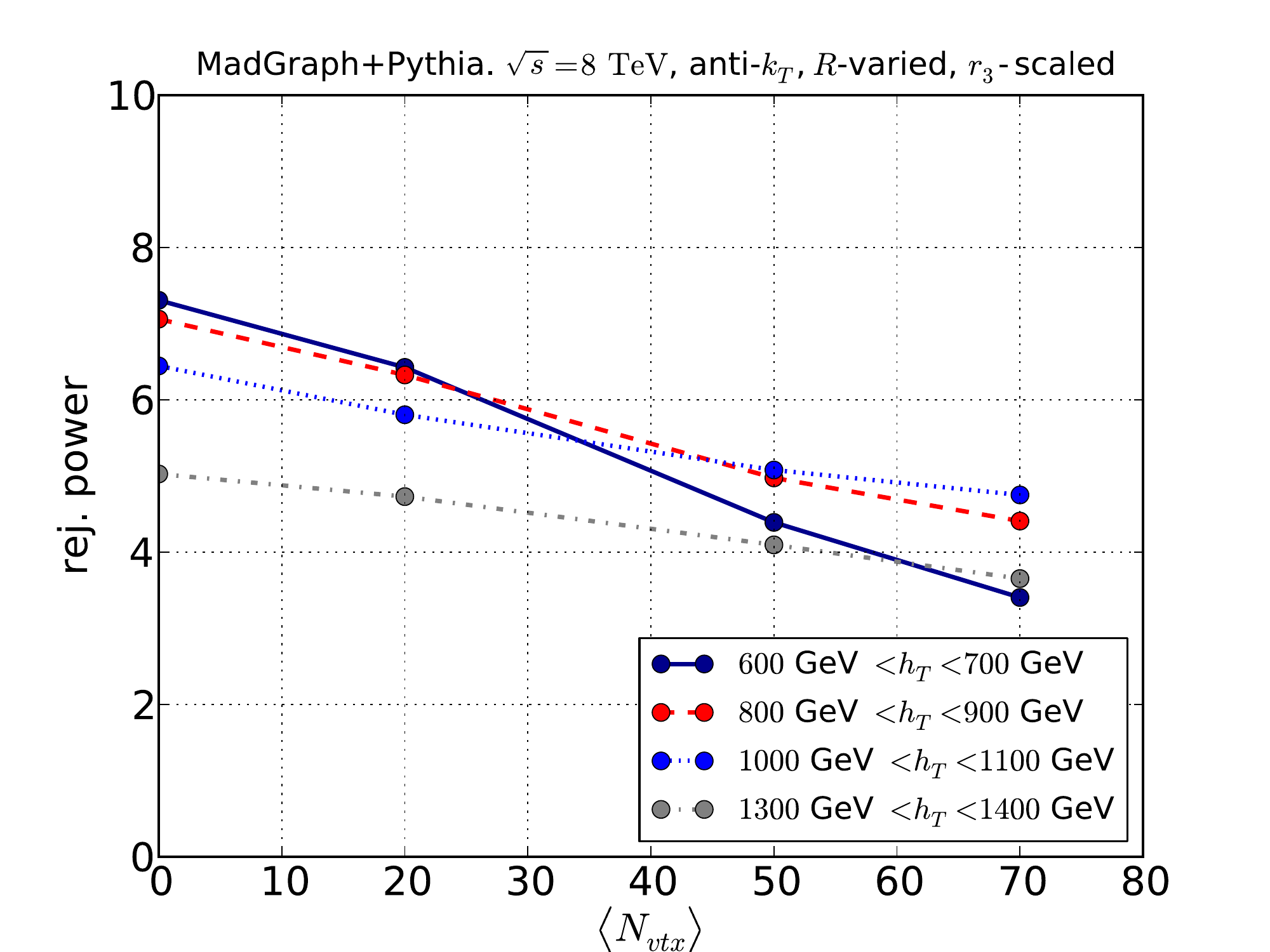}
\caption{Effects of pileup on the hadronic overlap analysis with a fixed overlap cut. The top left panel shows the signal efficiency for a fixed cut of $Ov^{had}_3 > 0.6$. Different curves represent different $h_T$ bins. The top right panel shows the $Ov_3^{had}$ distributions with various levels of pileup contamination and $600 \GeV < h_T < 700\GeV$.  The lower left panel shows the corresponding $W$+jets fake rate for a fixed  $Ov^{had}_3 > 0.6$. The bottom right panel shows the corresponding $W$+jets rejection power. The analysis does not assume a mass cut on the fat jet or $b$-tagging. The signal efficiency and background fake rate are measured relative to the Basic Cuts of Eq.~\eqref{eq:bc}.} 
\label{fig:sig_eff_pileup}
\end{center}
\end{figure}

We begin with the study of pileup effects on $Ov_3^{had}$. The benchmark points of $\langle N_{vtx} \rangle = {20, 50, 70}$ pileup events per bunch crossing serve to illustrate the performance of TOM in a pileup environment, while several $h_T$ bins serve to illustrate the effects of pileup at various jet transverse momenta.

The top left panel of Fig.~\ref{fig:sig_eff_pileup} shows the signal efficiency with a fixed $Ov_3^{had} > 0.6$ cut at various levels of pileup contamination. Each $h_T$ bin shows virtually no dependence on levels of pileup, with the signal efficiency staying constant at $\approx 80 \%$. For illustration we also show a $Ov_3^{had}$ signal distribution in the top right panel of Fig.~\ref{fig:sig_eff_pileup} at different levels of pileup for $600 \GeV < h_T < 700 \GeV$ (where the effects of pileup are supposed to be most severe due to $R=1.0$). Even at high pileup contamination of 70 interactions per bunch crossing, we do not encounter significant effects on the shape of the $Ov_3^{had}$ distribution of the signal. The top right panel of Fig. \ref{fig:sig_eff_pileup} thus illustrates very well that TOM indeed properly tags the spiky energy depositions within the fat jet.

Contrary to the behavior of the $t\bar{t}$ hadronic overlap distribution, the $W$+jets background analysis exhibits some dependence on pileup contamination. The lower left panel in Fig.~\ref{fig:sig_eff_pileup} shows the dependence of the $W$+jets fake rate at various levels of pileup and fixed $Ov_3^{had} > 0.6$. Addition of soft radiation and a shift in the fat jet mass allow for more efficient mis-tagging of the $W$+jets events. Naturally, the larger cones characteristic of low $h_T$ bins exhibit the strongest dependence on the level of pileup contamination, while the effects are softened at higher $h_T$ due to the use of both smaller fat jet cones and the smaller average size of the template sub-cone. Notice that the effects of $N_{vtx} = 20$ interactions per bunch crossing are mild in the entire $h_T$ range, consistent with our previous analysis in Ref.~\cite{Backovic:2012jj}. The result shows that TOM can perform well without significant pileup subtraction or correction on the current $8 \TeV$ data set, while alternative ways of dealing with pileup are likely to be needded beyond $\langle N_{vtx}\rangle  > 50 $.

\subsection{Pileup Effects on Leptonic Peak Overlap ($Ov_3^{lep}$)}
Effects of pileup contamination on $Ov_3^{lep}$ are less severe than in the case of hadronic overlap. The leptonic top $b$-quark, clustered with a small cone of $r=0.4$ displays limited sensitivity to soft contamination compared to the fat jet, while the hard lepton remains mostly unaffected by soft hadronic noise and the effects of pileup on missing energy can be efficiently corrected~\cite{Vichou:2012roa}.

\begin{figure}[htb]
\begin{center}
\includegraphics[width=3.2in]{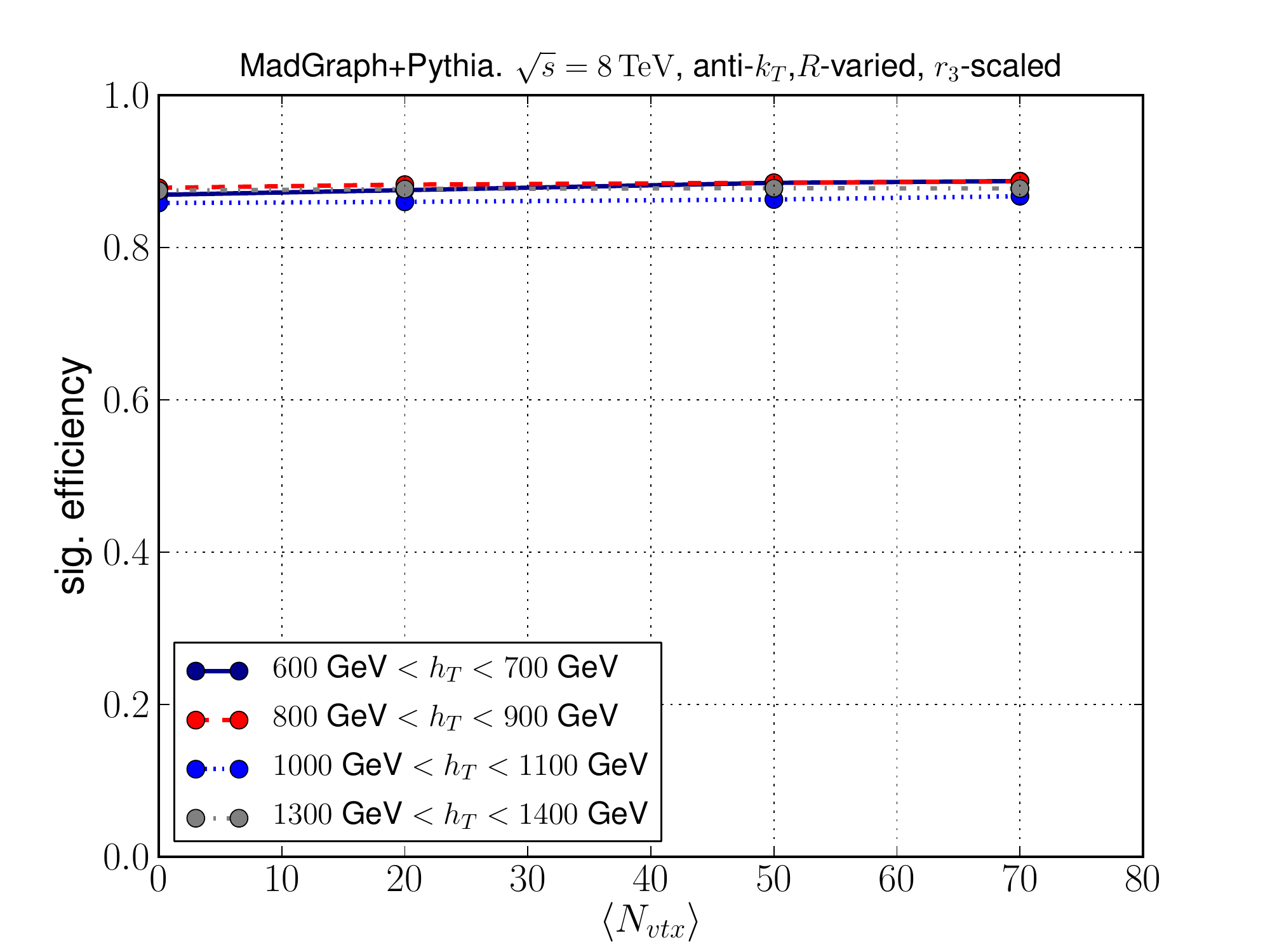}\includegraphics[width=3.2in]{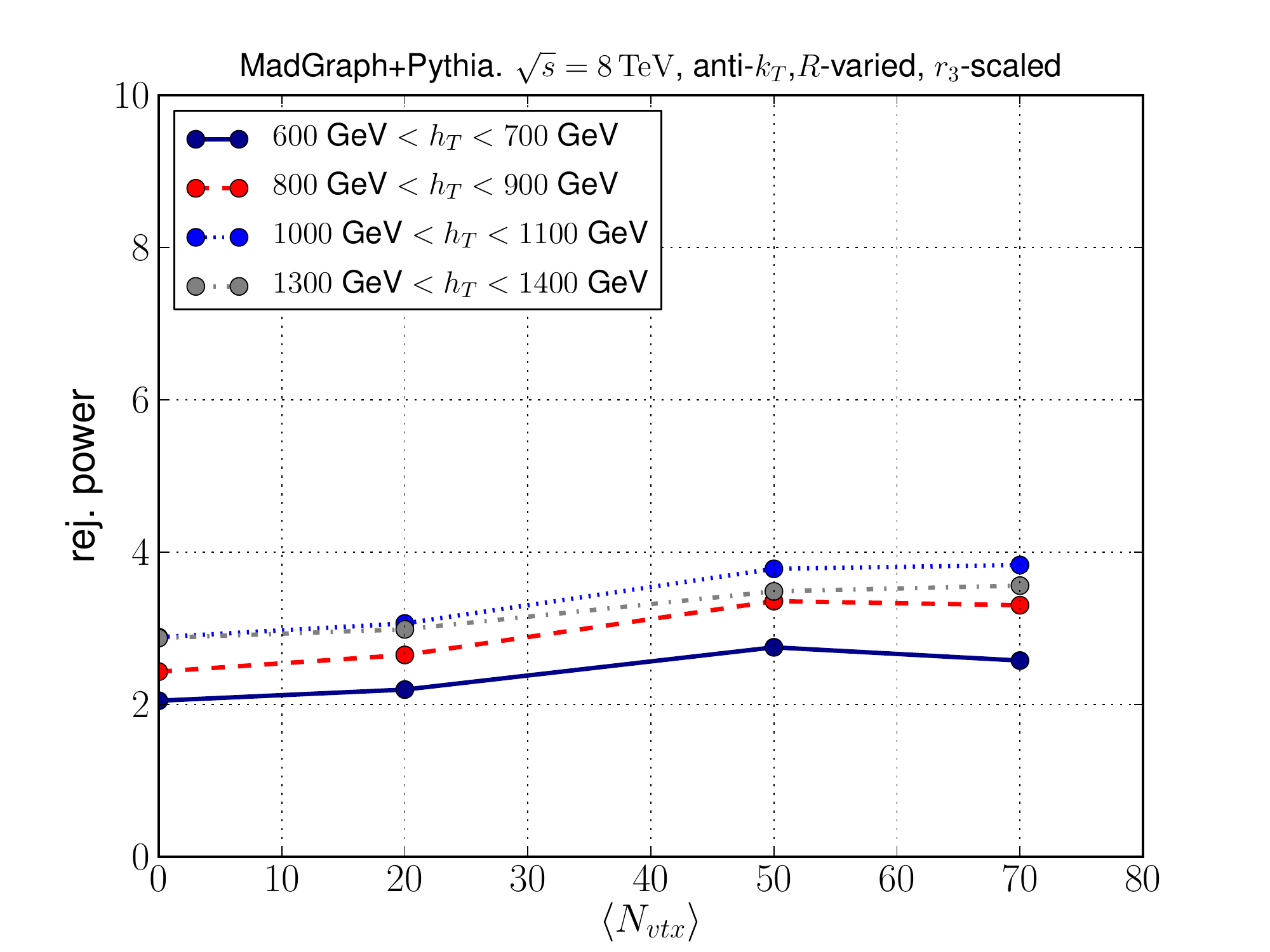}

\caption{Effects of pileup on background fake rate and rejection power of leptonic overlap for $Ov_3^{lep} > 0.6$. The left panel shows the $t\bar{t}$ signal efficiency, where different curves represent different $h_T$ bins at various levels of pileup contamination. The right panel shows the corresponding rejection power. The analysis does not assume a mass cut or $b$-tagging. All efficiencies are measured relative to the Basic Cuts of Eq.~\eqref{eq:bc}. } 
\label{fig:ov3lepPileup}
\end{center}
\end{figure}

Figure~\ref{fig:ov3lepPileup} shows dependence of the signal efficiency and background rejection power for $Ov_3^{lep} > 0.6$ and various levels of pileup contaminations. The signal efficiency relative to the Basic Cuts of Eq.~\eqref{eq:bc} remains constant, while the rejection power appears to slightly increase with the increased presence of pileup. This effect is fully due to the fact that Basic Cuts include a requirement that there is at least one anti$-k_T$ $r=0.4$ jet within $\Delta R = 1.5$ from the mini-isolated lepton. Consider for instance $W$+jets events with $ 600 \GeV < h_T < 700 \GeV$. The background to leptonic tops consists of a leptonically decaying $W$ and an uncorrelated jet which falls within $R=1.5$ from the lepton. 
 Table~\ref{tab:Ov3lpileup} shows the fraction of events which pass the Basic Cuts at various levels of pileup. If no pileup is present, 11$\%$  of background events pass the Basic Cuts. In the presence of pileup, the fraction increases to as much as $21 \%$ due to soft contamination being more likely to ``fake'' an un-correlated jet which happens to fall within $R=1.5$ from the lepton. After imposing a $Ov_3^{lep} > 0.6$, the fraction of surviving events goes down to $\approx 5 \%$ , regardless of levels of pileup, thus decreasing the overall fake rate, and increasing the rejection power. We conclude that leptonic top overlap remains impervious to pileup up to at least 70 interactions per bunch crossing.

\begin{table}
\begin{center}

\begin{tabular}{|c|c|c|}
\hline
$\langle N_{vtx} \rangle$ &  events after Basic Cuts$\, (\%)$ & events after $Ov_3^{lep} > 0.6 \,(\%)$ \\ \hline

0 & 11.0 & 5.0\\
20 & 13.0 & 5.0 \\
50 & 14.0  & 6.0 \\
70 & 21.0 & 6.0 \\

\hline

\end{tabular}

\label{tab:Ov3lpileup}
\end{center}
\caption{Fraction of all $W$+jets events which pass the Basic Cuts of Eq. \eqref{eq:bc} and the additional $Ov_3^{lep} > 0.6$ cut, at various levels of pileup contamination. Here we consider only events with $600 \GeV< h_T < 700 \GeV$. }
\end{table}

\section{Searches for Top-philic New Physics with TOM} \label{sec:BSM}

The results shown above suggest that TOM is able to discriminate against the SM $t\bar{t}$ reducible, and non-$t\bar t$ irreducible backgrounds while mantaining a relatively high signal efficiency. The  template top tagger is particularly useful in the search for NP in $t\bar t$ resonances, 
as it can efficiently remove contributions from asymmetric SM $t\bar t$ events.

Many models provide possibilities 
for new interactions with enhanced couplings to top quarks. To demonstrate the effectiveness of the TOM, we present a simple search 
for a massive spin-one, $t\bar t$ color octet  resonance in the lepton plus jets channel for the past run of the LHC at $\sqrt{s} = 8 \TeV$.  
We further study the case of heavy NP characterised by effective field theory (EFT).  Specifically, we add a four-fermion $u\bar u t \bar t$ operator capable of accommodating the discrepancy in the Tevatron $t\bar t$ forward-backward asymmetry.

Our analysis focuses on the kinematic range in which the $t\bar t$ system has sufficient energy for
the decay products of each top quark to be fully merged into fat jets of $R \sim 1$, leading to di-jet topologies in which
the events have one fully merged hadronic decaying top-quark candidate and one fully merged leptonic 
decaying top-quark candidate.  Event reconstruction follows the same steps as in Sec.~\ref{sec:preselection}. 
For events passing the Basic Cuts of Eq.~\eqref{eq:bc}, we further demand that the semileptonic $t\bar t$ candidate contain two top tags and satisfy
extra cut on template $m_{t\bar{t}}$ consistent with the decay of a heavy resonance.

\subsection{Benchmark models}

We consider new physics in two specific benchmark models.
In the first case, we consider Kaluza-Klein (KK) gluons from the bulk Randall-Sundrum model (RS)~\cite{Agashe:2006hk, Lillie:2007yh} with $\Gamma_{\rm KK}/M_{\rm KK} = 15 \%$. 
Neglecting effects related to Electro-weak symmetry breaking (EWSB), the left-handed ($g_L$) and right-handed
$(g_R)$ couplings to quarks in this model are 
\ba
g_L^{q\bar q} =-\frac15 g_s \, ,\,\,\,\,\,  & \,\,\,\,\,\,  g_R^{q\bar q} =-\frac15 g_s \nn\, ,\\
g_L^{b\bar b} =g_s\,, \,\,\,\,\,  & \,\,\,\, \,\, g_R^{b\bar b} = -\frac15 g_s\, ,\\
g_L^{t\bar t} = g_s \,,\,\,\,\,\,  & \,\,\,\, \,\, g_R^{t\bar t} = 4g_s\, , \nn
\ea
where $q=u,d,c,s$ and $g_s $ is the SM $SU(3)_C$ gauge coupling. Masses below $\approx 2\TeV$ for KK gauge particles are 
disfavored by precision tests~\cite{Carena:2006bn,Carena:2007ua}, while direct constraints from CMS limit the KK gluon to be heavier than roughly $2.5 \TeV$~\cite{Chatrchyan:2013lca}, assuming a signal K-factor of 1.3, derived from color singlet NLO analyses~\cite{Gao:2010bb}. As here we consider a color octet resonance, we conservatively do not apply this K-factor.
We consider two specific KK gluon masses: $M_{\rm KK} = 2.5 \TeV$  and $M_{\rm KK} = 3 \TeV$. In this mass regime, KK gluons decay dominantly to $t\bar t$ with a branching ratio of $\approx 95 \%$. 

As a second example, we consider a non-resonant top-philic NP model. Assuming that new physics is heavy enough, one can take an EFT approach to describe the NP by means of higher dimensional interactions among the SM fields. For simplicity, we focus on the operator
\begin{equation} \label{eq:eft}
 {\cal L}_{\rm EFT} = \frac{g_A}{\Lambda^2} (\bar u \gamma_{\mu} \gamma_5 T^a u)(\bar t \gamma^{\mu} \gamma_5 T^a t)  \, ,
\end{equation}
where $\Lambda$ is the scale of the new interaction, $T^a$ being an $SU(3)$ generator ($a=1...8$) and $g_A$ is the ``axigluon'' coupling. 
This  presence of this operator can be motivated by the anomalous top forward-backward asymmetry at the Tevatron (see {\it e.g.}~\cite{Jung:2009pi,Degrande:2010kt,Jung:2010yn,Blum:2011up,Delaunay:2011gv,AguilarSaavedra:2011vw,Delaunay:2012kf}). 
As a reference point, we chose $g_A/\Lambda^2\sim1.4/{\rm TeV}^2$, which can account for the observed asymmetry.  Since $\Lambda$ is relatively low, we expect a strong enhancement of the differential $t\bar t$ production cross section. It is worth noting that
the heavy NP described by the above EFT is already in a tension with the recent CMS search for anomalous $t\bar t$ production of Ref.~\cite{Chatrchyan:2013lca}.

\subsection{LHC signals}

We simulate the signal and background samples using the procedure described in Sec.~\ref{sec:preselection}. The events are required to satisfy the Basic Cuts described in Eq.~\eqref{eq:bc} with the fat jet transverse momentum $p_T > 500 \GeV$. 
In Table~\ref{tab:BSM_xs}, we summarize the cross sections considered after the Basic Cuts. 
While the signal cross sections are computed at LO, the background cross sections are  obtained with MadGraph normalized to the theoretical cross sections of Ref.~\cite{Czakon:2013goa} (for $t\bar t$ at NNLO),  and Ref.~\cite{Bern:2013zja} (for $Wjj$ at NLO).

\begin{table}[htb]
\begin{center}
\begin{tabular}{|c|c|}
\hline

 Sample & Cross-section $\times$ BR \\
 \hline
 $t\bar t \to b \bar b j j \ell \nu_{\ell}$  & 140 ${\rm \, fb}$ \\
  $W+ j j\to \ell \nu_{\ell} j j$  & 520 ${\rm  \, fb}$ \\
   ${\rm KK} (2.5 \TeV) \to t\bar t \to b \bar b j j \ell \nu_{\ell} $ & 5.8 $\rm \,fb$\\
   ${\rm KK} (3.0 \TeV) \to t\bar t \to b \bar b j j \ell \nu_{\ell} $ & 2.0 $\rm \,fb$ \\
   EFT $\to t\bar t \to b \bar b j j \ell \nu_{\ell}$  & 110 $\rm \, fb$
  \\  \hline

\end{tabular}

\end{center}
\caption{Signal and background cross sections times branching ratios considered in the analysis in the $\ell$+jets final state at $\sqrt{s}= 8 \TeV$ after  Basic Cuts of Eq.~\eqref{eq:bc}.
The listed numbers assume a leptonically decaying $W$ with first two generations of leptons included.}

\label{tab:BSM_xs}
\end{table}

\begin{figure}[tbp]
\begin{center}
\includegraphics[scale = 0.5]{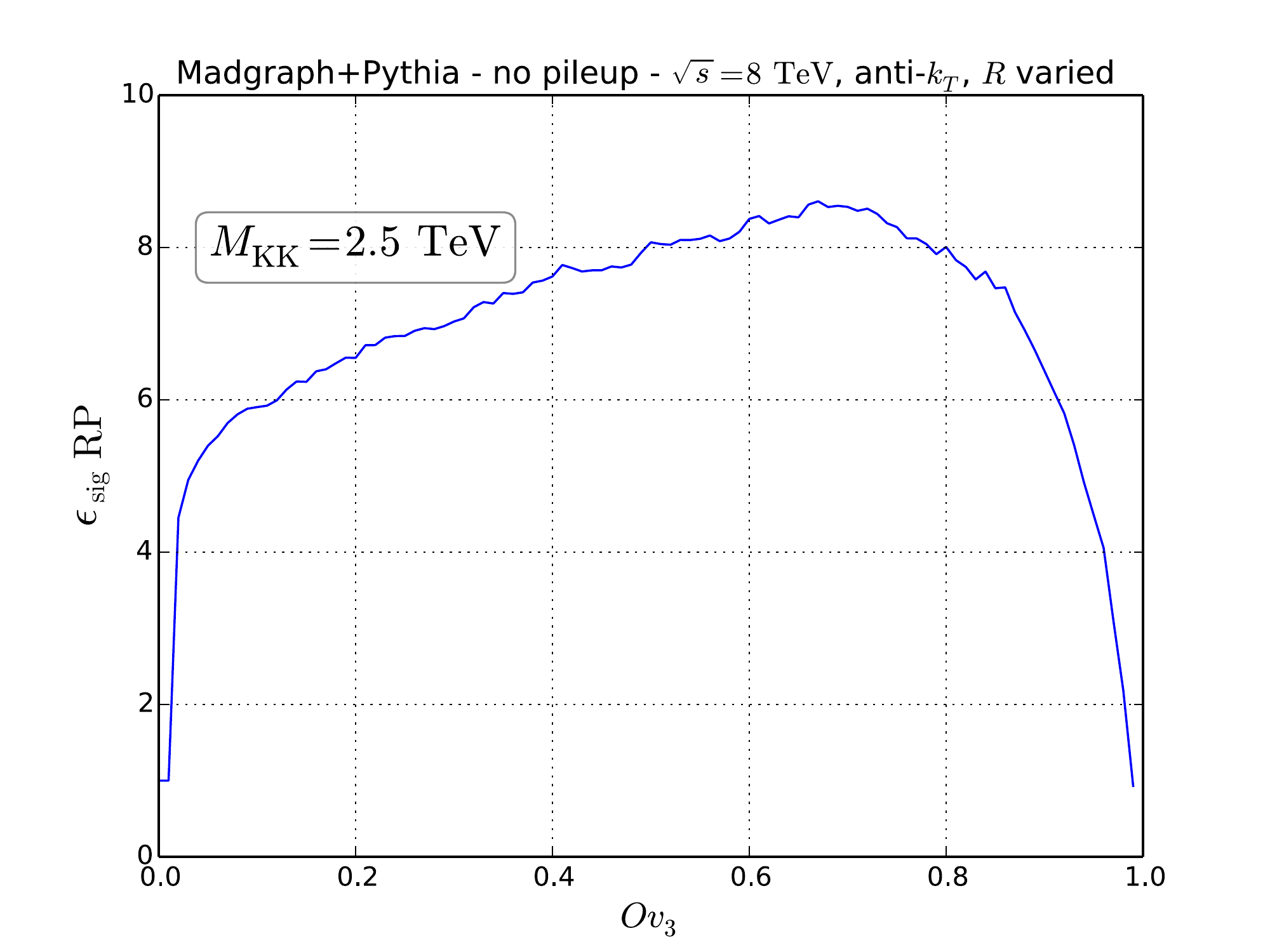}
\caption{Optimization of the $Ov_3$ cut to maximize $S/\sqrt{B}$. The curve shows a maximum ${\rm RP} \times \epsilon_{\rm sig}$ as a function of the lower cut on $Ov_3$. }
\label{fig:SBoptimal}
\end{center}
\end{figure}

For events passing the basic reconstruction, we further demand that the semi-leptonic $t\bar{t}$ candidates satisfy the overlap cuts,
\begin{equation}
 Ov_3^{had} >0.7\, , \,\,\,\,\,\,\,\,\,\,\,\,\,\,\,\,\,\,\,\,\,\,\,\,\,\,\,Ov_3^{lep} >0.7\, . \label{eq:NPcuts}
\end{equation}
The top-tagging algorithm for the hadronic and leptonic top-quark candidates was described in Sec.~\ref{Sec:Ov}. 
To optimize the choice of $Ov_3$ cut for the NP search, we turn to the expression for $S/\sqrt{B}$ as a function of background rejection power, RP, and the signal efficiency $\epsilon_{\rm sig}$\,: 
\be
	\frac{S}{\sqrt{B}} = \sqrt{\mathcal{L}} \times \sqrt{{\rm RP} \times \epsilon_{\rm sig}} \times \frac{\sigma(S)^{\rm BC}}{\sqrt{\sigma(B)^{\rm BC}}} \, ,
\ee
where $\epsilon_{\rm sig}$ is the efficiency of the $Ov_3$ cut relative to basic cuts of Eq.~(\ref{eq:bc}).  Luminosity, $\mathcal{L}$, and Basic Cuts are fixed by the experimental setup, hence the maximum $S/\sqrt{B}$ occurs at the value of $Ov_3^{\rm min}$ which maximizes ${\rm RP} \times \epsilon_{\rm sig}\,$. Figure~\ref{fig:SBoptimal} shows the result, with $Ov_3^{\rm min} = 0.7$ maximizing $S/\sqrt{B}\,$. 

\begin{figure}[htb]
\begin{center}
\begin{tabular}{cc}

\includegraphics[width=3.5in]{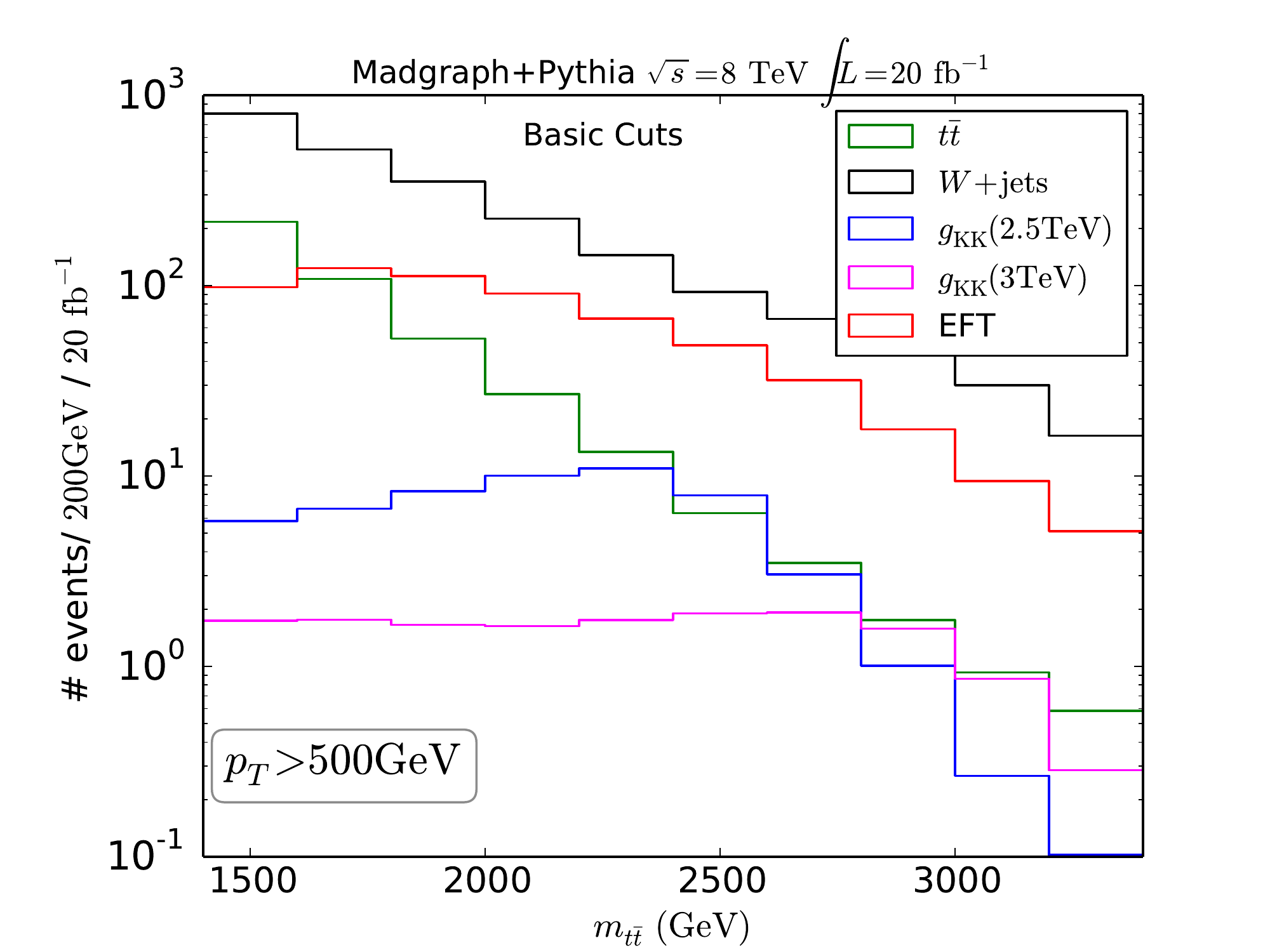} &
\includegraphics[width=3.5in]{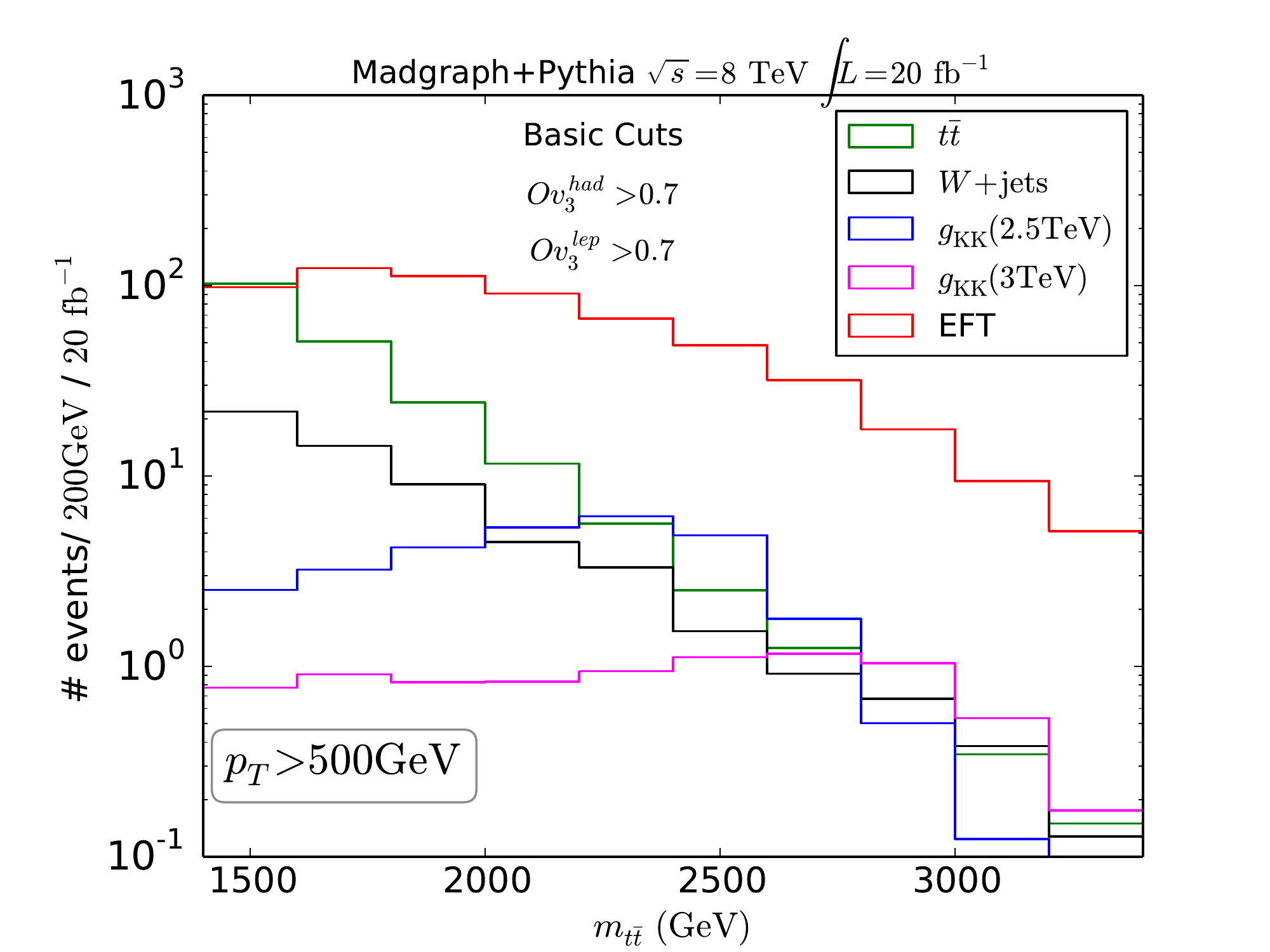}
\end{tabular}

\caption{The reconstructed $m_{t\bar t}$ for SM $t\bar t$ and $W+\rm jets$ backgrounds, and a bulk RS KK gluon with $M_{\rm KK}=2.5\TeV$, $M_{\rm KK}=3\TeV$ and the EFT model of Eq.~\eqref{eq:eft}.
We show the distributions before (left panel) and after (right panel) two top tags. The $t$ and $\bar t$ decay semi-leptonically, with no $b$-tagging.  The left panel shows the distributions after then basic cuts of Eq.~\eqref{eq:bc}, while the right panel contains additional cuts of Eq.~\eqref{eq:NPcuts}.  Note that ``reconstructed $m_{t\bar{t}}$'' implies that the di-jet invariant mass was calculated from peak template states.} 
\label{fig:mttbar_bsm}
\end{center}
\end{figure}

We show the reconstructed $m_{t\bar t}$ distribution after imposing the basic cuts in Fig.~\ref{fig:mttbar_bsm} (left panel) for 
the two dominant backgrounds 
and three different benchmark signal models described above. Here  $m_{t\bar t}$ denotes the invariant mass of the leptonic and hadronic \textit{peak templates}. 
The curves are for
SM $t\bar t$ production (green), SM $W+ \rm jets$ production (black), KK gluon production with $M_{\rm KK} = 2.5 \TeV $ (magenta) and $M_{\rm KK} = 3 \TeV$ (blue),
and the $t\bar t$ production via the EFT described in Eq.~\eqref{eq:eft} (red).
Both the SM $t\bar t $ and $W+\rm jets$ production rates fall steeply as a function of the $t\bar t$ mass.
From the left panel of Fig.~\ref{fig:mttbar_bsm} it is clear that the dominant background in this analysis in the absence of top tags 
is from $W$+jets events rather than from SM $t\bar t$ production. 

Further purification of the signal can be achieved by applying cuts on the hadronic and leptonic top jets. The right panel of  Fig.~\ref{fig:mttbar_bsm} shows that the SM non-$t\bar t$ background is significantly  reduced once an overlap cut is applied to the top jet candidates. 
 We see that SM $t\bar t$ dominates over $W$+jets for  $m_{t\bar t }< 2.5 \TeV$, while the long tail in the invariant mass distribution of the $W$+jets background is comparable to SM $t\bar t$ for $m_{t\bar t  }>2.5 \TeV$.  The absence of $b$-tagging in our current study does not allow direct comparisons with the ATLAS study~\cite{TheATLAScollaboration:2013kha}, where $b$-tagging significantly reduces the $W$+jets background. However, it is worth noting that the background composition in our study and the one performed by ATLAS are similar for  $m_{t\bar t  }<2.5 \TeV$, with SM $t\bar t$ being the dominant background.

Note that most of the events from the high mass KK gluon resonances do not appear as a sharp resonance but instead
are smeared over a wide range of the 
$m_{t\bar t}$ distribution.  This effect is due to the fact that the KK gluon is rather broad and more importantly due to the convolution of the rapidly falling parton distribution functions with the Born cross section.  
Furthermore, obviously the contributions from the EFT operator, being irrelevant, are increasing with energy.
 Therefore, the $t\bar t$ spectrum tends to be harder in presence of new physics.
To improve the signal to background ratio, we apply a sliding lower cut $m_{t\bar t}> m_{\rm min}$ for each resonance, conveniently adjusted to give an approximately flat $S/B$ ratio.

\begin{figure}[htb]
\begin{center}
\begin{tabular}{cc}
\includegraphics[width=4.0in]{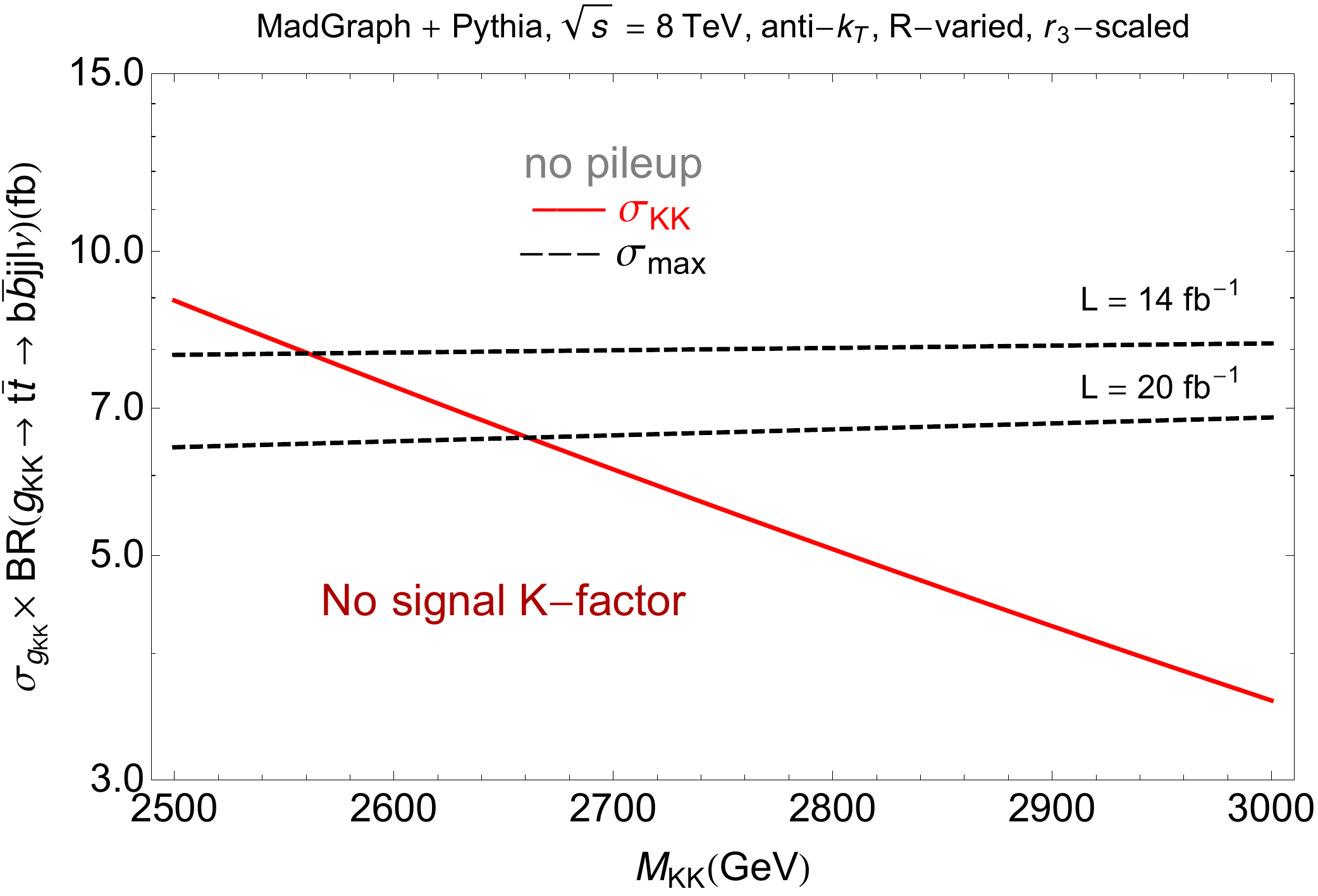} &
\end{tabular}
\caption{95$\%$ confidence level reach for the KK gluon search at $\sqrt{s} = 8 \TeV$ with $\mathcal{L} = 14 \,{\rm fb}^{-1}, \,20 \,{\rm fb}^{-1}$. The dashed line represents the upper limit on the cross section whereas the solid line is the leading order KK gluon cross section. The symbol $\ell = e, \mu$ in the axis label.} 
\label{fig:sensitivity}
\end{center}
\end{figure}

In order to determine the reach, we apply a simplified Bayesian approach using a flat prior distribution  and neglecting systematic uncertainties \cite{Beringer:1900zz}. We assume that the probability of measuring $n$ events is given by a Poisson distribution
\begin{equation}
P(n|S,B) = \frac{(S+B)^n}{n!} e^{-(S+B)}\,,
\end{equation}
where $B$ and $S \equiv  \sigma_{\rm sig}\epsilon_{\rm sig} \mathcal{L} $ are the number of expected background and signal events, respectively. Here we regard $\sigma_{\rm sig}$ as a free parameter in order to consider different signal production rates and fix $B$ and $\epsilon_{\rm sig}$ according to our expectations based on the Monte Carlo distributions. 
 An upper limit for $\sigma_{\rm sig}$  at confidence level ${\rm CL}=1-\alpha$ can be constructed by integrating the posterior probability, 
\begin{equation} \label{eq:bayes}
{\rm CL} = 1- \alpha =\frac{ \int_{0}^{\sigma^{ \rm CL}} P(n | \sigma_{\rm sig} \epsilon_{\rm sig} \mathcal{L} ,B) d{\sigma_{\rm sig}}}{ \int_{0}^\infty P(n|\sigma_{\rm sig} \epsilon_{\rm sig} \mathcal{L},B) d\sigma_{\rm sig} }\,.
\end{equation}
 We  assume that $n$ is equal to the integer closest to $B$, and solve Eq.~(\ref{eq:bayes}) for $\sigma^{ \rm CL}$ assuming $\alpha =0.05$ ($95\%$ exclusion). Figure~\ref{fig:sensitivity} shows the results for the projected $95\%$ CL exclusion of the KK gluon search at $\sqrt{s} = 8 \TeV$ and $\mathcal{L} = 14 \,{\rm fb}^{-1},\, 20 \, {\rm fb}^{-1}$. 
 We find  that KK gluon masses up to $\approx 2.6 \TeV$ can be excluded with $\mathcal{L} = 14 \, \rm fb^{-1}$, and masses up to $\approx 2.7 \TeV$ with $\mathcal{L} = 20\, \rm fb^{-1}$, assuming  no $b$-tagging, no pileup, no detector effects and no signal K-factor.

Table~\ref{tab:BSM} summarizes our results for the sensitivity to the KK gluon and EFT examples. In the KK gluon case, 
TOM is able to improve $S/B$ relative to the Basic Cuts by a  factor of $\approx 15$ at $M_{\rm KK} =2.5, \, 3.0 \TeV$, while the signal significance, although too low to claim discovery, 
improves roughly three-fold for $M_{\rm KK} =2.5 \TeV$ and two-fold at $M_{\rm KK} =3.0 \TeV$. The fact that the efficiency of the overlap cut on SM $t\bar{t}$   is somewhat lower than that of signal $t \bar{t}$ events is another indication that higher order effects are more significant in SM $t\bar t$ events, as discussed in Sec.~\ref{sec:NLO}. Notice that for $m_{t\bar {t}} > 2.55 \TeV$ case, the efficiency of $Ov_3$ cuts on both the SM $t\bar{t}$ and the signal events is comparable, while at lower $m_{t\bar{t}}$ the signal efficiency is clearly higher. The effect is in part due to the fact that  $A_{t\bar{t}}$ inversely scales with $m_{t\bar t}$ for fixed event $H_T$. Selecting only events with large template $m_{t\bar{t}}$ thus already selects both the SM and BSM events with small $A_{t\bar{t}}$ at which point the  further ability of TOM to distinguish SM $t\bar{t}$ from signal events diminishes. 

It is also important to mention that our simulation of NP scenarios includes only the real emissions through matching with no contributions from the virtual part of the NLO diagrams. Yet, we expect the asymmetry in BSM $s$-channel $t\bar t$ events not to  be particularly large due to the absence of diagrams with soft and collinear singularities.  Having used MadGraph+Pythia for the simulation of signal events, we found it inappropriate to compare the signal events to the background samples generated at full NLO with Powheg (which we used in some of the previous sections to study TOM in the context of $A_{t\bar{t}}$), and instead we opt for background samples generated with the same Monte Carlo tools as the signal. The choice of LO background samples means that only the real emissions will contribute to the background $A_{t\bar{t}}$, leading to a likely lower SM $t\bar{t}$  asymmetry.
Based on our NLO study of section~\ref{sec:NLO}, we expect that the power of Template Overlap to distinguish between SM and BSM $t\bar t$ events will increase in samples produced with NLO accuracy.  

\begin{table}[htb]
\begin{center}
\begin{tabular}{|c|c|c|c|c|c|c|}
\hline
Model & \multicolumn{2}{|c|}{$M_{\rm KK} = 2.5 \TeV$}& \multicolumn{2}{|c|}{$M_{\rm KK} = 3.0 \TeV$}& \multicolumn{2}{|c|}{EFT} \\ \hline
$m_{t\bar t}^{\rm min}$& \multicolumn{2}{|c|}{$ 2125 \GeV$}& \multicolumn{2}{|c|}{$ 2550 \GeV$}& \multicolumn{2}{|c|}{$ 2000 \GeV$} \\ \hline
$Ov_3^{\rm min}$  & $0 $& $0.7$&$0$&$0.7$ &$0$&$0.7$ \\
 \hline
 $\sigma_{t\bar t} $ (${\rm fb}$)& 1.8& 0.75 &0.43 &0.14 &2.7& 1.1 \\
  $\sigma_{W+\rm jets} $ (${\rm fb}$)& 30&  0.51 &  13 & 0.15 &38&0.67\\
   $\sigma_S$ (${\rm fb}$) & 1.4 & 0.82  &  0.46 & 0.16 &13.0 & 12.0\\  \hline
  
  $S/B$ & 0.04 & 0.65& 0.04&0.55 &0.3& 6.8\\ \hline
 $S/\sqrt{B}\,  (14.3\, {\rm fb}^{-1})$  & 0.9 & 2.8 & 0.5 & 1.1 &7.7&34\\ \hline
 $S/\sqrt{B} \,(20.0\, {\rm fb}^{-1})$ & 1.1 & 3.3 & 0.6 & 1.3 &9.1&40\\ \hline

\end{tabular}

\end{center}
\caption{Rejection power of $Ov_3^{had}$ and $Ov_3^{lep}$ at several benchmark efficiency points. The values in the column labeled by $Ov_3^{\rm min} =0$ assume the basic cuts of Eq.~\eqref{eq:bc}. }
\label{tab:BSM}
\end{table}

\section{Conclusions and Discussion}

In this paper we introduced a tagger for semi-leptonic $t\bar t$ events based on the Template Overlap Method (TOM). 
We demonstrated that at large boost the leptonic-top tagger leads to an additional rejection power of roughly 4. The tagger may serve to compensate or complement the rejection power lost due to the reduction of b-tagging efficiency. We showed that the semi-leptonic $t\bar{t}$ TOM tagger is by itself robust against pileup up to 50 interactions per bunch crossing, without the use of additional pileup correction techniques. The relative insensitivity of TOM to pileup may thus serve to study the systematic effects of other pileup correction techniques.

Furthermore, we demonstrated that TOM is able to efficiently reject events in which  $t\bar{t}$ pairs are produced in association with a hard gluon and hence single out the back to back $t\bar{t}$ events. Our results show that $Ov_3^{had}$ is able to provide an improvement of a factor of 2 in back to back $t\bar{t}$ signal purity compared to  the ATLAS tagger based on cuts on the $k_T$ splitting scale and the trimmed jet mass selection. Our method is able resolve the kinematic distributions of high energy top quark events to a reasonable degree, and better than the above-mentioned ATLAS tagger. The improvement in resolution is due to the fact that conventional approaches will often tag the extra hard jet as a hadronic-top candidate. 
The hadronic TOM rejects $W$+jets events at the rate of $\approx 10$ with the SM $t\bar{t}$ efficiency of $60 \%$ at $p_T \sim 500 \GeV$. The rejection power decreases with energy, due to the mentioned higher order effects that are characterized by hard and wide gluon emission and the gluon splitting function to a top quark pair. 

We performed a detailed study of pileup effects on TOM. To illustrate the performance of TOM in a high luminosity environment, we chose not to subtract pileup from our events. Instead, we introduce a simple approach to damp the effects of soft contamination on results of the overlap analysis. We introduced a method to estimate the $p_T$ of the hadronic top template based on the scalar sum of the leptonic top decay products' transverse momenta as a pileup insensitive alternative. In addition, we omitted the cut on the fat jet mass and instead relied on TOM's intrinsic mass filtering ability only. Our results revealed that the hadronic formulation of TOM is fairly robust agains soft contamination up to $\approx 50 $ interactions per bunch crossing, while the leptonic top TOM remaining weakly affected up to at least 70 interactions.

As a case study, we have investigated the performance of TOM  in the context of a KK gluon resonance and a non-resonant top-philic searches at a $8 \TeV$ LHC, with the major backgrounds consisting of SM $t\bar t $ continuum and $W+\rm jets$. The additional rejection power provided by our semi-leptonic top template tagger suggests that an analysis based on TOM can achieve a better sensitivity than previous analyses. In particular, we found that a KK gluon could be excluded up to masses of $ \approx 2.7 \TeV$ with $20 \, \rm fb^{-1}$ of data at 95\% CL. Non-resonant new physics contributions to $t\bar t$ production could in principle be excluded with the same efficiencies. 

Finally, we discussed many technical and experimental aspects of TOM in the Appendix. We showed that covering a wide range of top transverse momenta insists on a use of some sort of a template sub-cone scaling rule, while no single fixed value of sub-cone radii is adequate to provide a fixed efficiency for a fixed $Ov_3^{had}$ cut. Our study of missing energy resolution showed that TOM is mildly affected by smearing in $   \sl{E_T}\,\,\,  $  with the effects on $Ov_3^{had}$ being subdominant. The results on overlap analysis are also fairly insensitive to the angular resolution of the template momenta, with the rejection power at fixed efficiency remaining unaffected after $50$ steps in $\eta, \phi$.
Further analysis of TOM for $t\bar{t}$ tagging would benefit from inclusion of full detector simulation. 

\bf{Acknowledgments:} We thank Trisha Farooqe, Pekka Sinervo, Francesco Spano  and Ryan Underwood for giving us encouragement to work on this topic and providing valuable comments on technical aspects of TOM. We also thank Matteo Franchini and Matteo Negrini for contributing to the implementation of the TemplateTagger code into ATHENA, 
as well as Gavin Salam for useful discussions on higher order effects in $t\bar{t}$ production. 
We are very grateful to Seung J. Lee for stimulating discussions about TOM and comments on BSM searches. JJ and MB would like to thank the CERN theory group for their hospitality
where a portion of this work was completed.
GP is supported by GIF, Minerva, IRG, ISF grants and by the Gruber award.

\appendix

 \section{Techincal Details of TOM and Template Properties}
 
 \subsection{Template Generation}
 
In order to speed up the overlap calculation we generate template states at fixed jet $p_T$ in the boosted frame. 
This requires us to generate several sets of templates and dynamically determine which set to use on an event by event basis. 
In this paper, unless otherwise noted, we use twelve template libraries starting at $p_T = 550 \GeV$ in increments of $\delta p_T = 100 \GeV$. 
The 3-particle top templates are determined by two four momenta, $p_1$ and $p_2$, subject to the constraints, 
\begin{equation} \label{eq:temp_gen}
(P-p_1-p_2)^2=0\,, \,\,\,\,\,\,\,\,\,\,\,\,\,(p_1+p_2)^2 = m_W^2\,,\,\,\,\,\,\,\,\,\,\,\,P^2=m_t^2\,, 
\end{equation}
where $P$ is the top total momentum, while $m_t$ and $m_W$ are the top and $W$ mass respectively. 
The third four momentum, $p_3$, is determined by momentum conservation. By solving Eq.~(\ref{eq:temp_gen}), we can 
generate the templates  with a sequential scan over  $\eta, \phi$ of the first two template momenta.

\subsection{Template Subcone Scaling} \label{sec:temp_scaling} 

The shape of both the signal and background overlap distributions is dependent on the choice of the template sub-cone size. However, Ref.~\cite{Backovic:2012jj} showed that there typically exists a wide region of template sub-cone radii for which rejection power stays constant at a fixed signal efficiency. Previous implementations of TOM for top tagging utilized fixed template sub-cones which were optimized for a small range of fat jet $p_T$ values. Here we are interested in covering a range of $\mathcal{O}(1 \TeV)$ in fat jet $p_T$, and the question of whether fixed template sub-cones are an adequate approach remains open. 

\begin{figure}[htb]
\begin{center}
\includegraphics[width=3in]{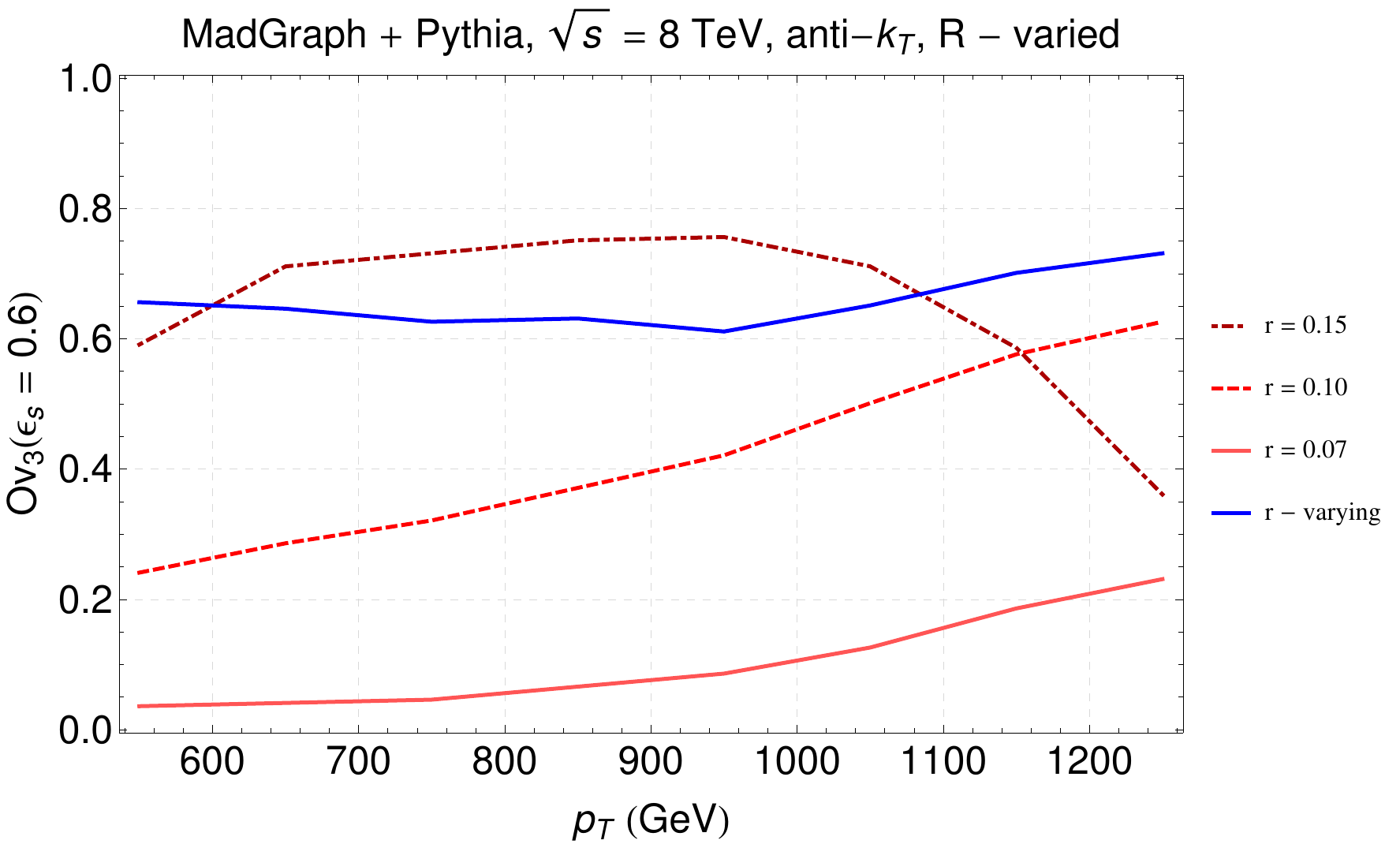}\includegraphics[width=3in]{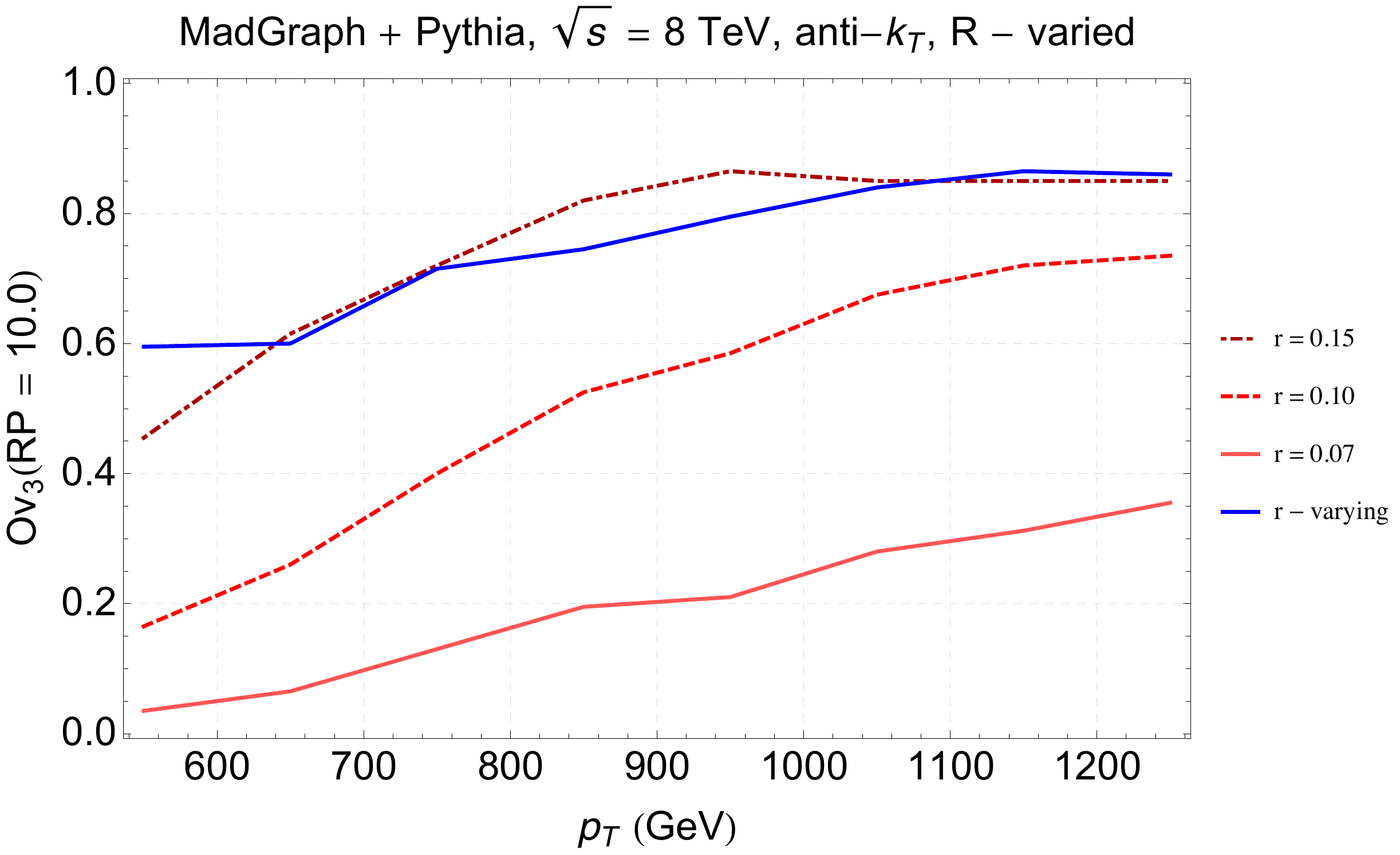} \\ \includegraphics[width=3in]{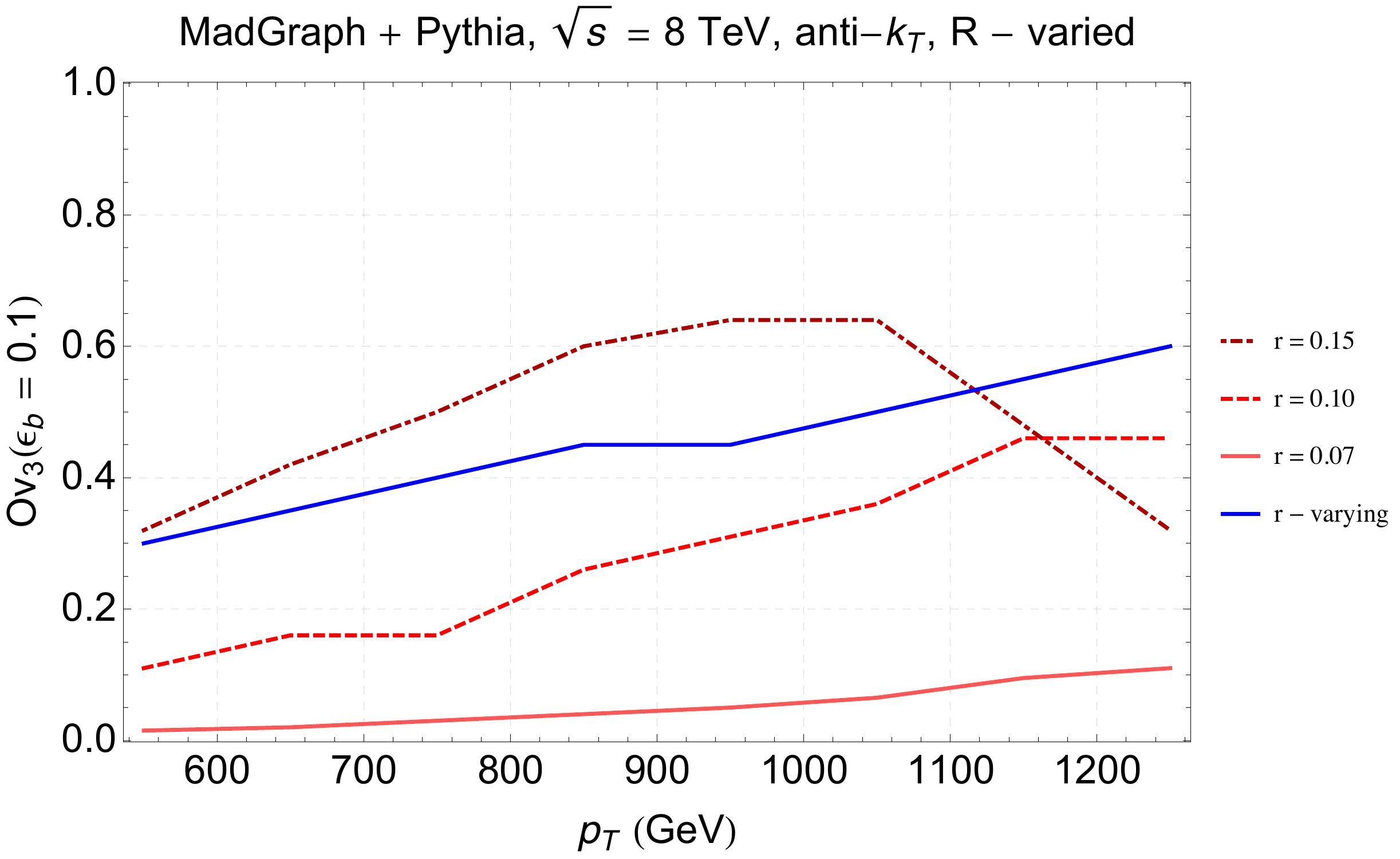}
\caption{Using fixed and varying template sub-cones for a wide range of fat jet $p_T$. Unless explicitly specified, the template sub-cones are fixed for every template parton. The varying sub-cones assume the rule of Eq. \eqref{eq:temprscale}. The curves show the location of the $Ov_3^{had}$ cut needed for the signal efficiency of $60 \%$  (top left panel), RP of 10 (top right panel) and background fake rate of $10\%$ (bottom panel), as a function of fat jet $p_T$.} 
\label{fig:fix_vs_scale}
\end{center}
\end{figure}

Naturally, one should expect the radiation pattern of a, say, $p_T = 100 \GeV$ quark to be wider that the radiation pattern of a $1 \TeV$ quark. Hence, the template sub-cone which is ``adequate'' to match the higher energy subjet could be too small to accurately capture most of the showering pattern of a lower energy one. In addition, how will the change of the adequate template sub-cone size affect the shape of the overlap distributions at different $p_T$? What effect will the change in shape of the distributions have on the signal efficiency and the rejection power of a fixed $Ov_3^{had}$ cut? The true understanding of the dependence of adequate template sub-cone size on the energy of the subjet is a topic in non-perturbative QCD as is beyond the scope of our analysis. We instead turn to a more data-driven approach, whereby we compare the properties of fixed template sub-cones over a wide range of fat jet $p_T$ to a polynomial fit to template sub-cone scaling rule of Ref.~\cite{Backovic:2012jj}:
\be
	r_a(p_T) =  0.041 + \frac{12.1}{p_{T,\, a}} - \frac{122.1}{p_{T,\, a}^2} \label{eq:temprscale}\, ,
\ee
where $p_{T,\, a}$ is the transverse momentum of a \textit{template parton}. We limit the template sub-cone sizes to be in the range $[0.05, 0.3]$, where the lower limit serves to take into account the detector resolution, while we set the upper limit to the value beyond which no data points exist (see Ref.~\cite{Backovic:2012jj} for more details).  

Varying sub-cones, while not necessarily providing an increase in rejection power at a fixed signal efficiency, have clear advantages over the fixed template sub-cones. Figure \ref{fig:fix_vs_scale} shows an example. The curves represent the $Ov_3^{had}$ cut which gives $60\%$ top tagging efficiency over a wide range of fat jet $p_T$. Our results show that an increase in the template sub-cone shifts the signal distribution to the higher value of overlap for a fixed value of fat jet $p_T$, until about $r \approx 0.15$ when the template sub-cones become too large to fit into the angular scale of a high $p_T$ top decay (due to the non-overlapping template constraint).  Hence, there is no  single value of fixed $r$  which is able to provide a fixed signal efficiency with a $Ov_3^{had}$ cut that is not strongly dependent both on the $p_T$ of the fat jet and the template sub-cone value. 
In contrast, we see that the varying cones of Eq. \eqref{eq:temprscale} (blue, solid line) provide a stable signal efficiency for a fixed $Ov_3^{had}$ cut, over the entire range of considered $p_T$ values.

\subsection{Selecting Template $p_T$ Bins} \label{app:HTCorr}

\begin{figure}[htb]
\begin{center}
\begin{tabular}{cc}
\includegraphics[width=3.3in]{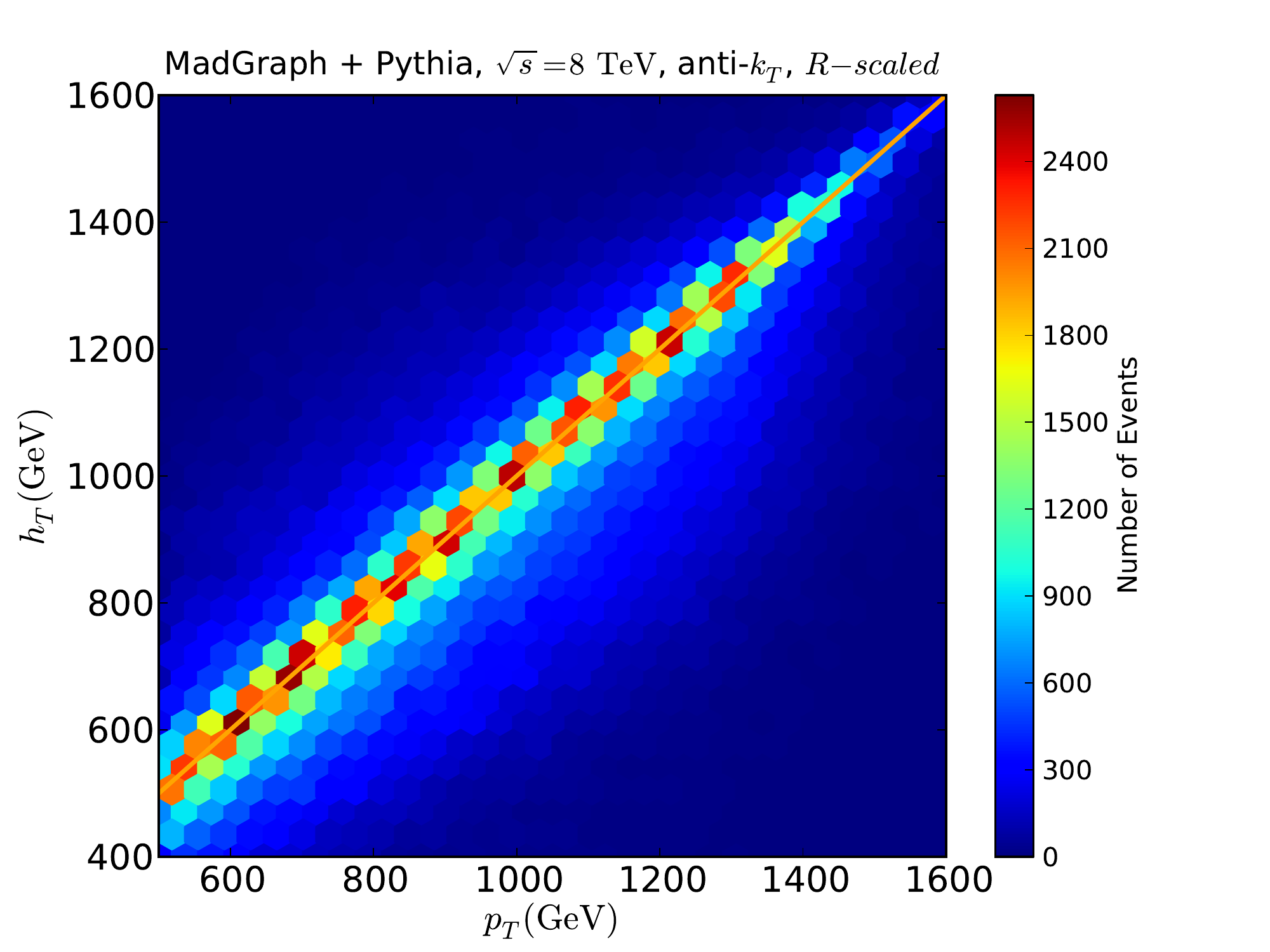}& \includegraphics[width=3.3in]{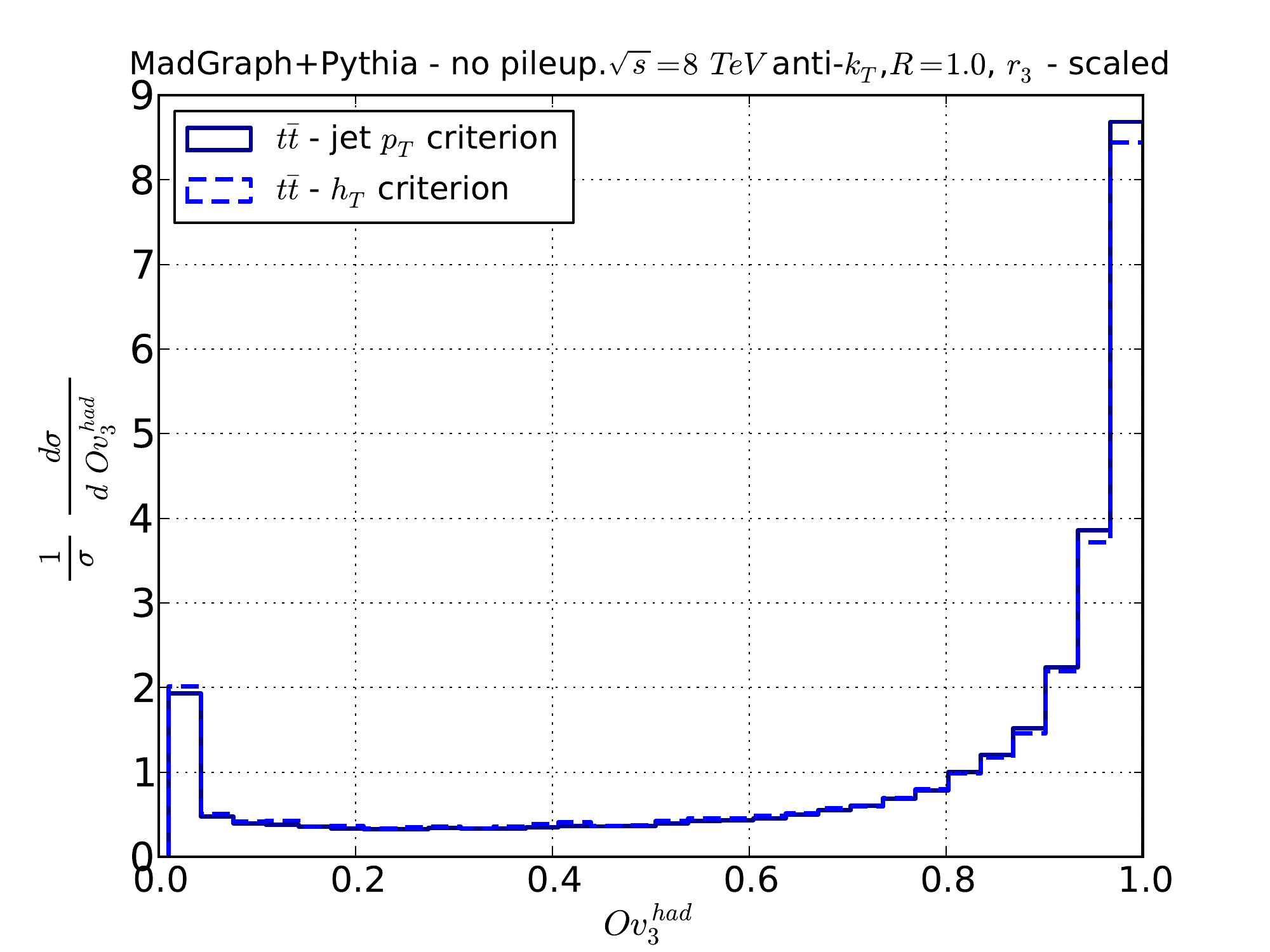}
\end{tabular}
\caption{The left panel shows the correlation between the fat jet $p_T$ and the scalar $p_T$ sum of Eq. \eqref{eq:ht}. The right plot shows the hadronic $Ov_3^{had}$ distributions using fat jet $p_T$ to select template bins (solid, dark blue) and $h_T$ (dashed, blue), where we excluded the asymmetric events discussed in Section \ref{sec:NLO}. Both plots assume Basic Cuts of Eq. \eqref{eq:bc}.}
\label{fig:pt_ht_corr}
\end{center}
\end{figure}

In our current work we opt not to use the fat jet transverse momentum as the estimator of template $p_T$  because of the susceptibility of jet $p_T$ to pileup.  Instead, we define the observable
\be
	h_T = \sum_{i = \ell, b, \nu} p^i_T\,, \label{eq:ht}
\ee
where $p^i_T$ is the transverse momentum of the leptonic top components (\textit{i.e.} the hardest lepton outside the fat jet with mini-ISO$>0.95$, the hardest anti-$k_T$, $r=0.4$ jet within $R=1.5$ from the lepton and the total $ \sl{E_T}\,\,\, $).  $h_T$ is correlated with the fat jet $p_T$, especially in events in which tops decay back to back. Figure \ref{fig:pt_ht_corr} shows an example. 
The high degree of correlation between $h_T$ and $p_T$ of the fat jet allows us to replace the criterion for template set selection based on pileup sensitive jet $p_T$ to a more pileup robust $h_T$. Notice that the $Ov_3^{had}$ distribution in Fig. \ref{fig:pt_ht_corr} remains unaffected by the choice of the template selection rule. 

The template $p_T$ estimation based on the $h_T$ of the leptonic top provides an additional discriminant of asymmetric $t\bar{t}$ events we discussed in Section~\ref{sec:NLO}. If an event is characterized by a large $A^{SV}_{t\bar{t}},$ the $h_T$ of the leptonic top will often not match the $p_T$ of the fat jet, even in the cases where the hadronic top jet is correctly pre-selected. The resulting peak overlap score will thus tend to small values, due to the mismatch between $p_T$ of the template and the transverse momentum of the hardest fat jet.

\subsection{Effects of MET Resolution on Template Selection Criteria} \label{app:MET}

So far we have taken the simplified approach to estimating $\MET$, where we assumed that the missing transverse momentum was simply a sum of the transverse components of 
all the neutrino four momenta in an event. Here we explore the effects of properly reconstructing the missing energy and the $\MET$ resolution on the TOM analysis. 

We follow the ATLAS prescription of Ref.~\cite{Aad:2012re} 
for reconstruction and smearing of missing energy. We begin by calculating the missing energy as the sum of $x$ and $y$ components of the clusters
\be
	E_{x,y} = \sum_i E_{x,y}^i\,,
\ee
where the sum over $i$ goes over all the final state particles in the event which are not neutrinos and satisfy the following criteria:
\ba 
	\eta_i < 5.0, & \,\,\,\,\,\,p_T^i > 0.8 \GeV\,.
	\label{eq:met_particles}
\ea
The requirement on pseudo-rapidity guarantees that the particle does not end up down the beam line, while the $p_T$ requirement is necessary to take into account the effects of charged particle's track bending in the strong magnetic field. 

Next, we smear $E_{x,y}$ individually by drawing a random number from a gaussian centered at $E_{x,y}$ with a width given by the $\MET$ resolution 
\be
	\sigma_{\MET} = 0.7 \sqrt{\sum_i \frac{E_T^i}{\GeV}} \GeV\,,
\ee
where $i$ runs over the non-neutrino event constituents satisfying the requirements of Eq. \ref{eq:met_particles}.

Finally, we calculate the missing energy from the smeared $E'_{x,y}$ as
\be
	\MET = \sqrt{E_x^{'2} + E_y^{'2}}\,.
\ee

\begin{figure}[htb]
\begin{center}
\begin{tabular}{cc}
\includegraphics[width=3.3in]{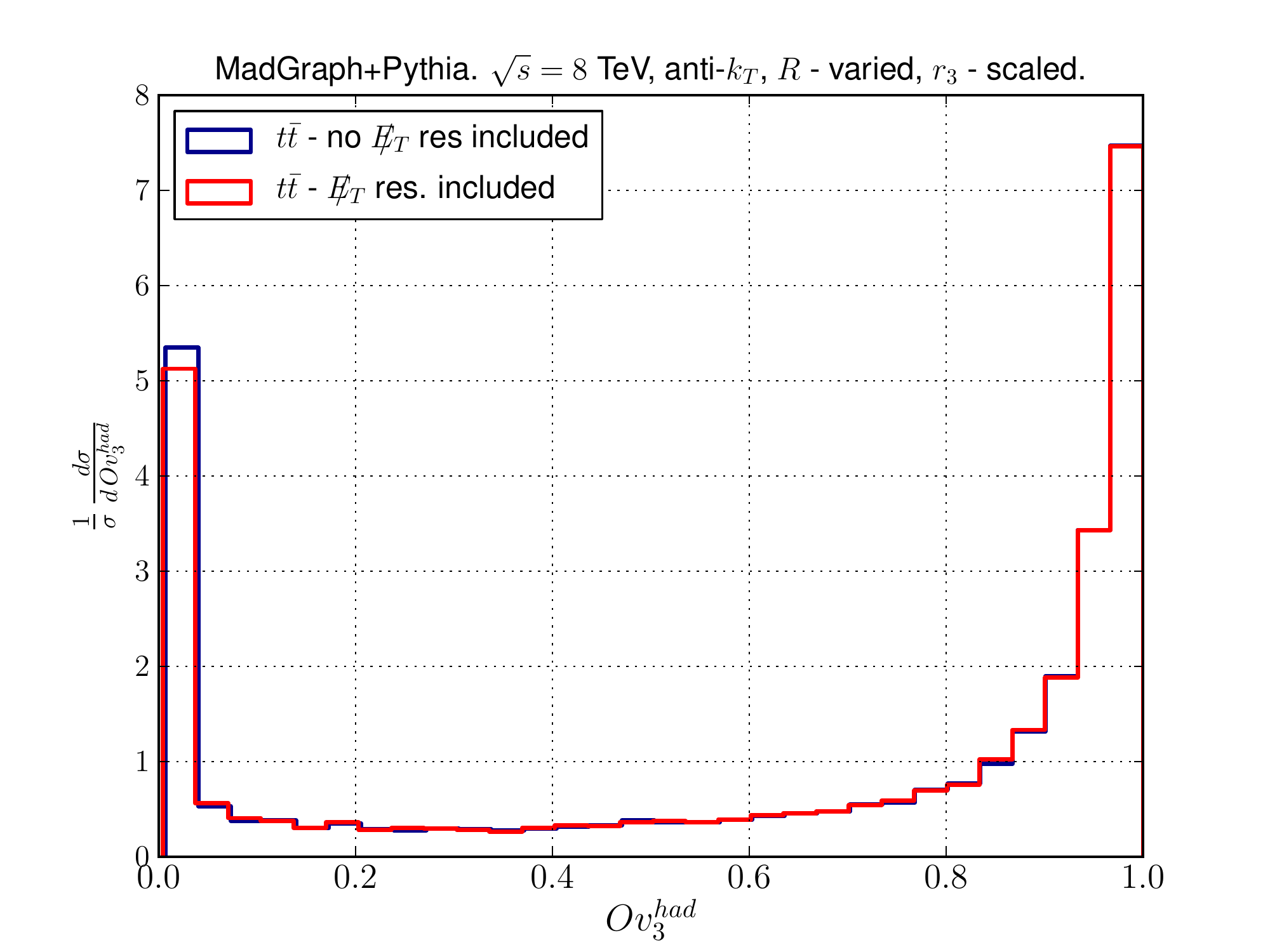}& \includegraphics[width=3.3in]{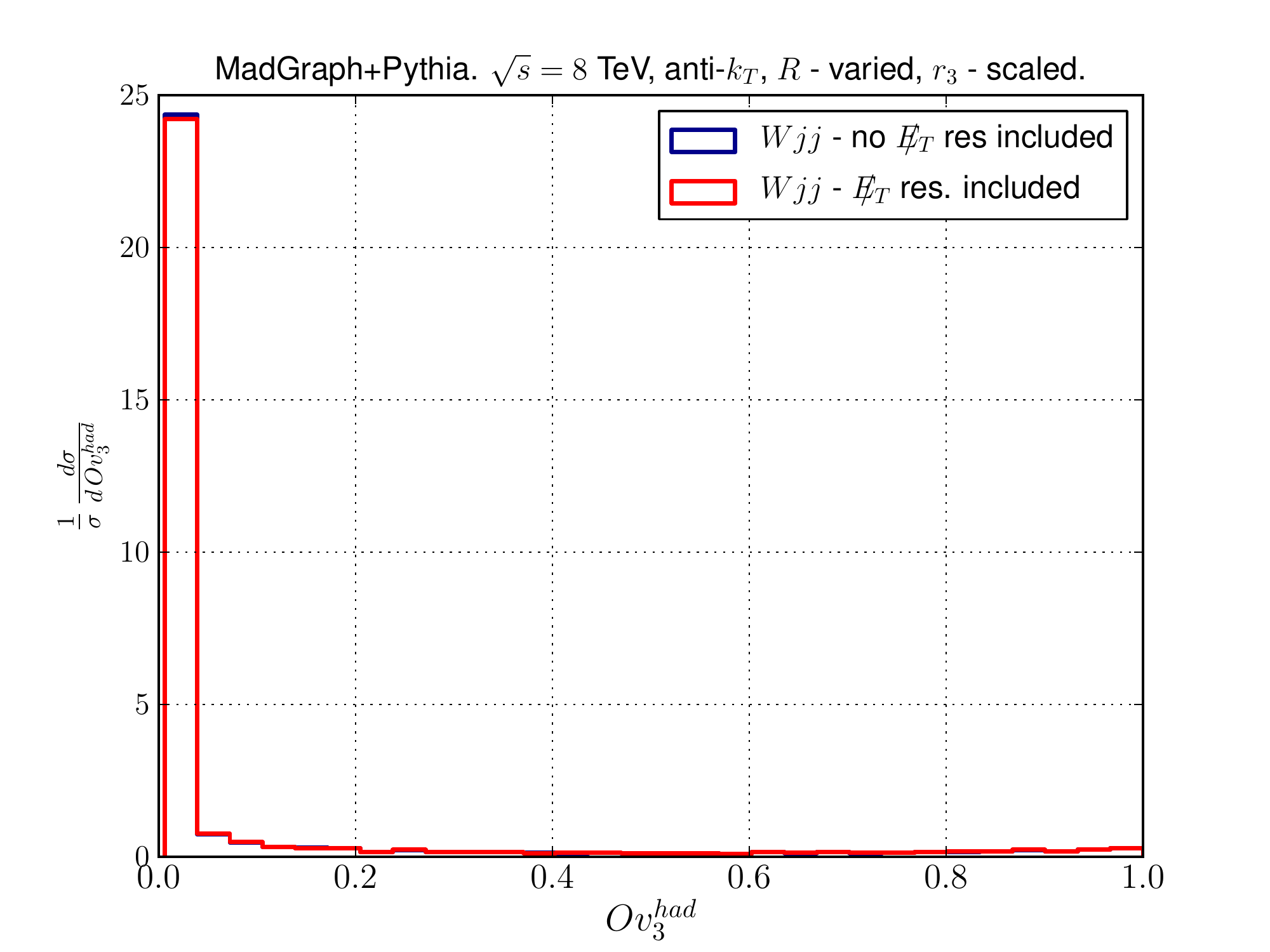}
\end{tabular}
\caption{Effects of missing energy resolution on $Ov_3^{had}$. The left panel shows the $Ov_3^{had}$ distribution for $t\bar{t}$ events with $500 \GeV < p_T < 600 \GeV.$ Right panel shows the same for $W$+jets. All events assume the Basic Cuts of Eq. \eqref{eq:bc}.   }
\label{fig:Ov3_missEt}
\end{center}
\end{figure}

Figure \ref{fig:Ov3_missEt} illustrates the effects of missing energy resolution on the results of the overlap analysis. The left panel shows the $Ov_3^{had}$ distributions for $t\bar{t}$, with the corresponding $W$+jets distributions on the right panel. For the purpose of illustration,  we include only events with fat jet $p_T$ between $500 \GeV$ and $600 \GeV$. Our results show that the hadronic overlap is nearly un-affected by missing energy resolution. 

\begin{figure}[htb]
\begin{center}
\begin{tabular}{cc}
\includegraphics[width=3.3in]{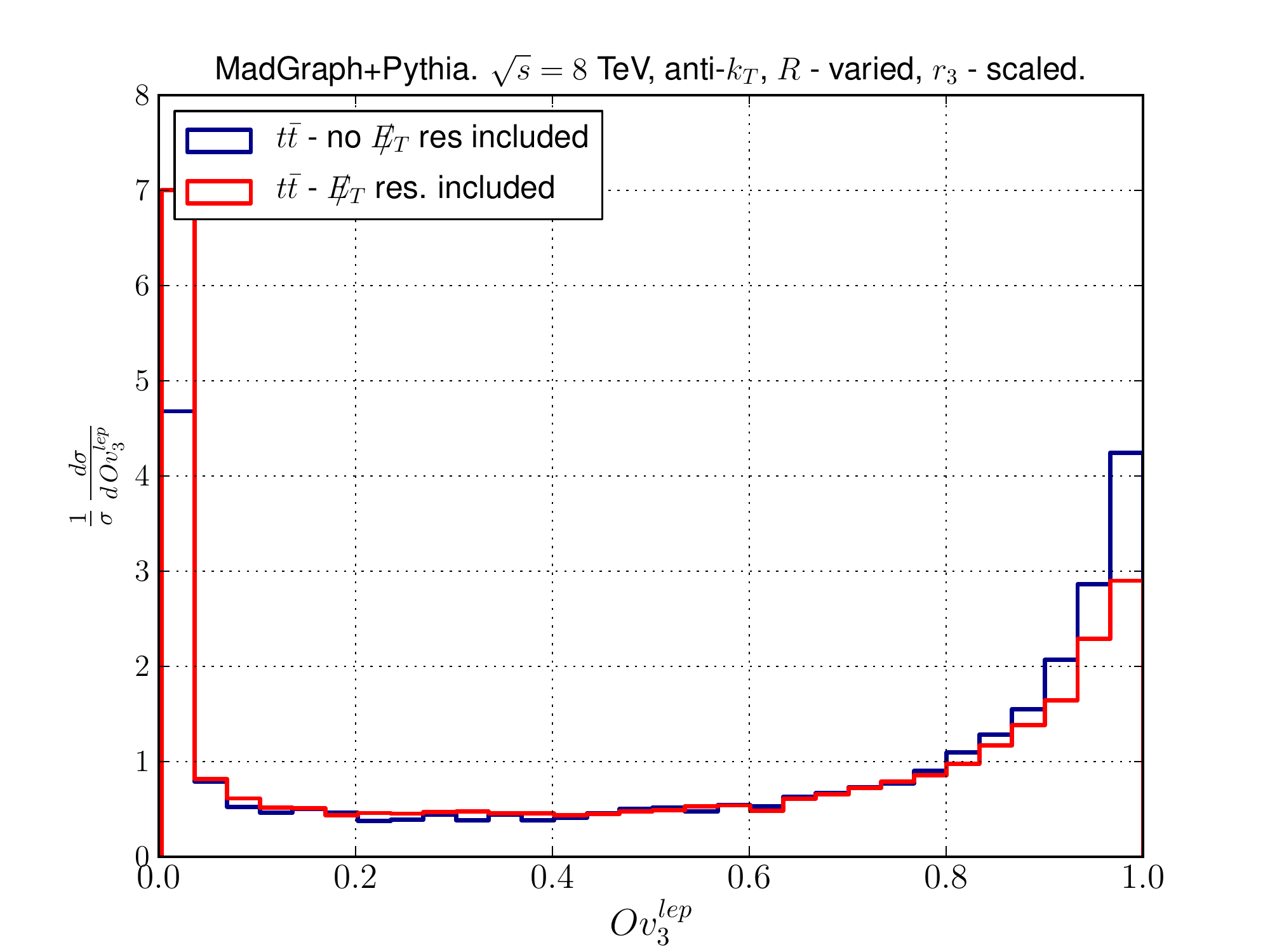}& \includegraphics[width=3.3in]{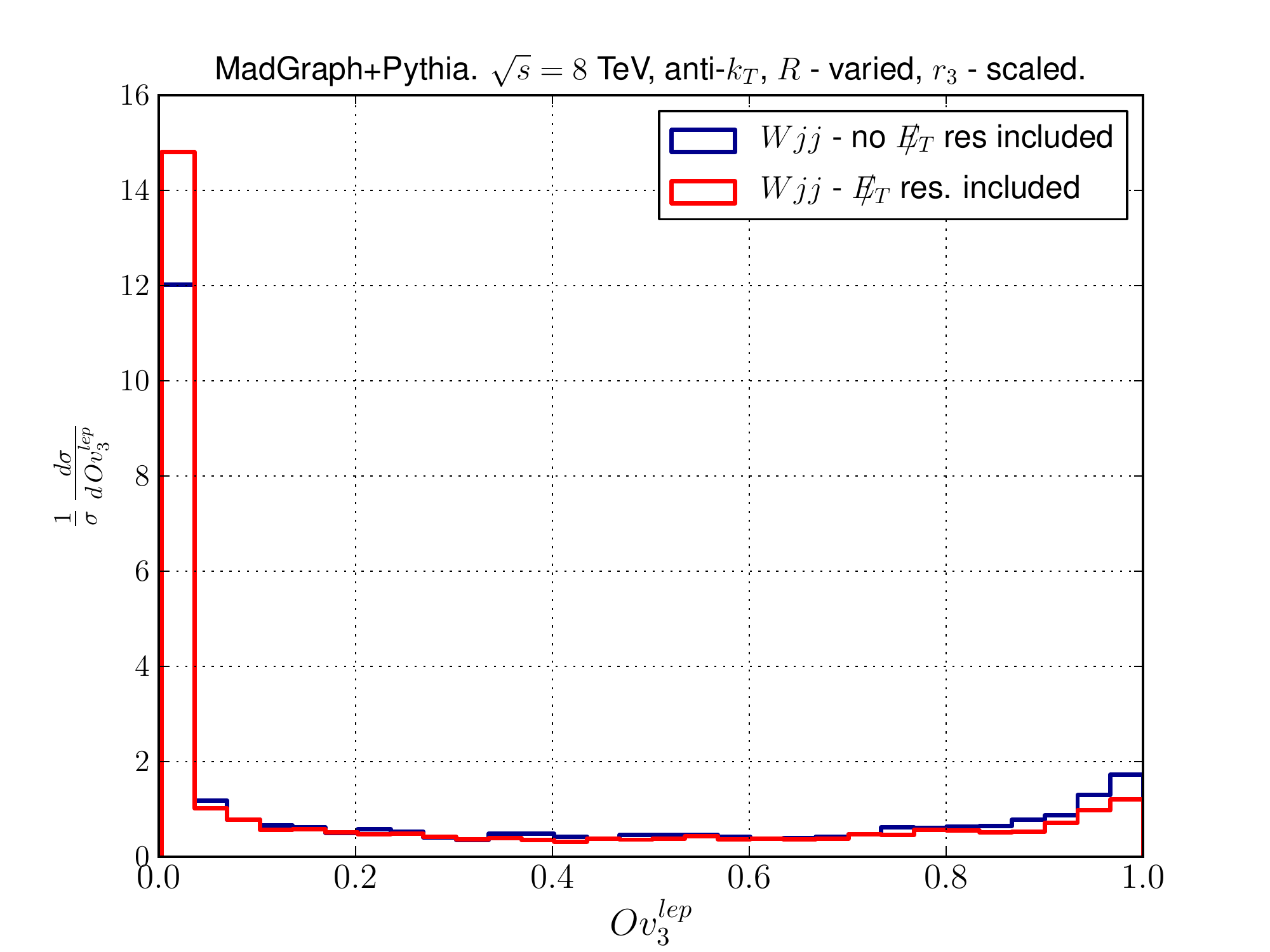}
\end{tabular}
\caption{Effects of missing energy resolution on $Ov_3^{lep}$. The left panel shows the $Ov_3^{lep}$ distribution for $t\bar{t}$ events with $500 \GeV < p_T < 600 \GeV.$ Right panel shows the same thing for $W$+jets. All events assume the Basic Cuts of Eq. \eqref{eq:bc}.   }
\label{fig:Ov3_missEtOv3l}
\end{center}
\end{figure}

Leptonic overlap shows somewhat more pronounced susceptibility to missing energy resolution. Figure \ref{fig:Ov3_missEtOv3l} shows an example distribution for $t\bar{t}$ and $W$+jets.  We find that $Ov_3^{lep}$ distributions are shifted slightly towards lower values of overlap if smearing of missing energy is included. This is not surprising, given that the missing energy goes directly into the computation of $Ov_3^{lep}$. Nonetheless, the effects are small enough to be concerning for the overall performance of the overlap analysis. 

\subsection{ Determining the Adequate Number of Templates}

How is the template analysis affected by the number of templates used in the calculation? In the limit of $N_{\rm templates} \rightarrow~\infty,$ we expect a perfect coverage of the top decay phase space. However, technical difficulties and processing time limit us to a finite number of template states,  where a large number of templates (typically of $\cOO(10^6)$) is declared ``adequate.'' So far, there has been no detailed analysis on the actual sensitivity of TOM to the number of used template states, as the problem requires a dedicated study at several template transverse momenta and various number of templates. For instance, a typical case covering a template range of $p_T = 500 - 1500 \GeV, $ in steps of $100 \GeV$ and $5$ different template sets (with different number of steps in angular variables,  $N_{\eta, \phi}$) would require $50$ runs of both the signal and background channels. 

In order to test the dependence of overlap results on the number of templates used, we vary the number of steps in $\eta, \phi$ used to generate templates from $50$ to $90$ steps in each, in increments on $10$ steps. This method gives template sets with roughly twice as many templates in each consecutive case. In the interest of time, we consider only two template $p_T$ values
\ba
{\rm Case\, 1}: &  \,\,\,\,\,\,\,500 \GeV < p_T <600 \GeV\,, \nn \\
{\rm Case\, 2}: & \,\,\,\,\,\,\,1400 \GeV <p_T <1500 \GeV\,. \nonumber
\ea
This particular choice of case studies looks at the extrema of the $p_T$ range of interest for the boosted top analyses of the near future and to first approximation we will consider the results valid for in-between values of template $p_T$.

\begin{figure}[htb]
\begin{center}
\begin{tabular}{cc}
\includegraphics[width=2.3in]{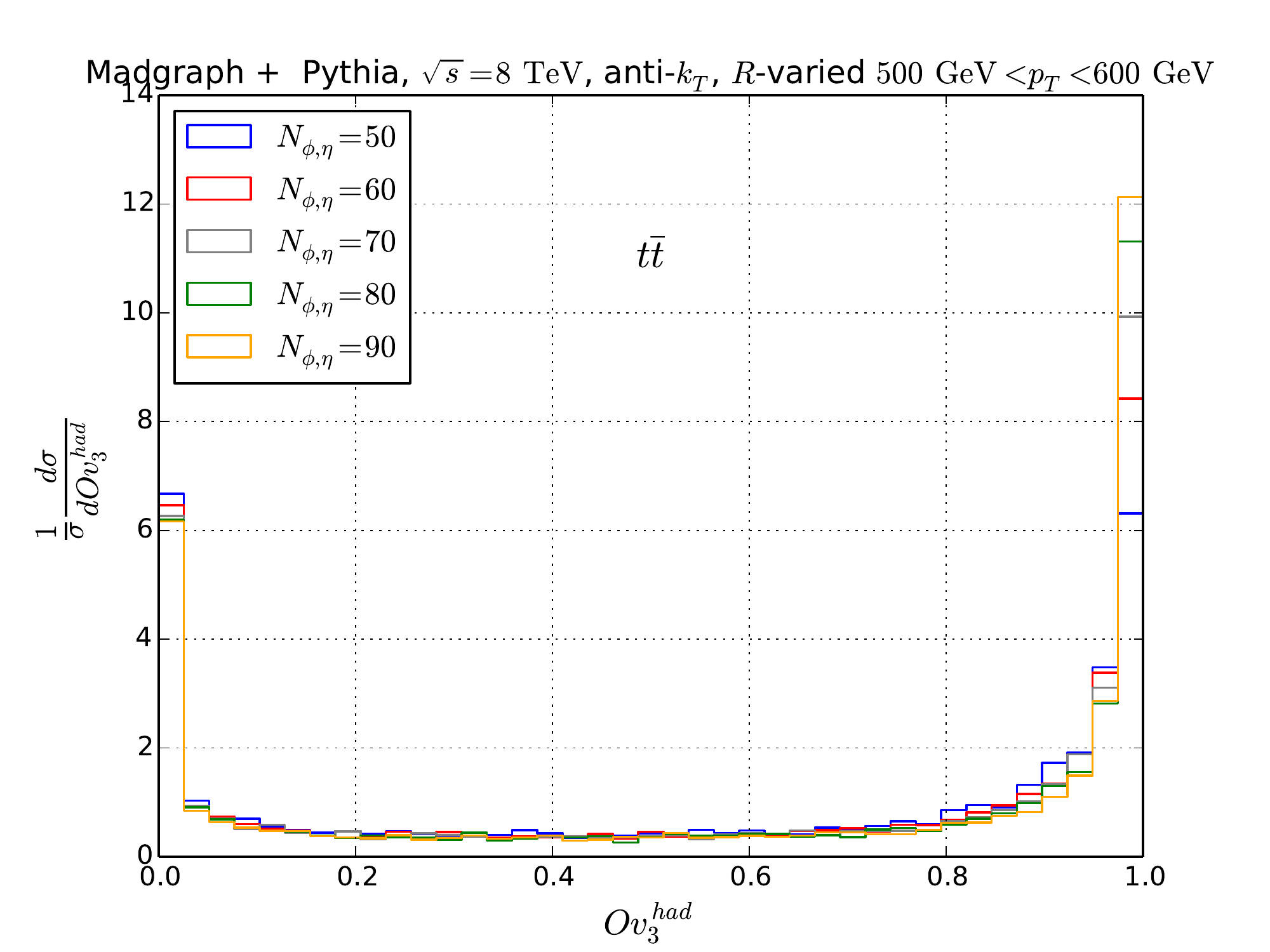}\includegraphics[width=2.3in]{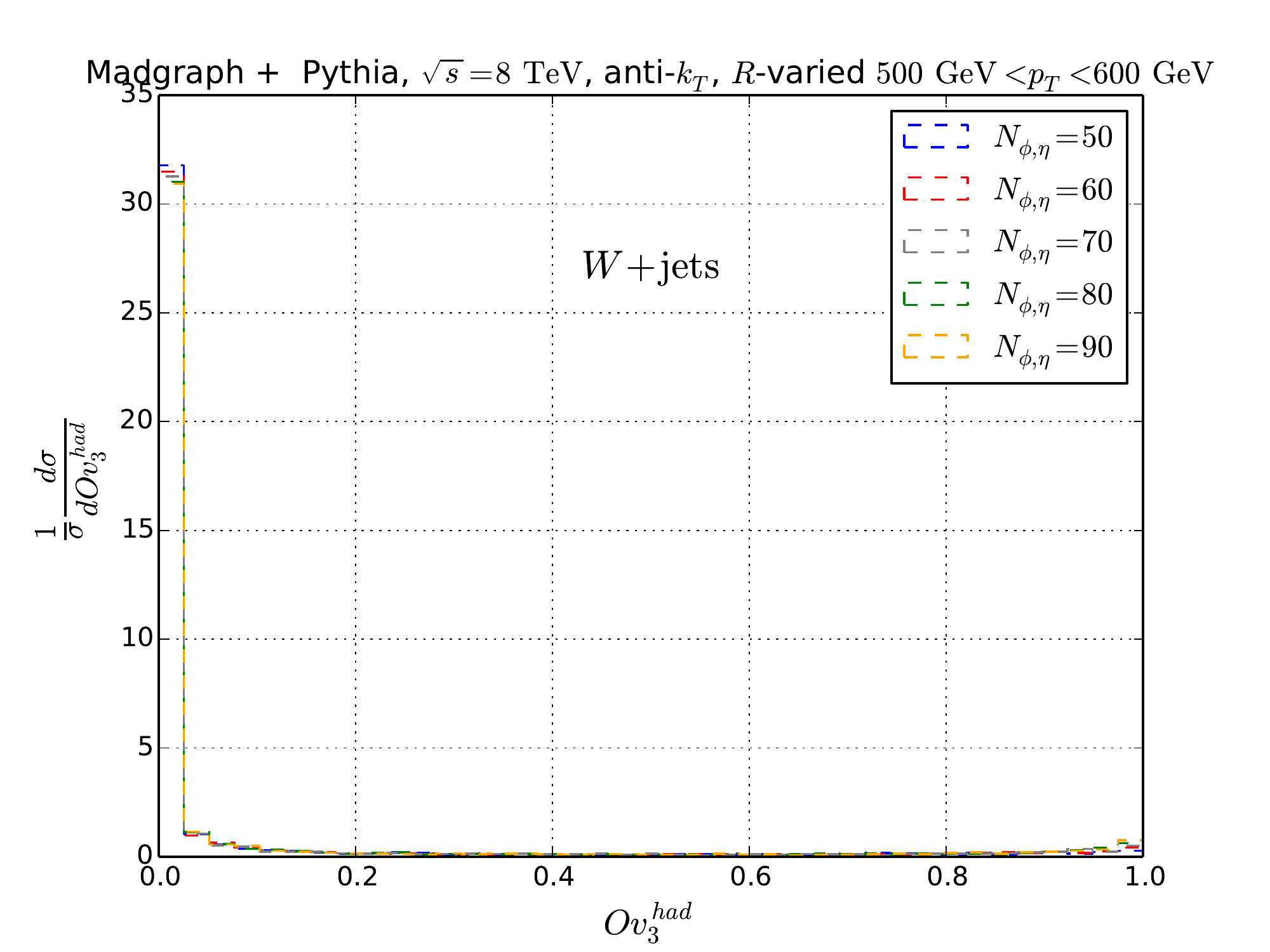} \includegraphics[width=2.3in]{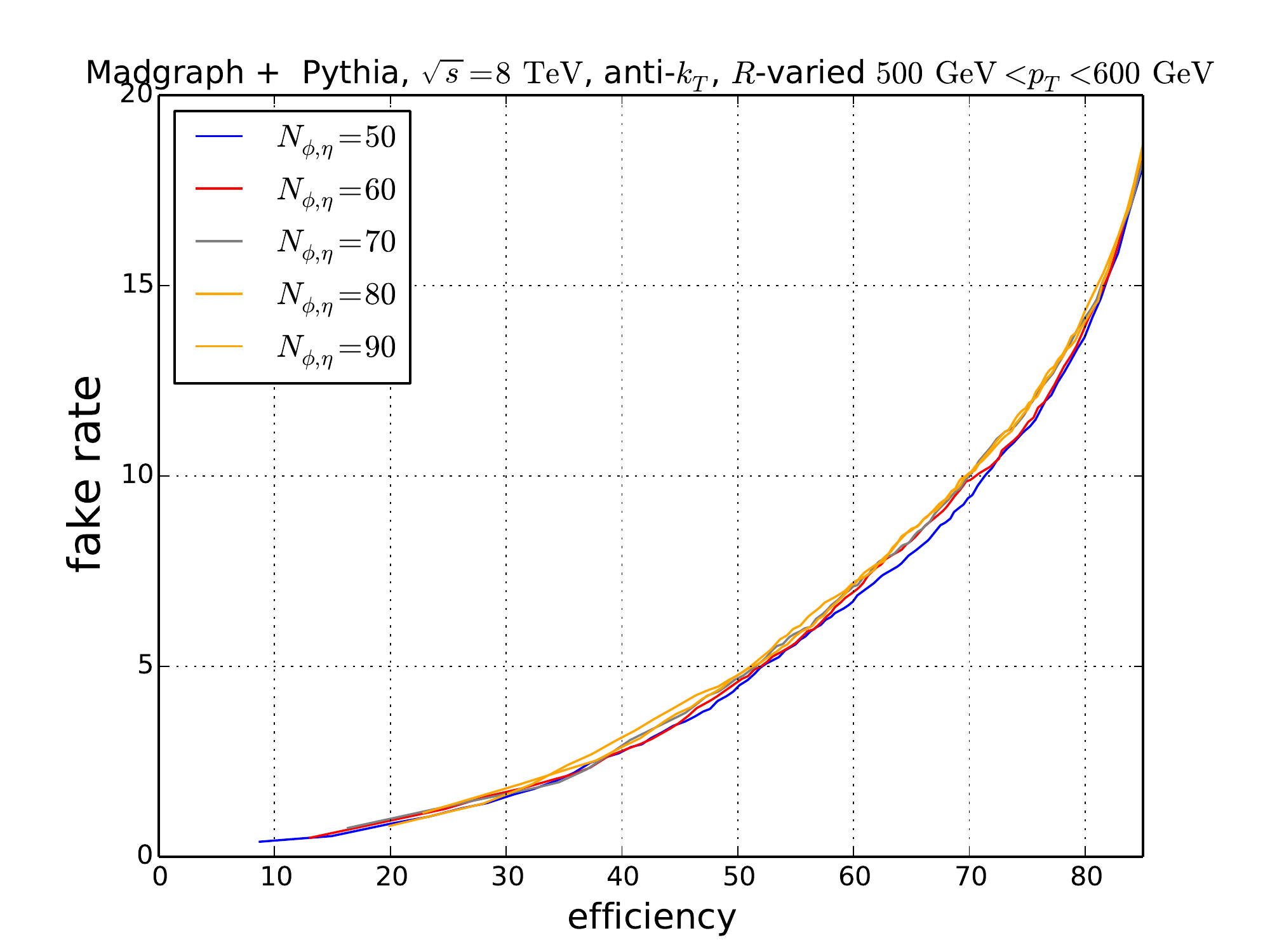}\\
\includegraphics[width=2.3in]{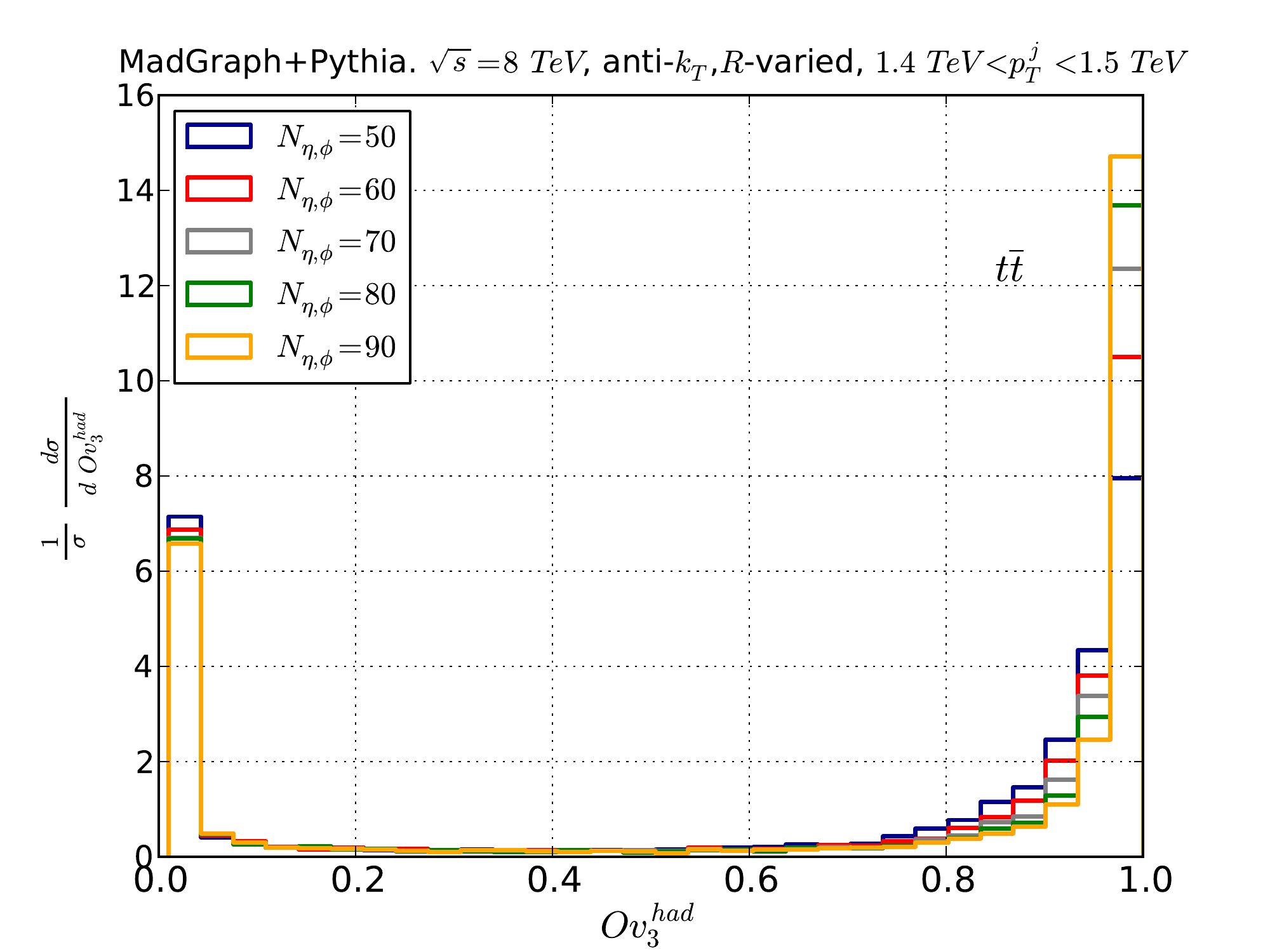}\includegraphics[width=2.3in]{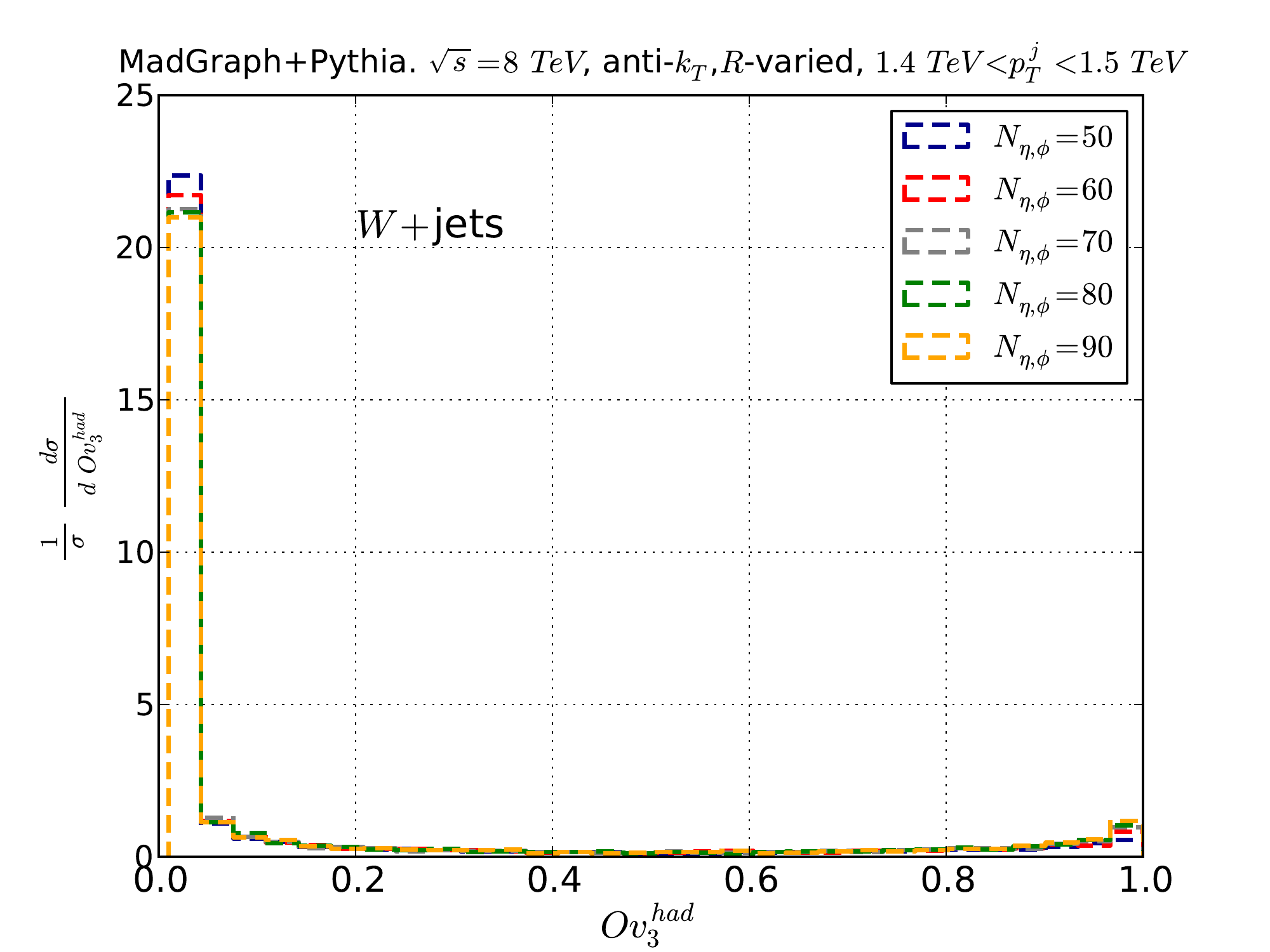} \includegraphics[width=2.3in]{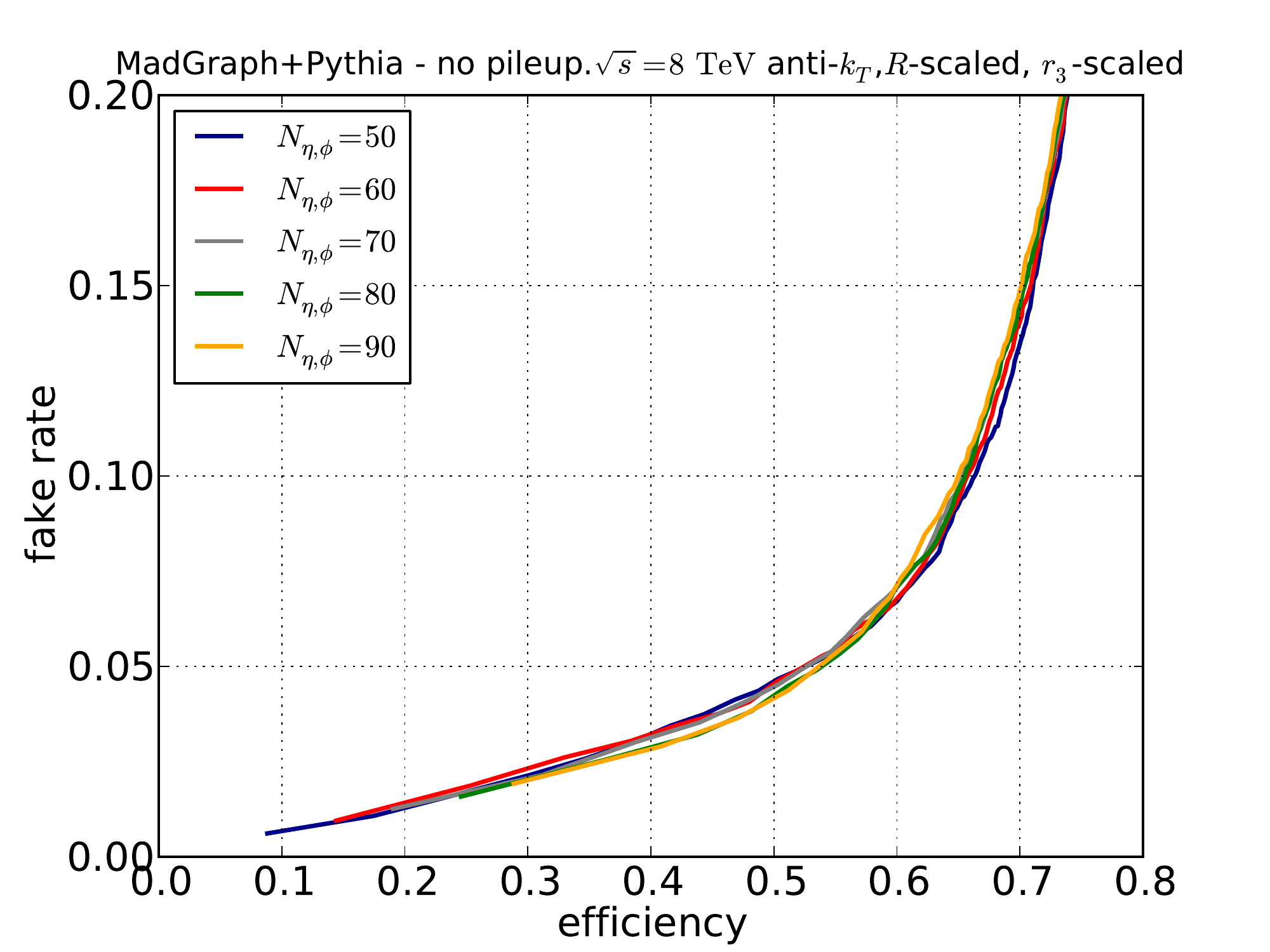}

\end{tabular}
\caption{Effects of varying the number of steps in $\eta, \phi$ during template generation on the TOM analysis. Top panels are for $500 \GeV < p_T < 600 \GeV$, while bottom plots are for $1400 \GeV < p_T < 1500 \GeV$. The left plots show the hadronic peak overlap distributions for the signal and $N_{\phi, \eta} = 50 - 90$. The plot in the middle shows the same for the $W$+jets background. The right panel shows the signal efficiency relative to the Basic Cuts of Eq. \eqref{eq:bc} vs the background fake rate. }
\label{fig:nsteps}
\end{center}
\end{figure}

Figure~\ref{fig:nsteps} shows the result. The background fake rate as a function of signal efficiency remains unaffected for all considered cases, signaling that even $N_{\eta, \phi} = 50$ adequately covers the top decay phase space. The effects of varying the number of templates are noticeable only in the high overlap region of the signal distribution where adding more templates naturally improves the resolution of subjets and thus slightly improves the peak overlap score . Notice however that the region of low overlap (i.e. $Ov^{had}_3 < 0.7$) remains unaffected, hence verifying that increasing the number of templates does not lead to an increase in the mis-tag rate of events which do not match the topology and substructure of a boosted semi-leptonic $t\bar{t}$ decay.

 \bibliography{draft}
 
\end{document}